\documentclass[12pt,letterpaper]{article}
\usepackage[T1]{fontenc}
\usepackage[margin=1in]{geometry}
\usepackage{amsmath,amssymb,amsthm,mathtools}
\usepackage{txfonts}
\usepackage{microtype}
\usepackage{xcolor,graphicx,booktabs,array}
\usepackage{tikz,pgfplots}
\usepackage{caption}
\usepackage{needspace,placeins}
\usepackage[numbers,sort&compress]{natbib}
\usepackage[colorlinks=true,allcolors=blue!65!black]{hyperref}
\usetikzlibrary{arrows.meta,calc,positioning,decorations.pathreplacing}
\usepgfplotslibrary{groupplots}
\pgfplotsset{compat=1.18}
\definecolor{erasureBlue}{RGB}{33,92,141}
\definecolor{erasureTeal}{RGB}{24,117,107}
\definecolor{erasureOrange}{RGB}{186,98,31}
\definecolor{erasureGray}{RGB}{100,106,112}

\hypersetup{pdftitle={Second-order asymptotics for the quantum and private capacities of the erasure channel},pdfauthor={Mark M. Wilde}}
\numberwithin{equation}{section}
\newtheorem{theorem}{Theorem}[section]
\newtheorem{lemma}[theorem]{Lemma}
\newtheorem{proposition}[theorem]{Proposition}
\newtheorem{corollary}[theorem]{Corollary}
\theoremstyle{definition}

\theoremstyle{remark}

\newcommand{\cE}{\mathcal E}
\newcommand{\cC}{\mathcal C}
\newcommand{\cD}{\mathcal D}
\newcommand{\cH}{\mathcal H}
\newcommand{\cT}{\mathcal T}
\newcommand{\bbC}{\mathbb C}
\newcommand{\bbR}{\mathbb R}
\newcommand{\bbN}{\mathbb N}
\newcommand{\bbZ}{\mathbb Z}
\newcommand{\E}{\mathbb E}
\newcommand{\Prb}{\operatorname{Pr}}
\newcommand{\Tr}{\operatorname{Tr}}
\newcommand{\id}{\operatorname{id}}

\newcommand{\Bin}{\operatorname{Bin}}
\newcommand{\Ner}{K}
\newcommand{\row}{\mathrm{row}}
\newcommand{\et}{\mathrm{et}}
\newcommand{\eg}{\mathrm{eg}}
\newcommand{\opt}{\mathrm{opt}}
\newcommand{\marg}{\mathrm{marg}}
\newcommand{\one}{\mathbf 1}
\newcommand{\eps}{\varepsilon}
\newcommand{\norm}[1]{\left\|#1\right\|}
\newcommand{\ket}[1]{\lvert #1\rangle}
\newcommand{\bra}[1]{\langle #1\rvert}
\newcommand{\braket}[2]{\langle #1\vert #2\rangle}

\newcommand{\PhiM}{\Phi_{R\widehat S}^{M}}
\allowdisplaybreaks[2]

\AtBeginDocument{%
  \setlength{\abovedisplayskip}{8pt plus 3pt minus 4pt}%
  \setlength{\belowdisplayskip}{8pt plus 3pt minus 4pt}%
  \setlength{\belowdisplayshortskip}{5pt plus 3pt minus 3pt}%
}

\begin{document}
\thispagestyle{plain}
\vspace*{0.35in}
\begin{center}
{\LARGE\bfseries Second-order asymptotics for the quantum\\[3pt]and private capacities of the erasure channel\par}
\vspace{1.3em}
{\large Mark M. Wilde\par}
\vspace{0.6em}
{\itshape School of Electrical and Computer Engineering, Cornell University\\
Ithaca, New York 14853, USA\par}
\vspace{1.2em}
{\large November 24, 2026\par}
\end{center}
\vspace{1em}
\begin{abstract}
An erasure channel either delivers a quantum system intact or tells the receiver that it was lost. How much quantum information, or secret classical information, can a finite block transmit without assistance? We determine the optimal second-order coding rates for input dimension $d\ge2$, erasure probability $0<p<1/2$, and fixed infidelity (one minus fidelity) $0<\varepsilon<1$. For $n$ channel uses, the optimal base-two logarithmic code size is
\begin{equation}\tag{A}\label{eq:abstract-rate}
n(1-2p)\log_2d
+2\sqrt{np(1-p)}\log_2d\,\Phi^{-1}(\varepsilon)
+O((\ln n)^2),
\end{equation}
where $\Phi$ is the standard normal cumulative distribution function and $\Phi^{-1}$ is its quantile function. The square-root term is the second-order correction. This formula holds for entanglement transmission, entanglement generation, and private classical communication. Quantum error measures failure to preserve entanglement. Private error compares the actual output with a correctly decoded uniform message independent of an optimized environment state. The achievable Gaussian term and a matching quantum expansion below infidelity one half were established recently. We prove matching upper bounds for every fixed infidelity, including private communication, with the displayed remainder. The proof bounds one code's average performance across all loss patterns of a given size; binomial fluctuations then determine the Gaussian correction. It combines tests of correct and secret communication with tensor-product averaging and a classical polynomial identity. Arbitrary mixed encoders and collective decoders are allowed. Requiring the private error bound for every message preserves the Gaussian term. Fixing the ideal environment state to the actual marginal instead changes the quantile to $\Phi^{-1}(1-\sqrt{1-\varepsilon})$. Thus the precise privacy criterion affects communication at second order.
\end{abstract}

\clearpage
\begingroup
\fontsize{10}{11}\selectfont
\tableofcontents
\endgroup

\section{Introduction}\label{sec:intro}

\subsection{Background and motivation}
The goal of quantum communication is to preserve an unknown state, while that for private classical communication is to convey a message without revealing it to the environment of a quantum channel. The erasure channel makes their comparison particularly direct: the receiver knows which systems arrived, and the environment receives the missing ones. One code must protect its information against many possible loss patterns.

For input dimension $d$ and erasure probability $p$, the unassisted quantum and private capacities of the erasure channel $\cE_{p,d}$ are
\begin{equation}\label{eq:capacity-intro}
Q(\cE_{p,d})=P(\cE_{p,d})=\max\{1-2p,0\}\log_2d.
\end{equation}
The quantum capacity was established in Ref.~\cite{BDS1997}. For $p\le1/2$, the channel is degradable: its environment output can be simulated from its receiver output. The coding theorem and coherent-information characterization then give the private-capacity equality~\cite{Devetak2005,Smith2008}. For $p\ge1/2$, the environment can simulate the receiver; this antidegradability forces the private capacity to zero. These capacities describe vanishing-error communication, but do not specify the finite-block correction at a prescribed nonzero error.

For $n$ channel uses, let $M$ be the protected quantum dimension or the number of private messages, so $\log_2M$ measures the information carried by the code. Achievable bounds give sizes attained by some code; converse bounds limit the size of every code. The erasure count $\Ner$ (i.e., number of erasures for $n$ transmissions) is binomial with mean $np$ and variance $np(1-p)$. Each loss removes one system from the receiver and gives one to the environment, suggesting the balance
\begin{equation}\label{eq:dimension-balance}
\log_dM\approx(n-\Ner)-\Ner=n-2\Ner.
\end{equation}
The standard deviation of $(n-2\Ner)\log_2d$ is $2\sqrt{np(1-p)}\log_2d$, suggesting a Gaussian correction. The obstacle is to justify \eqref{eq:dimension-balance} for an arbitrary code: a code can place all its information in positions surviving one chosen pattern. The converse must constrain its average performance across many patterns.

Strong converses force the relevant fidelity to vanish at rates strictly above capacity. Earlier quantum results include the pretty strong converse of Morgan and Winter~\cite{MorganWinter2014} and a random-code erasure-channel strong converse~\cite{WildeWinter2014}. For degradable and antidegradable channels, Kondra et al. proved a strong converse~\cite{Kondra2026}, and Wilde proved an exponential private strong converse~\cite{WildePrivate2026}. For general finite-dimensional memoryless channels, Beigi and Tomamichel proved exponential strong converses at the regularized quantum, classical, and private capacities~\cite{BeigiTomamichel2026}; Cheng and Tomamichel established them at the regularized quantum and classical capacities~\cite{ChengTomamichel2026}. Cheng, Hirche, and Tomamichel characterize the exact strong-converse exponent for joint private fidelity over general finite-dimensional wiretap channels~\cite[Theorem~2.6]{CHT2026}. These fixed-rate results leave open the transition on a rate interval of width $1/\sqrt n$.

Second-order quantum achievability was studied by Beigi, Datta, and Leditzky~\cite[Theorem~3.5]{BDL2016} and Tomamichel, Berta, and Renes~\cite[Theorem~6]{TBR2016}. The former directly concerns entanglement transmission; the conversion from entanglement generation to entanglement transmission in the latter's general argument is discussed in Ref.~\cite[Section~6.3]{CanWilde2026}. Let $\eps\in(0,1)$ be the tolerated infidelity and $\Phi$ the standard normal cumulative distribution function. Can and Wilde obtain the erasure-channel expansion
\begin{equation}\label{eq:gaussian-intro}
n(1-2p)\log_2d+2\sqrt{np(1-p)}\log_2d\,\Phi^{-1}(\eps)-O(1)
\end{equation}
with a constant-order achievable remainder~\cite[Section~6.5, Eq.~(6.24)]{CanWilde2026}. Their maximally entangled input also gives an entanglement transmission code~\cite[Corollary~4.3]{CanWilde2026}.

Cheng proved the matching unassisted quantum expansion for entanglement generation and entanglement transmission when $0<\eps<1/2$, with an $o(\sqrt n)$ remainder~\cite[Theorem~4.1 and Eq.~(78)]{ChengDiscrimination2026}, by pairing erasure counts across the coding threshold. Our contribution is a private converse for every $0<\eps<1$, with an $O((\ln n)^2)$ remainder. Lemma~\ref{lem:q-to-p} transfers it at the same error to entanglement transmission; the Bell tests in Lemma~\ref{lem:lift} give the entanglement generation converse. Both sharpen the remainder in the previously covered range.

The private error criterion is consequential. The joint requirement of correct decoding and secrecy compares the message, estimate, and environment with a correctly decoded message independent of one common environment state. Ref.~\cite[Section~4.1, Eq.~(4.3)]{WildePosition2017} uses trace distance. We use optimized joint fidelity, as in Refs.~\cite[Section~3.1, Eq.~(3.3)]{WTB2017} and~\cite[Section~2.1]{WildePrivate2026}, defined in \eqref{eq:private-opt}. Entanglement transmission then gives private communication at the same dimension and error~\cite[Section~3.4]{WTB2017}; Lemma~\ref{lem:q-to-p} states the exact comparison. Fixing the ideal environment to the actual marginal, as in Ref.~\cite[Appendix~B, Definition~27, Eq.~(B.4)]{WTB2017}, gives a different second-order quantile. These joint-fidelity expansions do not identify same-parameter rates under separate decoding and trace-distance secrecy constraints.

Classical communication assistance changes the resource model. For qubits, the assisted quantum~\cite[Result~2]{TBR2016} and private~\cite[Proposition~23]{WTB2017} expansions have first-order coefficient $1-p$. Our unassisted coefficient is $1-2p$ for $p<1/2$.

The quantum Singleton bound~\cite{KL1997} gives $\log_dM\le n-2k$ for exact correction of \emph{every} size-$k$ erasure pattern, where $0\le k\le n/2$; it need not hold for one chosen pattern. Grassl, Huber, and Winter study high-fidelity recovery for most fixed-size patterns, with continuity terms depending on blocklength and error~\cite[Section~VI]{GHW2022}. Related work constrains the probabilities of exact recovery after erasure~\cite{ALBSK2026} and studies average-subset monogamy in channel simulation~\cite{AhmedSmith2026}. Reliability below capacity for prime input dimension is studied in Ref.~\cite{MaoFang2026}. Here we require a bound for arbitrary codes, including large prescribed errors: exceeding $n-2k$ by $O((\ln n)^2)$ must force the average joint fidelity to vanish.

\subsection{Summary of results}
For fixed $d\ge2$, $0<p<1/2$, and $0<\eps<1$, define the capacity $C$ and variance $V$ by
\begin{equation}\label{eq:qv-intro}
C\coloneqq(1-2p)\log_2d,
\qquad
V\coloneqq4p(1-p)(\log_2d)^2.
\end{equation}
Let $M^*(n,\eps)$ be the largest code size for entanglement transmission, entanglement generation, or optimized joint private communication at infidelity at most $\eps$, as defined in \eqref{eq:operational-M} and \eqref{eq:operational-P}. Each task obeys
\begin{equation}\label{eq:main-intro}
nC+\sqrt{nV}\,\Phi^{-1}(\eps)-O(1)
\ \le\ \log_2M^*(n,\eps)
\ \le\ nC+\sqrt{nV}\,\Phi^{-1}(\eps)+O((\ln n)^2).
\end{equation}
The maximal-message private criterion has the same Gaussian coefficient and an $O(\log n)$ achievable loss. Fixing the ideal environment to the actual marginal instead gives the quantile
\begin{equation}\label{eq:marg-quantile-intro}
\Phi^{-1}\!\left(1-\sqrt{1-\eps}\right).
\end{equation}
Theorems~\ref{thm:main} and~\ref{thm:other-criteria} state the results precisely. Thus agreeing vanishing-error capacities need not imply agreeing fixed-error second-order rates.

The main estimate used for establishing our results is an averaged Singleton bound for privacy tests. Let $\overline f_{n,k}$ be the optimized private fidelity averaged uniformly over size-$k$ erased sets. Fix $p<p_0<1/2$. Theorem~\ref{thm:fixed} constructs $h_n=O((\ln n)^2)$ such that
\begin{equation}\label{eq:fixed-intro}
\log_dM\ge n-2k+h_n
\quad\Longrightarrow\quad
\overline f_{n,k}\le3n^{-2},
\end{equation}
for $k\le p_0n$ above the cutoff $t=O(\ln n)$ in \eqref{eq:filter-constants}. The omitted binomial counts have exponentially small total probability.

A coherent implementation bounds conditional joint private fidelity by privacy-test acceptance, and two such tests overlap by at most the dimension of their shared receiver input divided by $M$~\cite[Lemmas~3.1--3.2]{WildePrivate2026}. Local purification systems introduce no additional dimension factor. Building on signed averaging of decoder tests~\cite[Lemma~S1]{Kondra2026}, we combine a biased tensor-product construction~\cite{Friedgut2008,Filmus2013}, a classical Meixner polynomial identity~\cite{DLMF}, and restriction to patterns of fixed size. The resulting support and absolute-row-sum estimates yield \eqref{eq:fixed-intro}. Binomial fluctuations then give the Gaussian converse.

\subsection{Paper organization}
Section~\ref{sec:setting} defines the tasks and states their comparisons; Section~\ref{sec:main} gives the main results and proof guide. Sections~\ref{sec:recovery}--\ref{sec:filter} develop the converse ingredients and prove Theorem~\ref{thm:fixed}. Its proof explicitly combines Lemmas~\ref{lem:lift}, \ref{lem:overlap}, \ref{lem:average}, and~\ref{lem:filter}. Section~\ref{sec:normal-converse} derives the Gaussian bounds, and the conclusion (Section~\ref{sec:conclusion}) summarizes our findings and discusses several remaining open questions.

The appendices retain the complete supporting arguments. Appendix~\ref{app:operational} gives operational comparisons, attainment, and the proof of the privacy-test overlap bound. Appendix~\ref{app:tensor-proof} constructs the averaging matrix and proves its estimates; Appendix~\ref{sec:ach} gives achievability and the Haar calculation; Appendix~\ref{sec:interpretation} discusses interpretation and finite-block examples; Appendix~\ref{app:probability} supplies the binomial estimates; and Appendix~\ref{app:boundaries} treats the parameter boundaries. 

\section{Notation and the communication model}\label{sec:setting}
We use one error number for each task. For quantum communication it measures preservation of entanglement. For private communication it measures correct decoding and secrecy jointly. We first define these quantities, then compare them without changing the error parameter.

\subsection{States, fidelity, and ideal entanglement}
We take all Hilbert spaces in this paper to be finite dimensional. We use the usual state and channel formalism, as developed, for example, in Ref.~\cite{Wilde2017}. The arguments below require finite-dimensional linear algebra, purifications, quantum channels, and fidelity.  A state is a positive semidefinite operator of trace one. A quantum channel is a completely positive, trace-preserving linear transformation on operators. We write $\mathcal N\colon A\to B$ when its input and output systems are $A$ and $B$. The symbol $I_A$ denotes the identity operator on $A$, and $\pi_A\coloneqq I_A/\dim A$ is the maximally mixed state. Identity operators and identity channels on unaffected systems are sometimes suppressed.

For an operator $X$, the notation $\norm{X}$ denotes its operator norm, while $\norm{X}_1$ and $\norm{X}_2$ denote its trace and Hilbert--Schmidt norms. For a Hilbert-space vector we use ket notation and write $\norm{\ket\psi}$. A projector is a self-adjoint idempotent operator, $P=P^\dagger=P^2$. Different projectors need not have orthogonal ranges.

The fidelity of normalized states is~\cite{Uhlmann1976}
\begin{equation}\label{eq:fidelity-definition}
F(\rho,\sigma)\coloneqq\norm{\sqrt\rho\sqrt\sigma}_1^2.
\end{equation}
For a pure state $\ket\psi$, this is $F(\rho,\ket\psi\bra\psi)=\bra\psi\rho\ket\psi$. The sine distance is $\sqrt{1-F(\rho,\sigma)}$~\cite{Rastegin2006}. Our error parameter is infidelity, $1-F$, throughout. A sine-distance tolerance $\delta$ therefore corresponds to an infidelity tolerance $\delta^2$ for the same pair of states.

We use two standard properties of fidelity~\cite[Chapter~9]{Wilde2017}. First, applying the same channel to both states cannot decrease fidelity:
\begin{equation}\label{eq:fidelity-monotonicity}
F(\mathcal N(\rho),\mathcal N(\sigma))\ge F(\rho,\sigma).
\end{equation}
Second, Uhlmann's theorem~\cite{Uhlmann1976} identifies fidelity with the largest squared overlap of purifications. More precisely, if $\ket\psi_{XY}$ purifies $\rho_X$, then, after enlarging $Y$ if needed,
\begin{equation}\label{eq:uhlmann-form}
F(\rho_X,\sigma_X)=\max_{\ket\gamma_{XY}:\,\Tr_Y\ket\gamma\bra\gamma_{XY}=\sigma_X}
|\braket{\gamma}{\psi}|^2.
\end{equation}
The freedom is on the purifying system $Y$. We also use the matrix inequality $0\le A\le B\Rightarrow\sqrt A\le\sqrt B$, called operator monotonicity of the square root.

For systems $R$ and $S$ of dimension $M$, fix bases and define
\begin{equation}\label{eq:bell}
\ket{\Phi^M}_{RS}\coloneqq\frac1{\sqrt M}\sum_{j=1}^M\ket j_R\ket j_S,
\qquad
\Phi_{RS}^{M}\coloneqq\ket{\Phi^M}\bra{\Phi^M}_{RS}.
\end{equation}
The system labels distinguish the maximally entangled state from the normal cumulative distribution function
\begin{equation}\label{eq:normal-cdf}
\Phi(x)\coloneqq\int_{-\infty}^{x}\phi(y)\,dy,
\qquad
\phi(y)\coloneqq\frac{e^{-y^2/2}}{\sqrt{2\pi}}.
\end{equation}
All information quantities and rates use base-two logarithms. We retain $\ln$ for the natural logarithm and $\log_d$ when measuring a dimension in $d$-dimensional units. Constants hidden in asymptotic notation may depend on fixed channel parameters and the fixed error tolerance.

\subsection{The erasure channel and its flag sectors}
The erasure channel has input space $\cH_A\cong\bbC^d$ and output space $\cH_B\cong\bbC^d\oplus\operatorname{span}\{\ket e\}$. The flag $\ket e$ is orthogonal to the transmitted subspace. For $0\le p\le1$, the channel acts on a state as
\begin{equation}\label{eq:channel}
\cE_{p,d}(\rho_A)\coloneqq(1-p)\rho_B\oplus p\ket e\bra e_B.
\end{equation}
Here $\rho_B$ is the input matrix $\rho_A$ identified with the transmitted $d$-dimensional summand of $B$. Its linear action on an arbitrary operator has $p\Tr(\rho_A)$ in the flag block. An isometric extension is
\begin{equation}\label{eq:channel-isometry}
\ket\psi_A\longmapsto
\sqrt{1-p}\ket\psi_B\ket e_Z
+\sqrt p\ket e_B\ket\psi_Z,
\end{equation}
where $Z$ has the same direct-sum structure as $B$. Tracing out $B$ gives the complementary channel, which is $\cE_{1-p,d}$ after identifying the two output spaces. The environment $Z$ is not available to the receiver.

Let $[n]\coloneqq\{1,\ldots,n\}$ and $A^n\coloneqq A_1\cdots A_n$, with $\dim A_j=d$. For $E\subseteq[n]$, set $A_E\coloneqq\bigotimes_{j\in E}A_j$ and $E^c\coloneqq[n]\setminus E$. The empty tensor product is one dimensional. The $n$-fold receiver output is block diagonal in the erased set $E$. Its probability is
\begin{equation}\label{eq:pattern-probability}
w_p(E)\coloneqq p^{|E|}(1-p)^{n-|E|},
\end{equation}
independently of the input. Within this block, the receiver holds $A_{E^c}$ and the encoded state has been traced over $A_E$. Because different erased sets are orthogonally flagged, a decoder can first read $E$ without disturbing the surviving systems. Conversely, restricting a global decoder to a flag block gives a channel $\cD_E\colon A_{E^c}\to\widehat S$. Here $\widehat S$ denotes the decoded quantum system. Thus a collection of such channels describes the quantum decoding problem exactly. For private communication the joint state with the environment requires an additional flag argument, given in Lemma~\ref{lem:flags}.

\subsection{Entanglement transmission and entanglement generation}
An entanglement transmission code of dimension $M$ consists of an encoder $\cC\colon S\to A^n$ and decoder $\cD\colon B^n\to\widehat S$, where $\dim S=\dim\widehat S=M$. Its fidelity is
\begin{equation}\label{eq:transmission-fidelity}
F_{\et}(\cC,\cD)
\coloneqq F\!\left(
(\id_R\otimes\cD\circ\cE_{p,d}^{\otimes n}\circ\cC)(\Phi_{RS}^{M}),
\PhiM\right).
\end{equation}
The reference $R$ is retained and is not acted on during encoding. The use of a reference tests preservation of quantum information entangled with another system.

For entanglement generation, the sender may instead prepare an arbitrary state $\rho_{RA^n}$, with $\dim R=M$, and retain $R$. The fidelity is
\begin{equation}\label{eq:generation-fidelity}
F_{\eg}(\rho,\cD)
\coloneqq F\!\left((\id_R\otimes\cD\circ\cE_{p,d}^{\otimes n})(\rho_{RA^n}),\PhiM\right).
\end{equation}
The reference marginal in this model is not required to be maximally mixed. Entanglement transmission is a special case, obtained by choosing $\rho_{RA^n}\coloneqq(\id_R\otimes\cC)(\Phi_{RS}^{M})$.

Both tasks use $n$ parallel, independent channel instances. The encoder and decoder may act jointly on their respective input systems and may have arbitrary finite local auxiliary systems. There is no initial shared entanglement, no feedback, no auxiliary classical or quantum communication, and no postselection of successful transmissions. The target dimension is the generated or transmitted dimension itself; no entanglement resource is subtracted from it.

For a positive integer $n$, an infidelity $0\le\eps<1$, and $\star\in\{\et,\eg\}$, define
\begin{align}
M_\star^*(n,\eps;p,d)
&\coloneqq\max\left\{M\in\bbN:
\text{a code for task $\star$ has fidelity at least $1-\eps$}\right\},\label{eq:operational-M}\\
Q_{n,\star}^{\eps}(p,d)&\coloneqq\log_2M_\star^*(n,\eps;p,d).
\label{eq:operational-Q}
\end{align}
We omit $p,d$ when they are fixed and write $M^*\coloneqq M_\et^*$ when only entanglement transmission is discussed. For every $n$ and $\eps$, the inclusion of entanglement transmission among entanglement generation codes gives $M_\et^*\le M_\eg^*$.

These maxima are finite and attained; the dimension and compactness arguments appear in Appendix~\ref{app:quantum-dimension}.

\subsection{Private communication and its joint fidelity}\label{subsec:private-definition}
A private code of size $M$ consists of input states $(\rho_{A^n}^{m})_{m=1}^{M}$ and a receiver positive operator-valued measure (POVM) $(\Lambda_{B^n}^{j})_{j=1}^{M}$. Thus $\Lambda_{B^n}^j\ge0$ and $\sum_j\Lambda_{B^n}^j=I_{B^n}$. The sender chooses $m$ uniformly, records it in a classical system $K$, and the receiver records the estimate in $\widehat K$. Here $K$ in a system subscript labels the message register; the scalar $\Ner$ used for the erasure count is distinguished by context. Let $U\colon A\to BZ$ be the channel isometry in \eqref{eq:channel-isometry}. For a message $m\in\{1,\ldots,M\}$, define the joint receiver--environment state before decoding by
\begin{equation}\label{eq:private-channel-output}
\theta^m_{B^nZ^n}\coloneqq U^{\otimes n}\rho^m_{A^n}(U^\dagger)^{\otimes n}.
\end{equation}
The message--estimate--environment state is
\begin{align}
\omega_{K\widehat K Z^n}
&\coloneqq\frac1M\sum_{m,j=1}^{M}\ket m\bra m_K\otimes\ket j\bra j_{\widehat K}
\otimes\omega_{Z^n}^{m,j},\label{eq:private-output}\\
\omega_{Z^n}^{m,j}
&\coloneqq\Tr_{B^n}\!\left[(\sqrt{\Lambda_{B^n}^{j}}\otimes I_{Z^n})
\theta^m_{B^nZ^n}(\sqrt{\Lambda_{B^n}^{j}}\otimes I_{Z^n})\right].
\end{align}
The sandwich makes positivity of $\omega^{m,j}_{Z^n}$ explicit. Cyclicity within the traced receiver system also gives $\omega^{m,j}_{Z^n}=\Tr_{B^n}[(\Lambda^j\otimes I)\theta^m]$. These operators may have trace below one; their traces are the decoding probabilities conditional on $m$. The full state in \eqref{eq:private-output} is normalized. As in the quantum task, mixed inputs and collective operations are allowed, but no initial shared resource, additional communication, or postselection is available.

The ideal classical correlation is
\begin{equation}\label{eq:ideal-key}
\overline\Phi_{K\widehat K}^{M}
\coloneqq\frac1M\sum_{m=1}^{M}\ket{mm}\bra{mm}_{K\widehat K}.
\end{equation}
For a private code $\mathsf C$, define
\begin{equation}\label{eq:private-opt}
f_{\opt}(\mathsf C)
\coloneqq\max_{\sigma_{Z^n}}F\!\left(\omega_{K\widehat K Z^n},
\overline\Phi_{K\widehat K}^{M}\otimes\sigma_{Z^n}\right),
\qquad \eps_{\opt}(\mathsf C)\coloneqq1-f_{\opt}(\mathsf C),
\end{equation}
where the maximum is over normalized states. This is the fidelity form of the joint correctness-and-secrecy criterion in Ref.~\cite[Section~4.1, Eq.~(4.3)]{WildePosition2017}, whose distance is trace distance. The exact fidelity convention appears in Refs.~\cite[Section~3.1, Eq.~(3.3)]{WTB2017} and~\cite[Section~2.1, Eqs.~(2.10)--(2.11)]{WildePrivate2026}. The ideal state imposes both requirements at once: the message is decoded correctly and is independent of the environment. The same $\sigma_{Z^n}$ must work for every message.

The fidelity in \eqref{eq:private-opt} is not an arithmetic mean of messagewise fidelities. Let $\omega^m_{\widehat K Z^n}$ be the normalized output conditional on $m$, and define its root fidelity with the ideal output by
\begin{equation}\label{eq:message-root-fidelity}
r_m(\sigma_{Z^n})\coloneqq
\sqrt{F\!\left(\omega^m_{\widehat K Z^n},
\ket m\bra m_{\widehat K}\otimes\sigma_{Z^n}\right)}.
\end{equation}
The blocks indexed by $m$ are mutually orthogonal. Square roots and products preserve this block structure, and the trace norm of a block-diagonal operator is the sum of its block trace norms. Therefore
\begin{align}
\sqrt{F\!\left(\omega_{K\widehat K Z^n},\overline\Phi_{K\widehat K}^{M}\otimes\sigma_{Z^n}\right)}
&=\frac1M\sum_{m=1}^{M}r_m(\sigma_{Z^n}),\label{eq:message-root-average}\\
F\!\left(\omega_{K\widehat K Z^n},\overline\Phi_{K\widehat K}^{M}\otimes\sigma_{Z^n}\right)
&=\left[\frac1M\sum_{m=1}^{M}r_m(\sigma_{Z^n})\right]^2.\label{eq:message-fidelity}
\end{align}
Thus we average the square roots first and then square the result. We will preserve this distinction when conditioning on erasure flags.

Define the actual environment marginal by $\omega_{Z^n}\coloneqq\Tr_{K\widehat K}\omega_{K\widehat K Z^n}$. We also consider two variants discussed in Ref.~\cite[Section~2.1]{WildePrivate2026}:
\begin{subequations}\label{eq:private-marg}
\begin{align}
f_{\marg}(\mathsf C)
&\coloneqq F\!\left(\omega_{K\widehat K Z^n},
\overline\Phi_{K\widehat K}^{M}\otimes\omega_{Z^n}\right),\label{eq:private-marg-fidelity}\\
\eps_{\marg}(\mathsf C)&\coloneqq1-f_{\marg}(\mathsf C).\label{eq:private-marg-error}
\end{align}
\end{subequations}
\begin{align}
\eps_{\max}(\mathsf C)
&\coloneqq\min_{\sigma_{Z^n}}\max_{1\le m\le M}
\left\{1-F\!\left(\omega^m_{\widehat K Z^n},
\ket m\bra m_{\widehat K}\otimes\sigma_{Z^n}\right)\right\}.
\label{eq:private-max}
\end{align}
The first fixes the ideal environment state to the actual marginal. The second asks that each message satisfy the error tolerance with one common ideal environment state. Allowing optimization over $\sigma_{Z^n}$ can only increase fidelity, so $f_{\opt}\ge f_{\marg}$. For the maximal-message criterion, choose a common state $\sigma_{Z^n}$ attaining the minimum in \eqref{eq:private-max}. Every term of \eqref{eq:message-root-fidelity} is then at least $\sqrt{1-\eps_{\max}}$, and hence
\begin{equation}\label{eq:private-max-to-average}
f_{\opt}\ge\left[\frac1M\sum_{m=1}^{M}r_m(\sigma_{Z^n})\right]^2
\ge\left[\frac1M\sum_{m=1}^{M}\sqrt{1-\eps_{\max}}\right]^2
=1-\eps_{\max}.
\end{equation}
Taking one minus the fidelities proves
\begin{equation}\label{eq:private-ordering}
\eps_{\opt}\le\eps_{\marg},\qquad \eps_{\opt}\le\eps_{\max}.
\end{equation}
There is no assertion here that the errors agree at finite blocklength.

For a criterion label $\bullet\in\{\opt,\marg,\max\}$, define
\begin{equation}\label{eq:operational-P}
P_{n,\bullet}^{\eps}(p,d)
\coloneqq\log_2\max\{M\in\bbN:\text{a private code has }\eps_\bullet(\mathsf C)\le\eps\}.
\end{equation}
These maxima are finite and attained; see Appendix~\ref{app:private-dimension}.

\subsection{From entanglement transmission to private communication}\label{subsec:q-to-p}
The following comparison will supply all the private achievable bounds. The optimized same-error implication is established in Ref.~\cite[Section~3.4, Eq.~(3.18)]{WTB2017}. We give a direct proof that identifies the comparison environment state and also bounds the fidelity with the actual marginal. This explains exactly which quantum converse follows from a private converse at the same error.

\begin{lemma}[Two quantum-to-private comparisons]\label{lem:q-to-p}
Let $\mathcal N\colon A\to B$ be a finite-dimensional channel, and let $n,M\ge1$ be integers. Suppose an unassisted entanglement transmission code for $\mathcal N^{\otimes n}$ has dimension $M$ and fidelity $F_e$, defined as in \eqref{eq:transmission-fidelity} with $\mathcal N$ in place of $\cE_{p,d}$. There is an unassisted private code of size $M$ whose optimized and actual-marginal joint fidelities, defined in \eqref{eq:private-opt} and \eqref{eq:private-marg}, satisfy
\begin{equation}\label{eq:q-to-p-fidelity}
f_{\opt}\ge F_e,
\qquad
f_{\marg}\ge F_e^2.
\end{equation}
In particular, for the erasure channel $\cE_{p,d}$ in \eqref{eq:channel}, every $n\ge1$ and $0\le\eps<1$ obey
\begin{equation}\label{eq:one-shot-ordering}
Q_{n,\et}^{\eps}\le P_{n,\opt}^{\eps},
\qquad
Q_{n,\et}^{\eta(\eps)}\le P_{n,\marg}^{\eps},
\qquad
\eta(\eps)\coloneqq1-\sqrt{1-\eps},
\end{equation}
where the optimal logarithmic code sizes are defined in \eqref{eq:operational-Q} and \eqref{eq:operational-P}.
\end{lemma}
The proof is given in Appendix~\ref{app:q-to-p}.

The ordering in \eqref{eq:one-shot-ordering} concerns entanglement transmission, whose reference is maximally mixed. For entanglement generation, the reference can be arbitrary. We prove its converse directly using the Bell-test specialization of Lemma~\ref{lem:lift}, so no conversion of that reference to a uniform message is needed.

\section{Main results and proof guide}\label{sec:main}
The main result uses the joint private criterion in \eqref{eq:private-opt}. The related criteria in \eqref{eq:private-marg} and \eqref{eq:private-max} are stated separately because their finite-error comparisons are different.

\begin{theorem}[Second-order quantum and private capacities]\label{thm:main}
Fix an integer $d\ge2$, an erasure probability $0<p<1/2$, and an infidelity $0<\eps<1$. Consider $n$ unassisted parallel uses of the erasure channel $\cE_{p,d}$ in \eqref{eq:channel}. Define
\begin{equation}\label{eq:qv}
C\coloneqq(1-2p)\log_2d,
\qquad
V\coloneqq4p(1-p)(\log_2d)^2.
\end{equation}
Let $\Phi$ be the standard normal cumulative distribution function in \eqref{eq:normal-cdf}. Each of the three optimal logarithmic code sizes
\begin{equation}\label{eq:main-tasks}
L_n^\eps\in\{Q_{n,\et}^{\eps},\ Q_{n,\eg}^{\eps},\ P_{n,\opt}^{\eps}\}
\end{equation}
defined in \eqref{eq:operational-Q} and \eqref{eq:operational-P}, satisfies, for all sufficiently large $n$,
\begin{equation}\label{eq:main-bounds}
nC+\sqrt{nV}\,\Phi^{-1}(\eps)-O(1)
\le L_n^\eps
\le nC+\sqrt{nV}\,\Phi^{-1}(\eps)+O((\ln n)^2).
\end{equation}
The constants depend only on $p,d,\eps$. In particular,
\begin{equation}\label{eq:second-order-limit}
\lim_{n\to\infty}\frac{L_n^\eps-nC}{\sqrt n}
=\sqrt V\,\Phi^{-1}(\eps).
\end{equation}
\end{theorem}

The private converse and the same-error ordering \eqref{eq:one-shot-ordering} give the entanglement transmission upper bound. The Bell-test specialization in Lemma~\ref{lem:lift} gives the entanglement generation upper bound. For achievability, one entanglement transmission construction suffices for all three tasks. The estimates are for fixed parameters; they are not uniform as $p$ approaches one half or $\eps$ approaches zero or one.

Dividing \eqref{eq:main-bounds} by $n$ identifies the $n^{-1/2}$ coefficient in the rate. For example, $d=2$ and $p=1/4$ give $C=1/2$ and $V=3/4$. The Gaussian correction is negative below infidelity $1/2$ and positive above it; in either case the rate tends to $C$. The result identifies this coefficient, while the explicit converse remainder is not intended as a tight numerical estimate at moderate blocklengths.

\Needspace{5\baselineskip}
\begin{theorem}[Other joint private criteria]\label{thm:other-criteria}
Fix an integer $d\ge2$, an erasure probability $0<p<1/2$, and an infidelity $0<\eps<1$. Consider $n$ unassisted parallel uses of $\cE_{p,d}$ from \eqref{eq:channel}. Let $C,V$ be defined in \eqref{eq:qv} and let $\Phi$ be defined in \eqref{eq:normal-cdf}. The optimal logarithmic code sizes $P_{n,\max}^{\eps}$ and $P_{n,\marg}^{\eps}$ are defined in \eqref{eq:operational-P} using the error criteria in \eqref{eq:private-max} and \eqref{eq:private-marg}, respectively. The maximal-message criterion obeys
\begin{subequations}\label{eq:max-main}
\begin{align}
nC+\sqrt{nV}\,\Phi^{-1}(\eps)-\tfrac12\log_2n-O(1)
&\le P_{n,\max}^{\eps}\\
&\le nC+\sqrt{nV}\,\Phi^{-1}(\eps)+O((\ln n)^2).
\end{align}
\end{subequations}
For the actual-environment-marginal criterion, let $\eta(\eps)\coloneqq1-\sqrt{1-\eps}$, as in \eqref{eq:one-shot-ordering}. Then
\begin{subequations}\label{eq:marg-main}
\begin{align}
nC+\sqrt{nV}\,\Phi^{-1}(\eta(\eps))-O(1)
&\le P_{n,\marg}^{\eps}\\
&\le nC+\sqrt{nV}\,\Phi^{-1}(\eta(\eps))+O((\ln n)^2).
\end{align}
\end{subequations}
The constants depend only on $p,d,\eps$, and both bounds hold for all sufficiently large $n$.
\end{theorem}

The variance $V$ is the same in these formulas, but the required binomial probability changes. To see why, imagine a schematic code with perfect conditional fidelity on patterns of total probability $s$ and zero fidelity on the rest. Optimizing the ideal environment's pattern weights gives $f_{\opt}=s$. Fixing them to the actual weights instead gives $f_{\marg}=s^2$. Thus the two infidelity requirements ask for $s\ge1-\eps$ and $s\ge\sqrt{1-\eps}$, respectively. Only the comparison state changes; all physical outcomes remain in the protocol.

Lemma~\ref{lem:flags} proves the exact identities behind this picture, and Section~\ref{subsec:criteria-meaning} develops it in Figure~\ref{fig:privacy-criteria}. Table~\ref{tab:criteria} compares the error criteria. The code need not have a perfect step transition. The converse bounds the average above the threshold, and the achievable ensemble average supplies the matching threshold estimate. None of the results above identifies a third-order coefficient.

The complete one-use example, including the optimization in its erased sector, is given in Appendix~\ref{app:one-use}.

\subsection{Why conditioning on the erasure count helps}
The erasure flags separate the communication problem into conditional problems that share the same encoder. We define their fidelities explicitly before outlining the proof. For private communication, we may first replace the receiver measurement by one that reads the erasure flags. Lemma~\ref{lem:flags} proves that this replacement cannot decrease either joint private fidelity in \eqref{eq:private-opt} and \eqref{eq:private-marg}; it therefore suffices for an upper bound.

Fix an erased set $E\subseteq[n]$. Let $(\Lambda_{A_{E^c}}^{E,j})_{j=1}^M$ be the receiver POVM conditioned on $E$, acting on the surviving input systems $A_{E^c}$. Given the input states $(\rho_{A^n}^{m})_{m=1}^M$, define
\begin{align}
\omega_{A_E}^{E,m,j}
&\coloneqq\Tr_{A_{E^c}}\!\left[
(\sqrt{\Lambda_{A_{E^c}}^{E,j}}\otimes I_{A_E})\rho_{A^n}^{m}
(\sqrt{\Lambda_{A_{E^c}}^{E,j}}\otimes I_{A_E})\right],\label{eq:conditional-private-block}\\
\omega^E_{K\widehat K A_E}
&\coloneqq\frac1M\sum_{m,j=1}^M\ket{mj}\bra{mj}_{K\widehat K}\otimes\omega_{A_E}^{E,m,j}.
\label{eq:conditional-private-output}
\end{align}
Tensor factors are identified with their physical positions in $A^n$. The state $\omega^E_{K\widehat K A_E}$ is normalized because the conditional POVM sums to the identity. Its message is still uniform because the probability $w_p(E)$ in \eqref{eq:pattern-probability} is independent of the message. The conditional optimized private fidelity is
\begin{equation}\label{eq:conditional-private-definition}
f_E\coloneqq\max_{\sigma_{A_E}}
F\!\left(\omega^E_{K\widehat K A_E},
\overline\Phi_{K\widehat K}^{M}\otimes\sigma_{A_E}\right),
\end{equation}
where $\sigma_{A_E}$ ranges over normalized states and the ideal key state is defined in \eqref{eq:ideal-key}.

For a quantum code, let $\rho_{RA^n}$ be the prepared reference--input state and let $\cD_E\colon A_{E^c}\to\widehat S$ be the decoder conditioned on $E$. Its conditional entanglement fidelity is
\begin{equation}\label{eq:conditional-quantum-definition}
F_E\coloneqq\Tr\!\left[
\Phi_{R\widehat S}^{M}
(\id_R\otimes\cD_E)(\Tr_{A_E}\rho_{RA^n})\right].
\end{equation}
The target $\Phi_{R\widehat S}^{M}$ is defined in \eqref{eq:bell}. For entanglement transmission, $\rho_{RA^n}\coloneqq(\id_R\otimes\cC)(\Phi_{RS}^{M})$; for entanglement generation, it may be any state with $\dim R=M$.

Conditioned on the number $k$ of erasures, all size-$k$ erased sets have the same probability. Define the corresponding averages by
\begin{equation}\label{eq:conditional-fidelity}
\overline f_{n,k}\coloneqq\frac1{\binom nk}\sum_{|E|=k}f_E,
\qquad
\overline F_{n,k}\coloneqq\frac1{\binom nk}\sum_{|E|=k}F_E.
\end{equation}
For the flag-diagonal private code, Lemma~\ref{lem:flags} proves the first identity below. The second follows from the linearity of Bell-target fidelity:
\begin{equation}\label{eq:binomial-mixture}
f_{\opt}=\E_{\Ner}\overline f_{n,\Ner},
\qquad
F_\star=\E_{\Ner}\overline F_{n,\Ner},
\qquad \Ner\sim\Bin(n,p),\quad\star\in\{\et,\eg\}.
\end{equation}
Here $f_{\opt}$ is the optimized joint private fidelity from \eqref{eq:private-opt}, and $F_{\et},F_{\eg}$ are the entanglement fidelities from \eqref{eq:transmission-fidelity} and \eqref{eq:generation-fidelity}. The first identity in \eqref{eq:binomial-mixture} requires optimizing the ideal environment's flag weights; it does not follow merely from linearity. All dependence on $p$ is now in the scalar binomial weights. The conditional communication problem depends only on $n,k,d,M$.

A bound $\log_dM\le n-2k$ cannot hold separately for every well-corrected pattern. For example, with $n=4$, $k=1$, and $M=d^3$, an encoder can place its three logical systems in positions $2,3,4$ and fix position $1$ in a pure state. Erasing $E=\{1\}$ then permits perfect quantum recovery and perfect private communication, although $\log_dM=3>n-2k=2$. The same encoder must also face the other three single erasures. The averages in \eqref{eq:conditional-fidelity} express this requirement.

\subsection{The steps of the proof}
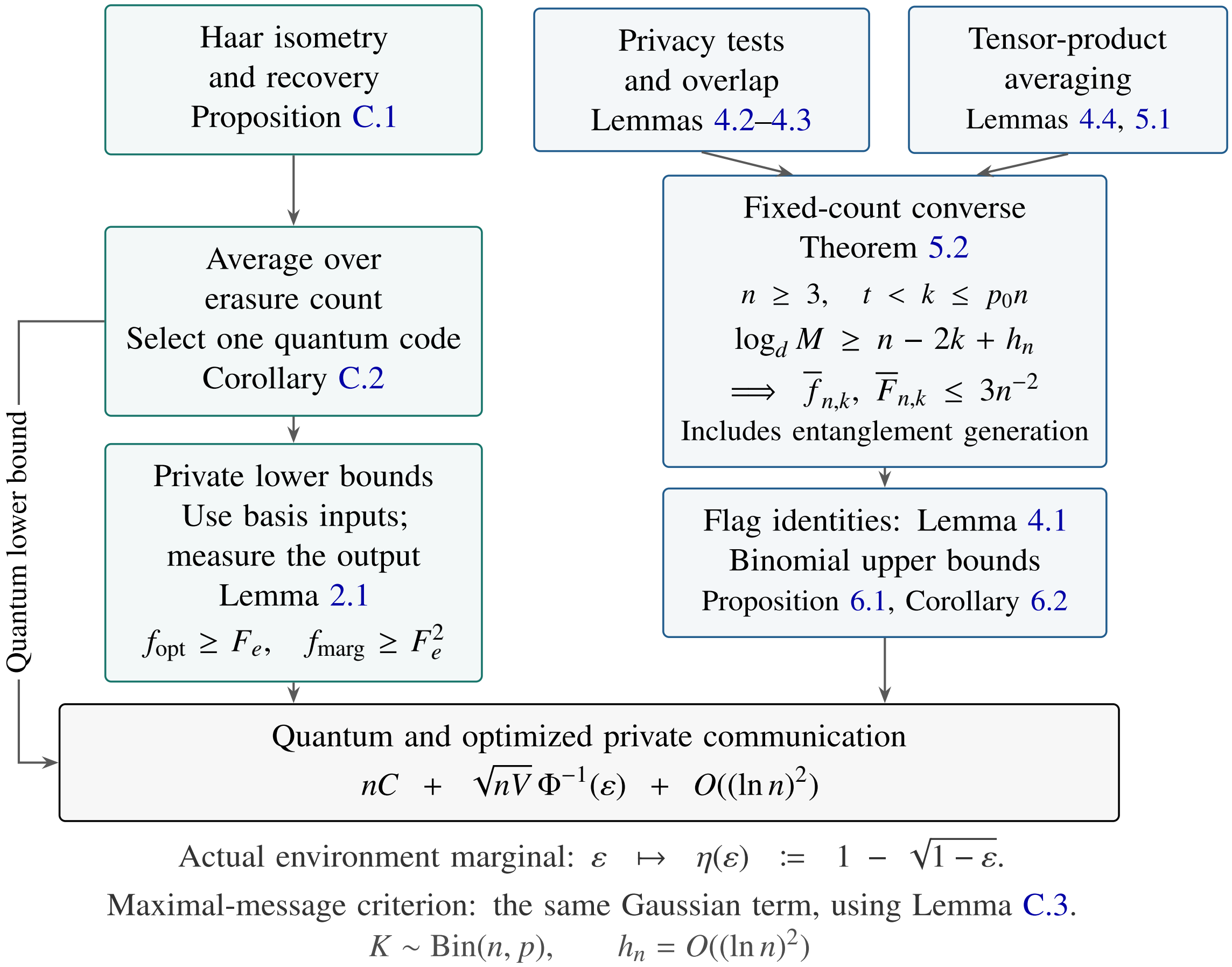
\begin{figure}[!t]
\centering
\resizebox{0.92\textwidth}{!}{
\begin{tikzpicture}[
  font=\small,
  box/.style={draw,rounded corners=2pt,line width=.65pt,align=center,
    inner xsep=6pt,inner ysep=7pt,text width=4.1cm},
  ach/.style={box,draw=erasureTeal,fill=erasureTeal!5},
  con/.style={box,draw=erasureBlue,fill=erasureBlue!5},
  arrow/.style={-{Stealth[length=2mm]},line width=.75pt,draw=black!65}
]
\node[ach] (haar) at (-4.5,7.4)
  {Haar isometry and recovery\\Proposition~\ref{prop:haar}};
\node[con,text width=3.6cm] (overlap) at (.4,7.4)
  {Privacy tests\\and overlap\\Lemmas~\ref{lem:lift}--\ref{lem:overlap}};
\node[con,text width=3.4cm] (filter) at (4.8,7.4)
  {Tensor-product\\averaging\\{\footnotesize Lemmas~\ref{lem:average}, \ref{lem:filter}}};

\node[ach] (lower) at (-4.5,4.5)
  {Average over erasure count\\Select one quantum code\\Corollary~\ref{cor:normal-lower}};
\node[con,text width=4.9cm] (countupper) at (2.6,4.5)
  {Fixed-count converse\\Theorem~\ref{thm:fixed}\\[2pt]
   {\footnotesize $n\ge3,\quad t<k\le p_0n$}\\[2pt]
   $\log_dM\ge n-2k+h_n$\\[4pt]
   $\Longrightarrow\ \overline f_{n,k},\ \overline F_{n,k}\le3n^{-2}$\\
   {\footnotesize Includes entanglement generation}};

\node[ach] (comparison) at (-4.5,1.6)
  {Private lower bounds\\Use basis inputs;\\measure the output\\Lemma~\ref{lem:q-to-p}\\[2pt]
   $f_{\mathrm{opt}}\ge F_e$,\quad
   $f_{\mathrm{marg}}\ge F_e^2$};
\node[con,text width=4.9cm] (upper) at (2.6,1.6)
  {Flag identities: Lemma~\ref{lem:flags}\\Binomial upper bounds\\
   {\footnotesize Proposition~\ref{prop:finite-converse}, Corollary~\ref{cor:normal-upper}}};

\node[box,fill=black!3,text width=12.3cm] (result) at (-.95,-.8)
  {Quantum and optimized private communication\\[3pt]
   $\displaystyle nC+\sqrt{nV}\,\Phi^{-1}(\varepsilon)
      +O((\ln n)^2)$};

\draw[arrow] (haar)--(lower);
\draw[arrow] (overlap.south)--([xshift=-11mm]countupper.north);
\draw[arrow] (filter.south)--([xshift=11mm]countupper.north);
\draw[arrow] (lower)--(comparison);
\draw[arrow] (lower.west)--(-7.8,4.5)--(-7.8,-.8)--(result.west);
\node[rotate=90,font=\footnotesize,fill=white,inner sep=2pt]
  at (-7.8,1.85) {Quantum lower bound};
\draw[arrow] (countupper)--(upper);
\draw[arrow] (comparison.south)--([xshift=-35.5mm]result.north);
\draw[arrow] (upper.south)--([xshift=35.5mm]result.north);

\node[align=center,font=\small,text=black!80,text width=13.4cm]
  at (-.95,-2.2)
  {Actual environment marginal:
   $\varepsilon\mapsto\eta(\varepsilon)\coloneqq1-\sqrt{1-\varepsilon}$.\\[3pt]
   Maximal-message criterion: the same Gaussian term, using Lemma~\ref{lem:expurgation}.};
\node[align=center,font=\small,text=black!70] at (-.95,-3)
  {$\Ner\sim\operatorname{Bin}(n,p)$,\qquad
   $h_n=O((\ln n)^2)$};
\end{tikzpicture}}
\caption{Proof guide for fixed $d\ge2$, $0<p<p_0<1/2$, and $0<\eps<1$. The private converse controls arbitrary mixed encoders through privacy tests and a signed average over erasure patterns. Bell tests give the same bound for entanglement generation. The fixed-count implication is \eqref{eq:fixed-threshold}; $t,h_n$ are defined in \eqref{eq:filter-constants} and \eqref{eq:hn}; Theorem~\ref{thm:fixed} applies for $n\ge3$. Entanglement transmission codes give the optimized and actual-marginal private lower bounds through Lemma~\ref{lem:q-to-p}; Lemma~\ref{lem:expurgation} gives the maximal-message bound. Both sides meet at the fluctuations of the erasure count. Here $C,V$ are defined in \eqref{eq:qv}, and $\Phi$ is the standard normal cumulative distribution function in \eqref{eq:normal-cdf}. The actual-marginal criterion uses the square root of fidelity in the converse and $f_{\marg}\ge F_e^2$ in achievability.}
\label{fig:proof-guide}
\end{figure}

Figure~\ref{fig:proof-guide} shows how the private converse and quantum achievability meet. There are four steps in the converse.

First, retain the sender's and receiver's local purification systems. Lemma~\ref{lem:lift} represents every private code by one pure reference--input state and a family of projectors $P_E$, defined in \eqref{eq:decoder-projector}, such that $f_E$ in \eqref{eq:conditional-private-definition} is no larger than the acceptance probability of $P_E$. Each projector is a mathematical privacy test; no additional joint measurement is supplied to the communication protocol. For quantum communication, the same construction uses a Bell-state test; its acceptance probability equals the conditional entanglement fidelity.

Second, compare two erased sets $E,F$ of size $k$. Their possible receivers share exactly the physical systems surviving both patterns. There are $n-2k+|E\cap F|$ such systems. Lemma~\ref{lem:overlap} gives
\begin{equation}\label{eq:guide-overlap}
\norm{P_EP_F}\le\frac{d^{n-2k+|E\cap F|}}M
\qquad(E\ne F).
\end{equation}
This estimate is independent of the dimensions of the local purification systems, so mixed encoders incur no extra loss.

Third, use these pairwise restrictions to bound the performance of one code averaged over all patterns. The difficulty is that the overlap estimate is strongest for disjoint erased sets, whereas the full average involves pairs with every possible intersection. We therefore seek a weighted collection of pairs that nearly reproduces uniform averaging while using only pairs with few common erased positions. We record the weights in a matrix $W$. Lemma~\ref{lem:average} turns this comparison into a bound on the average of the tests. It controls both the error in replacing the uniform average and the size of the weights. Some weights may be negative; they are coefficients in the proof, with no change to the physical communication protocol.

The construction first allows each position to be selected independently, so its spectral calculation reduces to diagonalizing one two-dimensional matrix and taking tensor products. A position can belong to at most one of the two initial patterns. The nonconstant one-position eigenvalue has magnitude strictly below one, so components involving many positions are already suppressed. A degree-$t$ polynomial improves the averaging while permitting at most $t$ changes of individual positions. The final pattern can therefore share at most $t$ positions with the initial one.

Lemma~\ref{lem:filter} then restricts this construction to patterns of size $k$. Choosing the auxiliary selection probability to be $k/n$ makes that layer sufficiently probable: its probability is at least $1/(n+1)$. Restriction and normalization therefore preserve the approximation with a controlled loss. Appendix~\ref{app:tensor-proof} proves the polynomial identity and this normalization step explicitly. The degree is $t=O(\ln n)$, and the weights and additional intersections cost only $O((\ln n)^2)=o(\sqrt n)$ in $\log_dM$. Combining these estimates with \eqref{eq:guide-overlap} proves Theorem~\ref{thm:fixed}, stated informally in \eqref{eq:fixed-intro}.

Finally, average the fixed-count bound over $\Ner\sim\Bin(n,p)$. Proposition~\ref{prop:finite-converse} shows that meeting fidelity $1-\eps$ forces a binomial cumulative probability to be at least $1-\eps-o(1)$. Corollary~\ref{cor:normal-upper} evaluates this probability and gives the quantile $\Phi^{-1}(\eps)$. For the actual-marginal fidelity in \eqref{eq:private-marg}, the argument instead controls the square root of fidelity, so the required probability is $\sqrt{1-\eps}-o(1)$. This accounts for the quantile in \eqref{eq:marg-main}.

The achievable argument is shorter. Choose an encoding isometry uniformly over all orientations of an $M$-dimensional subspace; this is the Haar ensemble used in Proposition~\ref{prop:haar}. Its mean conditional recovery fidelity admits a smooth lower bound near $\log_dM=n-2k$. Averaging over all erasure patterns selects one encoder with the required overall performance. Corollary~\ref{cor:normal-lower} evaluates the binomial quantile, and Lemma~\ref{lem:q-to-p} supplies both private lower bounds. Lemma~\ref{lem:expurgation} restricts the code to a suitable smaller message set and gives the maximal-message result.

\section{Erasure flags, privacy tests, and averaging}\label{sec:recovery}
We now establish the common converse mechanism. First, we justify conditioning the private criterion on the erasure pattern. Next, we represent its conditional fidelity by a projector on one common state. A bound on the overlap of two such projectors then leads to a bound on their average. The local purification systems retained along the way are mathematical devices; they provide no additional communication resource.

\subsection{Conditioning the private criterion on flags}
The receiver can read the orthogonal erasure flags without disturbing the surviving quantum systems. For a private code, however, an arbitrary receiver measurement can create off-diagonal environment-flag blocks when one also records its outcome. The unconditional environment marginal remains block diagonal. Lemma~\ref{lem:flags} shows that dephasing the receiver's measurement in the flags removes these conditional coherences without decreasing either $f_{\opt}$ in \eqref{eq:private-opt} or $f_{\marg}$ in \eqref{eq:private-marg}.

\Needspace{10\baselineskip}
\begin{lemma}[Flag reduction and exact fidelity identities]\label{lem:flags}
Let $n\ge1$, $d\ge2$, and $M\ge1$ be integers, let $0\le p\le1$, and consider a private code of size $M$ for $\cE_{p,d}^{\otimes n}$, with output state \eqref{eq:private-output}. There is a code with the same input states and a receiver POVM diagonal in the erasure flags whose optimized fidelity \eqref{eq:private-opt} and actual-marginal fidelity \eqref{eq:private-marg} are at least those of the original code. For the resulting code,
\begin{equation}\label{eq:flag-output-decomposition}
\omega_{K\widehat K Z^n}=\bigoplus_{E\subseteq[n]}w_p(E)\,\omega^{E}_{K\widehat K A_E},
\end{equation}
where $w_p(E)$ is defined in \eqref{eq:pattern-probability} and $\omega^E_{K\widehat K A_E}$ is the normalized conditional state in \eqref{eq:conditional-private-output}. The environment flag identifies each summand. Sectors of zero probability can be assigned arbitrary normalized conditional states. Define $\omega^E_{A_E}\coloneqq\Tr_{K\widehat K}\omega^E_{K\widehat K A_E}$. Recall the conditional optimized fidelity $f_E$ from \eqref{eq:conditional-private-definition}, and define its actual-marginal counterpart by
\begin{equation}\label{eq:conditional-private}
g_E\coloneqq F\!\left(\omega^E_{K\widehat K A_E},
\overline\Phi^M_{K\widehat K}\otimes\omega^E_{A_E}\right).
\end{equation}
Then
\begin{subequations}\label{eq:flag-identities}
\begin{align}
f_{\opt}&=\sum_Ew_p(E)f_E,\label{eq:flag-optimized-identity}\\
\sqrt{f_{\marg}}&=\sum_Ew_p(E)\sqrt{g_E},\label{eq:flag-marginal-identity}\\
g_E&\le f_E\qquad(E\subseteq[n]).\label{eq:flag-conditional-order}
\end{align}
\end{subequations}
\end{lemma}
\begin{proof}
We first relate dephasing the receiver's POVM to dephasing the environment in the final state. We then optimize the weights of the ideal environment's flag sectors.

Let $\Pi_{B^n}^{E}$ and $\Pi_{Z^n}^{E}$ denote the receiver and environment projectors for the same erased set $E$, respectively. The isometry $U$ in \eqref{eq:channel-isometry} correlates these labels perfectly:
\begin{equation}\label{eq:correlated-flags}
(\Pi_{B^n}^{E}\otimes I_{Z^n})U^{\otimes n}
=(I_{B^n}\otimes\Pi_{Z^n}^{E})U^{\otimes n}.
\end{equation}
Define the new POVM and the environment dephasing channel by
\begin{align}
\Lambda_{B^n}^{j\prime}&\coloneqq\sum_E\Pi_{B^n}^{E}\Lambda_{B^n}^{j}\Pi_{B^n}^{E},
\label{eq:flag-dephasing-channels}\\
\Delta_{Z^n}(X)&\coloneqq\sum_E\Pi_{Z^n}^{E}X\Pi_{Z^n}^{E}.
\label{eq:environment-dephasing}
\end{align}
Every term defining $\Lambda_{B^n}^{j\prime}$ is positive semidefinite. The new operators also sum to the identity:
\begin{align}
\sum_j\Lambda_{B^n}^{j\prime}
&=\sum_E\Pi_{B^n}^{E}\left(\sum_j\Lambda_{B^n}^{j}\right)\Pi_{B^n}^{E}
\label{eq:dephased-povm-sum}\\
&=\sum_E(\Pi_{B^n}^{E})^2=I_{B^n}.
\label{eq:dephased-povm-normalization}
\end{align}
Thus they form a POVM. For each message $m$, recall the joint channel output $\theta^m_{B^nZ^n}$ defined in \eqref{eq:private-channel-output}. Using \eqref{eq:correlated-flags} on both sides of this state and cyclicity within the partial trace gives
\begin{align}
\omega_{Z^n}^{m,j\prime}
&=\sum_E\Tr_{B^n}\!\left[
(\Pi_{B^n}^{E}\Lambda_{B^n}^{j}\Pi_{B^n}^{E}\otimes I_{Z^n})
\theta_{B^nZ^n}^{m}\right]
\label{eq:dephased-block-pullback}\\
&=\sum_E\Tr_{B^n}\!\left[
(\Lambda_{B^n}^{j}\otimes\Pi_{Z^n}^{E})
\theta_{B^nZ^n}^{m}(I_{B^n}\otimes\Pi_{Z^n}^{E})\right]
\label{eq:dephased-block-correlated}\\
&=\sum_E\Pi_{Z^n}^{E}\omega_{Z^n}^{m,j}\Pi_{Z^n}^{E}
=\Delta_{Z^n}(\omega_{Z^n}^{m,j}).
\label{eq:dephased-private-blocks}
\end{align}
In particular, if $\mathcal F\coloneqq\id_{K\widehat K}\otimes\Delta_{Z^n}$, then
\begin{equation}\label{eq:dephased-private-state}
\omega'_{K\widehat K Z^n}=\mathcal F(\omega_{K\widehat K Z^n}).
\end{equation}
The environment marginal already satisfies $\Delta_{Z^n}(\omega_{Z^n})=\omega_{Z^n}$, because the complementary erasure channel has orthogonal flag sectors. For a normalized environment state $\sigma_{Z^n}$, write
\begin{equation}\label{eq:flag-ideal-comparison}
\tau_{K\widehat K Z^n}(\sigma)
\coloneqq\overline\Phi^M_{K\widehat K}\otimes\sigma_{Z^n}.
\end{equation}
Fidelity monotonicity \eqref{eq:fidelity-monotonicity} gives, for every such $\sigma_{Z^n}$,
\begin{align}
F\!\left(\omega'_{K\widehat K Z^n},\tau_{K\widehat K Z^n}(\Delta_{Z^n}(\sigma))\right)
&=F\!\left(\mathcal F(\omega_{K\widehat K Z^n}),\mathcal F(\tau_{K\widehat K Z^n}(\sigma))\right)
\label{eq:flag-monotonicity-input}\\
&\ge F\!\left(\omega_{K\widehat K Z^n},\tau_{K\widehat K Z^n}(\sigma)\right).
\label{eq:flag-monotonicity-output}
\end{align}
Choosing $\sigma_{Z^n}=\omega_{Z^n}$ proves that the actual-marginal fidelity cannot decrease. Choosing an optimizer in \eqref{eq:private-opt} proves the same for the optimized fidelity: $\Delta_{Z^n}(\sigma_{Z^n})$ is a normalized state and hence remains admissible in that optimization. An upper bound for the new code therefore also bounds the original code.

We now work with the flag-diagonal output in \eqref{eq:flag-output-decomposition}. Applying \eqref{eq:flag-monotonicity-output} to this invariant output shows that a block-diagonal ideal environment state attains the optimum. Write
\begin{equation}\label{eq:ideal-flag-decomposition}
\sigma_{Z^n}=\bigoplus_Eq_E\sigma^E_{A_E},
\qquad q_E\ge0,\quad\sum_Eq_E=1,\quad\Tr\sigma^E_{A_E}=1.
\end{equation}
For brevity, define the conditional root fidelity
\begin{equation}\label{eq:conditional-root-comparison}
r_E(\sigma^E)\coloneqq
\sqrt{F\!\left(\omega^E_{K\widehat K A_E},
\overline\Phi_{K\widehat K}^{M}\otimes\sigma^E_{A_E}\right)}.
\end{equation}
The square roots of block-diagonal operators are block diagonal, and the trace norm is additive on direct sums. Applying these facts to \eqref{eq:fidelity-definition} gives
\begin{equation}\label{eq:flag-root-fidelity}
\sqrt{F\!\left(\omega_{K\widehat K Z^n},\tau_{K\widehat K Z^n}(\sigma)\right)}
=\sum_E\sqrt{w_p(E)q_E}\,r_E(\sigma^E).
\end{equation}
Each $\sigma^E_{A_E}$ can be optimized separately, giving $\max_{\sigma^E}r_E(\sigma^E)=\sqrt{f_E}$. The remaining optimization is over the probability distribution $(q_E)_E$. Cauchy--Schwarz implies
\begin{align}
\sum_E\sqrt{q_E}\sqrt{w_p(E)f_E}
&\le\left(\sum_Eq_E\right)^{1/2}
\left(\sum_Ew_p(E)f_E\right)^{1/2}
\label{eq:flag-weights-cauchy}\\
&=\left(\sum_Ew_p(E)f_E\right)^{1/2}.
\label{eq:flag-weights-bound}
\end{align}
When the sum on the right is positive, equality is attained by
\begin{equation}\label{eq:optimal-flag-weights}
q_E\coloneqq\frac{w_p(E)f_E}{\sum_Fw_p(F)f_F}.
\end{equation}
If the sum is zero, every summand on the left is zero as well. Squaring the optimized root fidelity proves \eqref{eq:flag-optimized-identity}.

For the actual-marginal criterion, the comparison state is fixed. Taking the marginal of \eqref{eq:flag-output-decomposition} gives
\begin{equation}\label{eq:actual-marginal-flag-decomposition}
\omega_{Z^n}=\bigoplus_Ew_p(E)\omega^E_{A_E}.
\end{equation}
Thus \eqref{eq:flag-root-fidelity} applies with $q_E=w_p(E)$ and $\sigma^E_{A_E}=\omega^E_{A_E}$. By \eqref{eq:conditional-private}, this gives \eqref{eq:flag-marginal-identity}. Finally, the state $\omega^E_{A_E}$ is one candidate in the maximization defining $f_E$, which proves \eqref{eq:flag-conditional-order}.
\end{proof}

Notice that \eqref{eq:flag-optimized-identity} does not replace the messagewise expression \eqref{eq:message-fidelity} by an arithmetic mean. It is a different identity, obtained by optimizing the environment's \emph{erasure-pattern} weights. This is the step that retains the original error parameter in the private converse.

\subsection{A common state and conditional privacy tests}
We use the purified privacy-test representation from Ref.~\cite[Lemma~3.1]{WildePrivate2026}, with an explicit proof given here in order to fix the systems and the error convention. Privacy tests arise from the private-state description of secret correlations~\cite{HHHO2005,HHHO2009,WTB2017}.

Let $K,\widehat K$ be $M$-dimensional key systems with fixed orthonormal bases, and let $L,T$ be local systems retained by the sender and receiver. These retained systems are called \emph{shields}; here they simply keep the purifications of local preparations and measurements. The sender holds $K,L$, the receiver holds $\widehat K,T$, and the channel environment is held separately. Given unitaries $V_1,\ldots,V_M$ on $LT$, define
\begin{equation}\label{eq:privacy-isometry}
J\colon LT\longrightarrow K\widehat KLT,
\qquad
J\ket\xi\coloneqq\frac1{\sqrt M}\sum_{m=1}^{M}\ket{mm}_{K\widehat K}V_m\ket\xi.
\end{equation}
Orthogonality of the key labels and unitarity of the $V_m$ imply
\begin{equation}\label{eq:privacy-isometry-check}
J^\dagger J=\frac1M\sum_{m,j=1}^M
\braket{mm}{jj}V_m^\dagger V_j
=\frac1M\sum_{m=1}^MV_m^\dagger V_m=I_{LT}.
\end{equation}
Thus $J$ is an isometry. Its range projector is
\begin{equation}\label{eq:privacy-test}
\Pi\coloneqq JJ^\dagger
=\frac1M\sum_{m,j=1}^{M}\ket{mm}\bra{jj}_{K\widehat K}\otimes V_mV_j^\dagger.
\end{equation}
In particular, $\Pi^\dagger=\Pi$ and $\Pi^2=J(J^\dagger J)J^\dagger=\Pi$. This operator tests a purified version of correct and secret key correlations. Its unitaries can act jointly on the two shields, so it need not be a local measurement. We use it only in a converse.

The encoded input state will be the same for every erasure pattern; only the test changes. We place the alternative decoder realizations on one enlarged Hilbert space so that the norm of their average bounds every input state.

\Needspace{5\baselineskip}
\begin{lemma}[Conditional tests on one state]\label{lem:lift}
Fix integers $n\ge1$, $d\ge2$, and $M\ge1$. Let $(\rho_{A^n}^{m})_{m=1}^M$ be a private input ensemble, with a conditional receiver POVM on $A_{E^c}$ for every erased set $E\subseteq[n]$. Let $f_E$ and $\overline f_{n,k}$ be the conditional and averaged optimized fidelities in \eqref{eq:conditional-private-definition} and \eqref{eq:conditional-fidelity}. There are a single pure state $\ket\Psi_{KLA^n}$, a separate auxiliary system $G_E$ initialized in $\ket0_{G_E}$ for each pattern, unitary decoder realizations
\begin{equation}\label{eq:decoder-unitary}
D_E\colon A_{E^c}G_E\longrightarrow\widehat K T_E,
\end{equation}
and privacy-test projectors $\Pi_E$ of the form \eqref{eq:privacy-test}, on $K\widehat KLT_E$, such that the operators pulled back through the decoders,
\begin{equation}\label{eq:decoder-projector}
P_E\coloneqq(I_{KL}\otimes D_E^\dagger)\Pi_E(I_{KL}\otimes D_E),
\end{equation}
are projectors acting on the same physical input $A^n$. Their common domain also contains the reference system $K$, Alice's purification system $L$, and the decoder auxiliaries. Write $G_{\mathrm{all}}\coloneqq\bigotimes_{F\subseteq[n]}G_F$ for the composite auxiliary system. To compare the projectors, we use the common Hilbert space
\begin{equation}\label{eq:common-space}
\cH_K\otimes\cH_L\otimes\cH_{A^n}\otimes
\bigotimes_{F\subseteq[n]}\cH_{G_F}.
\end{equation}
In \eqref{eq:decoder-projector}, identities on $A_E$ and on every $G_F$ with $F\ne E$ are implicit. With
\begin{equation}\label{eq:common-state}
\widetilde\Psi_{KLA^nG_{\mathrm{all}}}\coloneqq\ket\Psi\bra\Psi_{KLA^n}\otimes
\bigotimes_{F\subseteq[n]}\ket0\bra0_{G_F},
\end{equation}
these projectors satisfy
\begin{subequations}\label{eq:layer-operator}
\begin{align}
f_E&\le\Tr(P_E\widetilde\Psi_{KLA^nG_{\mathrm{all}}}),
\label{eq:layer-acceptance}\\
\overline f_{n,k}&\le\norm{\frac1{\binom nk}\sum_{|E|=k}P_E}.
\label{eq:layer-average-bound}
\end{align}
\end{subequations}
For an arbitrary entanglement generation state $\rho_{RA^n}$ with $\dim R=M$, and conditional decoders $\cD_E\colon A_{E^c}\to\widehat S$ with $\dim\widehat S=M$, Bell tests have the same representation with $K=R$, $\widehat K=\widehat S$, and $L$ trivial. The common state is then $\rho_{RA^n}\otimes\bigotimes_F\ket0\bra0_{G_F}$ and need not be pure. Equation~\eqref{eq:layer-acceptance} is an equality with $f_E$ replaced by $F_E$ from \eqref{eq:conditional-quantum-definition}; \eqref{eq:layer-average-bound} holds with $\overline f_{n,k}$ replaced by $\overline F_{n,k}$ from \eqref{eq:conditional-fidelity}.
\end{lemma}
\begin{proof}
The construction has three parts. We purify the preparation and the receiver measurement, use Uhlmann's theorem to obtain a privacy test, and finally pull that test back to a common input space. The last part is necessary because the operators for different patterns must be compared on the same Hilbert space.

\emph{Purifying the preparation and the measurement.}
Let $L_0$ be an $M$-dimensional copy of the message register $K$, and choose a common purifying system $L_1$ of dimension $d^n$. Every message state admits a purification $\ket{\phi_m}_{L_1A^n}$. Set $L\coloneqq L_0L_1$ and define
\begin{equation}\label{eq:coherent-encoder}
\ket\Psi_{KLA^n}
\coloneqq\frac1{\sqrt M}\sum_{m=1}^{M}\ket m_K\ket m_{L_0}\ket{\phi_m}_{L_1A^n}.
\end{equation}
This state is fixed once for every pattern. Tracing out $L_0$ removes the terms with unequal message labels, and tracing out $L_1$ gives
\begin{equation}\label{eq:purified-ensemble-marginal}
\Tr_L\ket\Psi\bra\Psi_{KLA^n}
=\frac1M\sum_{m=1}^{M}\ket m\bra m_K\otimes\rho^m_{A^n}.
\end{equation}
Thus the purification represents the original private input ensemble.

Fix a pattern $E$. Let $T_0$ be an $M$-dimensional copy of the estimate register $\widehat K$, and let $T_1$ be a copy of $A_{E^c}$. Put $T\coloneqq T_0T_1$. For the conditional POVM appearing in \eqref{eq:conditional-private-block}, define
\begin{equation}\label{eq:conditional-measurement-isometry}
U_E\coloneqq\sum_{j=1}^M\ket j_{\widehat K}\ket j_{T_0}\otimes
\sqrt{\Lambda_{A_{E^c}}^{E,j}}.
\end{equation}
The square-root factor maps $A_{E^c}$ to its copy $T_1$. Orthogonality of the outcome labels gives
\begin{equation}\label{eq:measurement-isometry-check}
U_E^\dagger U_E
=\sum_j\Lambda_{A_{E^c}}^{E,j}=I_{A_{E^c}}.
\end{equation}
Apply this isometry to the surviving share of \eqref{eq:coherent-encoder}, obtaining
\begin{equation}\label{eq:conditional-purified-output}
\ket{\psi_E}_{K\widehat K LTA_E}
\coloneqq(I_{KLA_E}\otimes U_E)\ket\Psi_{KLA^n},
\end{equation}
with tensor factors reordered as indicated. Tracing out $L_0$ removes off-diagonal message terms, and tracing out $T_0$ removes off-diagonal outcome terms. The remaining trace over $L_1T_1$ is exactly the partial trace in \eqref{eq:conditional-private-block}. Hence
\begin{equation}\label{eq:conditional-output-marginal}
\Tr_{LT}\ket{\psi_E}\bra{\psi_E}_{K\widehat K LTA_E}
=\frac1M\sum_{m,j}\ket{mj}\bra{mj}_{K\widehat K}\otimes\omega_{A_E}^{E,m,j}
=\omega^E_{K\widehat K A_E}.
\end{equation}

\emph{Constructing a test from the ideal comparison state.}
Let $\sigma^E_{A_E}$ attain the conditional optimum in \eqref{eq:conditional-private-definition}. Uhlmann's theorem \eqref{eq:uhlmann-form}, with fixed system $K\widehat K A_E$ and purifying system $LT$, gives a purification of $\overline\Phi^{M}_{K\widehat K}\otimes\sigma^E_{A_E}$ with the following properties:
\begin{align}
\ket{\gamma_E}_{K\widehat K LTA_E}
&=\frac1{\sqrt M}\sum_m\ket{mm}_{K\widehat K}\ket{\chi_m}_{LTA_E},
\label{eq:ideal-key-purification}\\
|\braket{\gamma_E}{\psi_E}|^2&=f_E,
\label{eq:ideal-coherent-key}\\
\Tr_{LT}\ket{\chi_m}\bra{\chi_j}_{LTA_E}
&=\delta_{mj}\sigma^E_{A_E}.
\label{eq:ideal-purification-blocks}
\end{align}
Here $\delta_{mj}$ is the Kronecker delta. The last equality follows by matching the $(mm,jj)$ block of the reduced state to the corresponding block of the ideal classical key state. For $m=j$, it says that each $\ket{\chi_m}$ is a normalized purification of the same environment state. For $m\ne j$, it records the vanishing ideal off-diagonal block. The chosen spaces are large enough: the ideal state's rank is at most $Md^{|E|}$, whereas
\begin{equation}\label{eq:purifying-space-size}
\dim(LT)=M^2d^{2n-|E|}\ge Md^{|E|}.
\end{equation}
Uhlmann's theorem therefore applies without enlarging the common sender system $L$. It chooses a comparison purification, not an operation that the communicating parties must perform.

Purifications of one state differ by a unitary on a sufficiently large common purifying system. Consequently, there are unitaries $V_m$ on $LT$ such that
\begin{equation}\label{eq:shield-purification-unitaries}
\ket{\chi_m}_{LTA_E}
=(V_m\otimes I_{A_E})\ket{\chi_1}_{LTA_E}.
\end{equation}
Let $J_E$ be the isometry \eqref{eq:privacy-isometry} formed from these unitaries, and let $\Pi_E\coloneqq J_EJ_E^\dagger$. Equations~\eqref{eq:ideal-key-purification} and \eqref{eq:shield-purification-unitaries} give
\begin{equation}\label{eq:ideal-key-in-range}
\ket{\gamma_E}_{K\widehat K LTA_E}
=(J_E\otimes I_{A_E})\ket{\chi_1}_{LTA_E}.
\end{equation}
Since $J_E^\dagger J_E=I_{LT}$, it follows that
\begin{align}
(\Pi_E\otimes I_{A_E})\ket{\gamma_E}_{K\widehat K LTA_E}
&=(J_EJ_E^\dagger J_E\otimes I_{A_E})\ket{\chi_1}_{LTA_E}
\label{eq:ideal-test-range-identity}\\
&=\ket{\gamma_E}_{K\widehat K LTA_E}.
\label{eq:ideal-test-acceptance}
\end{align}
We can thus insert the projector into the overlap in \eqref{eq:ideal-coherent-key}. Cauchy--Schwarz and normalization of $\ket{\gamma_E}$ imply
\begin{align}
f_E
&=\left|\bra{\gamma_E}(\Pi_E\otimes I_{A_E})\ket{\psi_E}\right|^2
\label{eq:privacy-acceptance-overlap}\\
&\le\norm{\ket{\gamma_E}}^2
\norm{(\Pi_E\otimes I_{A_E})\ket{\psi_E}}^2
\label{eq:privacy-acceptance-cauchy}\\
&=\Tr\!\left[(\Pi_E\otimes I_{A_E})\ket{\psi_E}\bra{\psi_E}_{K\widehat K LTA_E}\right].
\label{eq:privacy-acceptance}
\end{align}
The final equality uses $\Pi_E^\dagger\Pi_E=\Pi_E$.

\emph{Pulling the test back to a common input space.}
The isometry $U_E$ in \eqref{eq:conditional-measurement-isometry} has input dimension $d^{n-|E|}$ and output dimension $M^2d^{n-|E|}$. Choose $\dim G_E=M^2$, fix $\ket0_{G_E}$, and prescribe
\begin{equation}\label{eq:unitary-decoder-extension}
D_E(\ket\xi_{A_{E^c}}\otimes\ket0_{G_E})
\coloneqq U_E\ket\xi_{A_{E^c}}.
\end{equation}
The full input and output dimensions are equal. Completing orthonormal bases therefore extends this prescription to a unitary $D_E\colon A_{E^c}G_E\to\widehat K T_E$, where $T_E$ is the copy of $T$ associated with this pattern. Give each pattern its own auxiliary factor $G_E$.

For the following calculation, let $\widehat D_E\coloneqq I_{KLA_E}\otimes D_E$ denote the unitary extended to the sender and erased systems. Let
\begin{equation}\label{eq:single-decoder-input-state}
\Xi_{KLA^nG_E}\coloneqq\ket\Psi\bra\Psi_{KLA^n}\otimes\ket0\bra0_{G_E}.
\end{equation}
Equations~\eqref{eq:conditional-purified-output} and \eqref{eq:unitary-decoder-extension} imply
\begin{equation}\label{eq:unitary-output-identity}
\widehat D_E\Xi_{KLA^nG_E}\widehat D_E^\dagger
=\ket{\psi_E}\bra{\psi_E}_{K\widehat K LT_EA_E}.
\end{equation}
For this calculation only, write $P_E^{(0)}$ for $\widehat D_E^\dagger(\Pi_E\otimes I_{A_E})\widehat D_E$, before extending it to the other auxiliary factors. The acceptance probability is then
\begin{align}
\Tr\!\left[(\Pi_E\otimes I_{A_E})\ket{\psi_E}\bra{\psi_E}\right]
&=\Tr\!\left[(\Pi_E\otimes I_{A_E})\widehat D_E\Xi_{KLA^nG_E}\widehat D_E^\dagger\right]
\label{eq:privacy-pullback-output}\\
&=\Tr\!\left[\widehat D_E^\dagger(\Pi_E\otimes I_{A_E})\widehat D_E\Xi_{KLA^nG_E}\right]
\label{eq:privacy-pullback-cyclicity}\\
&=\Tr(P_E^{(0)}\Xi_{KLA^nG_E})
=\Tr(P_E\widetilde\Psi_{KLA^nG_{\mathrm{all}}}).
\label{eq:privacy-pullback-common-state}
\end{align}
The second line uses cyclicity of trace. For the last equality, $P_E$ is $P_E^{(0)}$ tensored with the identity on all $G_F$ with $F\ne E$, and every initial state $\ket0\bra0_{G_F}$ has trace one. Combining \eqref{eq:privacy-acceptance} with \eqref{eq:privacy-pullback-common-state} proves \eqref{eq:layer-acceptance}.

We also verify that these are projectors on the full space. Unitarity of $\widehat D_E$ and self-adjointness of $\Pi_E$ give
\begin{equation}\label{eq:decoder-projector-selfadjoint}
(P_E^{(0)})^\dagger
=\widehat D_E^\dagger(\Pi_E^\dagger\otimes I_{A_E})\widehat D_E
=P_E^{(0)}.
\end{equation}
Their squares satisfy
\begin{align}
(P_E^{(0)})^2
&=\widehat D_E^\dagger(\Pi_E\otimes I_{A_E})
\widehat D_E\widehat D_E^\dagger
(\Pi_E\otimes I_{A_E})\widehat D_E
\label{eq:decoder-projector-square}\\
&=\widehat D_E^\dagger(\Pi_E^2\otimes I_{A_E})\widehat D_E
=P_E^{(0)}.
\label{eq:decoder-projector-idempotent}
\end{align}
Tensoring with identities preserves both properties, so $P_E^\dagger=P_E^2=P_E$ on \eqref{eq:common-space}.

To obtain the average bound, set
\begin{equation}\label{eq:average-test-operator}
A_{n,k}\coloneqq\frac1{\binom nk}\sum_{|E|=k}P_E.
\end{equation}
This is a positive semidefinite operator and hence $0\le A_{n,k}\le\norm{A_{n,k}}I$. The same normalized state $\widetilde\Psi$ occurs for every $E$. Therefore
\begin{align}
\overline f_{n,k}
&=\frac1{\binom nk}\sum_{|E|=k}f_E
\label{eq:average-private-acceptance-start}\\
&\le\frac1{\binom nk}\sum_{|E|=k}
\Tr(P_E\widetilde\Psi_{KLA^nG_{\mathrm{all}}})
\label{eq:average-private-acceptance}\\
&=\Tr(A_{n,k}\widetilde\Psi_{KLA^nG_{\mathrm{all}}})
\label{eq:average-private-linearity}\\
&\le\norm{A_{n,k}}\Tr\widetilde\Psi_{KLA^nG_{\mathrm{all}}}
=\norm{A_{n,k}}.
\label{eq:average-private-norm}
\end{align}
The first line is \eqref{eq:conditional-fidelity}, the second uses \eqref{eq:layer-acceptance}, and the third is linearity of trace. The last line proves \eqref{eq:layer-average-bound}.

\emph{The Bell-test specialization.}
For entanglement generation, identify $K$ with $R$ and $\widehat K$ with $\widehat S$, take $L$ to be one dimensional, and choose $V_m\coloneqq I_T$. Then \eqref{eq:privacy-test} becomes
\begin{equation}\label{eq:bell-test-specialization}
\Pi_E=\Phi^M_{R\widehat S}\otimes I_{T_E}.
\end{equation}
To obtain a unitary realization for an arbitrary decoder $\cD_E$, set $b_E\coloneqq\dim A_{E^c}$. A channel from a $b_E$-dimensional input to an $M$-dimensional output has a Stinespring isometry with environment dimension at most $Mb_E$: this follows, for example, by taking one Kraus operator for each nonzero eigenvalue of its Choi operator, whose rank is at most $Mb_E$. Pad that environment to dimension $Mb_E$ and choose $\dim G_E=M^2$. The input and output dimensions are both $M^2b_E$, so the orthonormal-basis extension used in \eqref{eq:unitary-decoder-extension} applies again.

Define
\begin{equation}\label{eq:quantum-common-state}
\widetilde\rho_{RA^nG_{\mathrm{all}}}
\coloneqq\rho_{RA^n}\otimes\bigotimes_F\ket0\bra0_{G_F}.
\end{equation}
Pulling back the Bell test and then tracing out its retained decoder system gives
\begin{align}
\Tr(P_E\widetilde\rho_{RA^nG_{\mathrm{all}}})
&=\Tr\!\left[\Phi^M_{R\widehat S}
(\id_R\otimes\cD_E)(\Tr_{A_E}\rho_{RA^n})\right]
\label{eq:bell-test-pullback}\\
&=F_E.
\label{eq:bell-test-conditional-fidelity}
\end{align}
The final equality is \eqref{eq:conditional-quantum-definition}. Averaging these equalities yields
\begin{equation}\label{eq:quantum-layer-operator}
\overline F_{n,k}
=\Tr(A_{n,k}\widetilde\rho_{RA^nG_{\mathrm{all}}})
\le\norm{A_{n,k}}.
\end{equation}
Only positivity and unit trace of $\widetilde\rho$ are used in the last inequality. No condition on $\rho_R$ is needed, and the prepared reference--input state may be mixed.
\end{proof}

The label $\widehat K$ in \eqref{eq:decoder-unitary} denotes the output of whichever conditional decoder is being considered. These are alternative output identifications; $\widehat K$ is not a shared receiver input in \eqref{eq:common-space}. Keeping the full auxiliary inputs is important: compressing $G_E$ to its initial vector can turn a projector into a general positive measurement effect. All operator norms below are taken before this compression. The distinct $G_E$ also ensure that two decoder placements share only their physical surviving input, not an artificial auxiliary system.

\paragraph{Relation to the earlier reduction.}
Lemma~\ref{lem:lift} specializes the coherent reduction in Ref.~\cite[Lemma~3.1, Eqs.~(3.4)--(3.12)]{WildePrivate2026}, together with its unitary-decoder and common-space constructions in Sections~3.2 and~4.1. The separate Lemma~3.2 of that reference is the overlap estimate used below. Both reductions allow mixed message states and unrestricted local purification systems. The present proof makes the dependence on the erasure pattern explicit: the decoder, optimizing environment state, and privacy test may depend on $E$, while the preparation $\ket\Psi_{KLA^n}$ and the enlarged Hilbert space stay fixed. This gives the average in \eqref{eq:layer-average-bound} even when the conditional fidelities differ from one pattern to another. The Bell-test construction also records the argument for arbitrary, possibly mixed entanglement generation inputs. These facts allow the fixed-count converse to be applied directly to the family of conditional erasure decoders.

\subsection{Overlap depends only on the common receiver input}
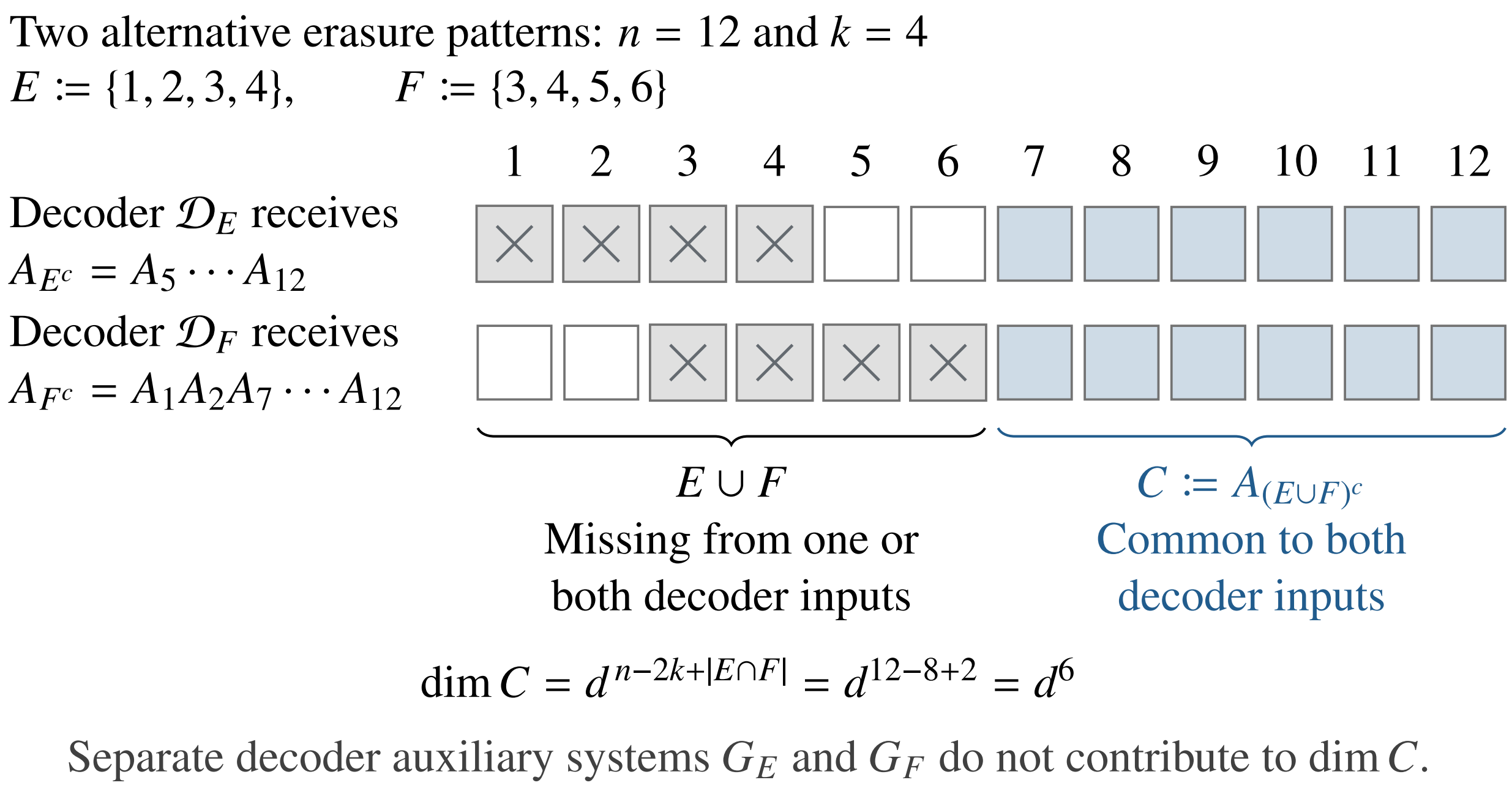
\begin{figure}
\centering
\resizebox{\textwidth}{!}{
\begin{tikzpicture}[font=\small,x=1cm,y=1cm,
  cell/.style={draw=black!55,line width=.55pt,minimum width=.62cm,
    minimum height=.62cm,inner sep=0pt},
  brace/.style={decorate,decoration={brace,mirror,amplitude=4pt},line width=.6pt}
]
\node[anchor=west] at (0,3.0)
  {Two alternative erasure patterns: $n=12$ and $k=4$};
\node[anchor=west] at (0,2.55)
  {$E\coloneqq\{1,2,3,4\},\qquad F\coloneqq\{3,4,5,6\}$};
\foreach \i in {1,...,12}{
  \pgfmathsetmacro{\xx}{4.4+.73*(\i-1)}
  \node at (\xx,1.95) {$\i$};
  \ifnum\i>6
    \node[cell,fill=erasureBlue!22] (a\i) at (\xx,1.25) {};
    \node[cell,fill=erasureBlue!22] (b\i) at (\xx,.25) {};
  \else
    \node[cell,fill=white] (a\i) at (\xx,1.25) {};
    \node[cell,fill=white] (b\i) at (\xx,.25) {};
  \fi
  \ifnum\i<5
    \fill[black!12] (a\i.south west) rectangle (a\i.north east);
    \draw[black!55,line width=.55pt] (a\i.south west) rectangle (a\i.north east);
    \draw[erasureGray,line width=.7pt] ([xshift=1.7mm,yshift=1.7mm]a\i.south west)
      --([xshift=-1.7mm,yshift=-1.7mm]a\i.north east);
    \draw[erasureGray,line width=.7pt] ([xshift=1.7mm,yshift=-1.7mm]a\i.north west)
      --([xshift=-1.7mm,yshift=1.7mm]a\i.south east);
  \fi
  \ifnum\i>2
    \ifnum\i<7
      \fill[black!12] (b\i.south west) rectangle (b\i.north east);
      \draw[black!55,line width=.55pt] (b\i.south west) rectangle (b\i.north east);
      \draw[erasureGray,line width=.7pt] ([xshift=1.7mm,yshift=1.7mm]b\i.south west)
        --([xshift=-1.7mm,yshift=-1.7mm]b\i.north east);
      \draw[erasureGray,line width=.7pt] ([xshift=1.7mm,yshift=-1.7mm]b\i.north west)
        --([xshift=-1.7mm,yshift=1.7mm]b\i.south east);
    \fi
  \fi
}
\node[anchor=west,align=left] at (0,1.25)
  {Decoder $\mathcal D_E$ receives\\$A_{E^c}=A_5\cdots A_{12}$};
\node[anchor=west,align=left] at (0,.25)
  {Decoder $\mathcal D_F$ receives\\$A_{F^c}=A_1A_2A_7\cdots A_{12}$};
\draw[brace] (4.09,-.3)--(8.36,-.3)
  node[midway,below=5pt,align=center] {$E\cup F$\\Missing from one or\\both decoder inputs};
\draw[brace,draw=erasureBlue] (8.47,-.3)--(12.74,-.3)
  node[midway,below=5pt,align=center,text=erasureBlue]
  {$C\coloneqq A_{(E\cup F)^c}$\\Common to both\\decoder inputs};
\node[align=center] at (6.37,-2.4)
  {$\dim C=d^{\,n-2k+|E\cap F|}=d^{12-8+2}=d^6$};
\node[align=center,text=black!75] at (6.37,-3.1)
  {Separate decoder auxiliary systems $G_E$ and $G_F$ do not contribute to $\dim C$.};
\end{tikzpicture}}
\caption{The physical input shared by two possible receivers. Crossed cells are erased and blue cells survive both patterns. Here each erased set has four elements and their intersection has two, leaving six common surviving systems. Their dimension gives the numerator in Lemma~\ref{lem:overlap}. The sender key and shield belong to both mathematical tests, but introduce no dimension factor. The diagram compares alternative erasure outcomes, not simultaneous recoveries.}
\label{fig:shared}
\end{figure}

The following bound is Ref.~\cite[Lemma~3.2]{WildePrivate2026}; its Bell-test specialization is Ref.~\cite[Lemma~S4]{Kondra2026}. We state the bound here and give its complete proof in Appendix~\ref{app:overlap-proof}. The erasure-specific application and the explanation of why the sender system contributes no dimension factor remain below. Figure~\ref{fig:shared} identifies the common physical receiver input.

\Needspace{5\baselineskip}
\begin{lemma}[Privacy-test overlap]\label{lem:overlap}
Let $M\ge1$, and let $P$ and $Q$ be two size-$M$ privacy-test projectors of the form \eqref{eq:privacy-test}, pulled back through receiver unitaries as in \eqref{eq:decoder-projector}. Both tests share sender systems $K,L$, with $\dim K=M$, and use the same orthonormal basis on $K$. Suppose the receiver input spaces are $XC$ and $CY$, respectively, where $X,C,Y$ are distinct tensor factors. Each test acts as the identity on the other receiver's exclusive factor; receiver auxiliary inputs are included separately in $X$ and $Y$. Then
\begin{equation}\label{eq:overlap-general}
\norm{PQ}\le\min\!\left\{1,\frac{\dim C}{M}\right\}.
\end{equation}
In particular, for $n\ge1$, $d\ge2$, and two distinct size-$k$ sets $E,F\subseteq[n]$, the projectors constructed in Lemma~\ref{lem:lift} satisfy
\begin{equation}\label{eq:overlap}
\norm{P_EP_F}\le\min\left\{1,\frac{d^{n-2k+|E\cap F|}}M\right\}.
\end{equation}
\end{lemma}
\begin{proof}[Application to erasure patterns]
The general inequality \eqref{eq:overlap-general} is proved in Appendix~\ref{app:overlap-proof}. We apply it with exclusive receiver inputs $X\coloneqq A_{F\setminus E}G_E$ and $Y\coloneqq A_{E\setminus F}G_F$.
For size-$k$ erasure patterns $E,F$, the common receiver input is $C=A_{(E\cup F)^c}$. Inclusion--exclusion gives $|E\cup F|=2k-|E\cap F|$, and therefore
\begin{equation}\label{eq:common-dimension}
\dim C=d^{n-|E\cup F|}=d^{n-2k+|E\cap F|}.
\end{equation}
The factors $G_E$ and $G_F$ belong to the two exclusive receiver inputs. The unused physical factor $A_{E\cap F}$ and the auxiliaries $G_H$ with $H\notin\{E,F\}$ carry identity operators, which do not change the operator norm. Substituting \eqref{eq:common-dimension} into \eqref{eq:overlap-general} proves \eqref{eq:overlap}.
\end{proof}

\paragraph{Why the sender system contributes no dimension factor.}
The privacy-test structure treats the shared sender system $L$ differently from a shared receiver input. For disjoint receiver inputs, denote the output shield systems by $T_1,T_2$. Let $V_j$ and $W_j$ be the two tests' shield unitaries on $LT_1$ and $LT_2$, extended by identities to $LT_1T_2$, for a key label $j\in\{1,\ldots,M\}$. Their ordered product in \eqref{eq:disjoint-privacy-overlap} is unitary because
\begin{equation}\label{eq:sender-shield-unitarity}
(V_j^\dagger W_j)^\dagger(V_j^\dagger W_j)
=W_j^\dagger V_jV_j^\dagger W_j=I_{LT_1T_2}.
\end{equation}
No commutation on $L$ is required. In \eqref{eq:disjoint-privacy-overlap}, each key label selects a block $M^{-1}V_j^\dagger W_j$, so every block has norm $1/M$. Equations~\eqref{eq:disjoint-overlap-norm-square}--\eqref{eq:disjoint-projector-lower} give the same norm for the product of the projectors, independently of $\dim L$. Correlations with $L$ are unrestricted. For a common receiver input $C$, the identity \eqref{eq:swap-link} introduces a formal copy $C'$ and traces out only $C'$. The estimate \eqref{eq:partial-trace-operator-bound} costs $\dim C$, while $L$ remains inside the operator throughout. No partial trace over $L$ or bound proportional to $\dim L$ is used. This absence of a sender-shield factor is already part of Ref.~\cite[Lemma~3.2, Eq.~(3.19)]{WildePrivate2026}.

For erasures, the entire common receiver input is $A_{(E\cup F)^c}$, whose dimension is exactly \eqref{eq:common-dimension}. The separate auxiliary inputs $G_E,G_F$ are exclusive and contribute no factor. This exact physical dimension, together with the bound's independence of the sender shield, is what permits the private converse. The general degradable-channel application in Ref.~\cite[Lemma~5.1 and Section~6]{WildePrivate2026} instead estimates a common receiver dimension using a linear transformation, normalization, and approximate recovery. The erasure argument requires no such state-dependent estimate.

\subsection{From pairwise overlap to an average}
Selected pairwise constraints do not yet bound an average over every pattern. The next lemma adapts the signed-averaging argument of Ref.~\cite[Lemma~S1]{Kondra2026} and sharpens Ref.~\cite[Lemma~4.2]{WildePrivate2026}. Keeping the norms of the projected vectors gives a linear dependence on the overlap bound. We explain the precise comparison after the proof.

\Needspace{5\baselineskip}
\begin{lemma}[Signed averaging of projectors]\label{lem:average}
Let $\Omega$ be a finite nonempty index set, and let $(P_x)_{x\in\Omega}$ be projectors on one finite-dimensional Hilbert space. Let $\one\in\bbR^\Omega$ be the all-ones vector and define $U\coloneqq |\Omega|^{-1}\one\one^{\mathsf T}$, the matrix that replaces a vector by its coordinate average. Suppose a real symmetric matrix $W$ on $\bbR^\Omega$ and nonnegative constants $\zeta,\ell,\eta$ satisfy
\begin{subequations}\label{eq:average-hypotheses}
\begin{align}
\norm{W-U}&\le\zeta,
\label{eq:average-spectral-hypothesis}\\
\max_x\sum_y|W_{xy}|&\le\ell,
\label{eq:average-row-hypothesis}\\
W_{xy}\ne0&\ \Longrightarrow\ \norm{P_xP_y}\le\eta.
\label{eq:average-overlap-hypothesis}
\end{align}
\end{subequations}
Then
\begin{equation}\label{eq:average-conclusion}
\norm{\frac1{|\Omega|}\sum_xP_x}\le\zeta+\ell\eta.
\end{equation}
\end{lemma}
\begin{proof}
The quantity to be bounded is the largest eigenvalue of the average of the projectors. We evaluate that average on one maximizing vector, put all its projections into a single vector indexed by $\Omega$, and compare the quadratic forms of $U$ and $W$ on that vector. The support condition on $W$ is what allows the pairwise overlap bound to enter.

Define
\begin{equation}\label{eq:average-operator-definition}
A\coloneqq\frac1{|\Omega|}\sum_xP_x,
\qquad a\coloneqq\norm A.
\end{equation}
Since $A$ is positive semidefinite, there is a unit eigenvector $\ket\psi$ with $A\ket\psi=a\ket\psi$. If $a=0$, \eqref{eq:average-conclusion} is immediate. Assume henceforth that $a>0$. Introduce an auxiliary index space with orthonormal basis $(\ket x)_{x\in\Omega}$ and define
\begin{equation}\label{eq:average-projected-vectors}
\ket{v_x}\coloneqq P_x\ket\psi,
\qquad
\ket v\coloneqq\sum_{x\in\Omega}\ket x\otimes\ket{v_x}.
\end{equation}
The real symmetric matrices $U$ and $W$ act on this complex index space through the same matrix entries. Their eigenvalues, and hence their operator norms, are unchanged.
Orthogonality of the index labels and the projector identities give
\begin{align}
\norm{\ket v}^2
&=\sum_x\norm{\ket{v_x}}^2
=\sum_x\bra\psi P_x^\dagger P_x\ket\psi
\label{eq:average-vector-norm-expansion}\\
&=\sum_x\bra\psi P_x\ket\psi
=|\Omega|\bra\psi A\ket\psi=|\Omega|a.
\label{eq:average-vector-norm}
\end{align}
The uniform matrix has every entry equal to $1/|\Omega|$. Consequently,
\begin{align}
\bra v(U\otimes I)\ket v
&=\frac1{|\Omega|}\sum_{x,y}\braket{v_x}{v_y}
\label{eq:average-uniform-form}\\
&=\frac1{|\Omega|}\norm{\sum_x\ket{v_x}}^2
=\frac1{|\Omega|}\norm{|\Omega|A\ket\psi}^2
\label{eq:average-vector-sum}\\
&=|\Omega|a^2.
\label{eq:average-vector-identities}
\end{align}
The identity operator here acts on the physical Hilbert space of the projectors; $U$ acts only on the index space.

The perturbation from $U$ to $W$ is controlled by the operator norm in \eqref{eq:average-spectral-hypothesis}. For any vector $\ket v$ and operator $B$, Cauchy--Schwarz gives $|\bra vB\ket v|\le\norm{\ket v}\norm{B\ket v}\le\norm B\norm{\ket v}^2$. Applying this with $B=(W-U)\otimes I$ yields
\begin{align}
\left|\bra v((W-U)\otimes I)\ket v\right|
&\le\norm{(W-U)\otimes I}\norm{\ket v}^2
\label{eq:average-perturbation-norm}\\
&=\norm{W-U}\,|\Omega|a
\le\zeta |\Omega|a.
\label{eq:average-perturbation}
\end{align}
The equality uses \eqref{eq:average-vector-norm} and the fact that tensoring an operator with an identity does not change its operator norm. Combining \eqref{eq:average-vector-identities} and \eqref{eq:average-perturbation} gives
\begin{equation}\label{eq:average-perturbation-comparison}
|\Omega|a^2\le\left|\bra v(W\otimes I)\ket v\right|+\zeta |\Omega|a.
\end{equation}

It remains to estimate the quadratic form of $W$. Since $P_x\ket{v_x}=\ket{v_x}$ and $P_y\ket{v_y}=\ket{v_y}$, self-adjointness of the projectors gives
\begin{equation}\label{eq:average-supported-inner-product}
\braket{v_x}{v_y}=\bra{v_x}P_xP_y\ket{v_y}.
\end{equation}
On an entry with $W_{xy}\ne0$, \eqref{eq:average-overlap-hypothesis} therefore implies
\begin{align}
|\braket{v_x}{v_y}|
&\le\norm{\ket{v_x}}\norm{P_xP_y\ket{v_y}}
\label{eq:average-inner-product-cauchy}\\
&\le\norm{P_xP_y}\norm{\ket{v_x}}\norm{\ket{v_y}}
\le\eta\norm{\ket{v_x}}\norm{\ket{v_y}}.
\label{eq:average-inner-product-bound}
\end{align}
Let $r_x\coloneqq\norm{\ket{v_x}}$, so $r_x\ge0$ and $\sum_xr_x^2=|\Omega|a$ by \eqref{eq:average-vector-norm}. Expanding the quadratic form and then using $2r_xr_y\le r_x^2+r_y^2$ gives
\begin{align}
\left|\bra v(W\otimes I)\ket v\right|
&=\left|\sum_{x,y}W_{xy}\braket{v_x}{v_y}\right|
\le\sum_{x,y}|W_{xy}|\,|\braket{v_x}{v_y}|
\label{eq:average-quadratic-expansion}\\
&\le\eta\sum_{x,y}|W_{xy}|r_xr_y
\le\frac\eta2\sum_{x,y}|W_{xy}|(r_x^2+r_y^2)
\label{eq:average-quadratic-squares}\\
&=\frac\eta2\left[
\sum_xr_x^2\sum_y|W_{xy}|
+\sum_yr_y^2\sum_x|W_{xy}|\right]
\le\frac\eta2\left[\ell\sum_xr_x^2+\ell\sum_yr_y^2\right]
=\ell\eta |\Omega|a.
\label{eq:average-quadratic-bound}
\end{align}
The overlap estimate is needed only when $W_{xy}\ne0$, because all other terms vanish. For the last inequality, \eqref{eq:average-row-hypothesis} bounds every absolute row sum, and symmetry gives the same bound for every absolute column sum: $\sum_x|W_{xy}|=\sum_x|W_{yx}|\le\ell$.

Substituting \eqref{eq:average-quadratic-bound} into \eqref{eq:average-perturbation-comparison} proves
\begin{equation}\label{eq:average-final-quadratic}
|\Omega|a^2\le(\zeta+\ell\eta)|\Omega|a.
\end{equation}
Because $|\Omega|a>0$, division gives $a\le\zeta+\ell\eta$, which is \eqref{eq:average-conclusion}.
\end{proof}

\paragraph{Comparison of the averaging estimates.}
Under the same three hypotheses \eqref{eq:average-hypotheses}, Ref.~\cite[Lemma~4.2, Eqs.~(4.27)--(4.28)]{WildePrivate2026} gives $\norm A\le\zeta+\sqrt{\ell\eta}$, where $A$ is the average in \eqref{eq:average-operator-definition}. Its proof bounds each supported inner product by $\eta$, giving $\norm A^2\le\zeta\norm A+\ell\eta$. Here \eqref{eq:average-inner-product-bound} retains the factors $\norm{P_x\ket\psi}$ and $\norm{P_y\ket\psi}$. The sum in \eqref{eq:average-quadratic-bound} consequently has an extra factor $\norm A$, and division gives $\norm A\le\zeta+\ell\eta$. This is sharper whenever $0<\ell\eta<1$. Writing an arbitrary finite index set instead of binary strings makes the subset application explicit; the prior proof also uses no property of the labels beyond their number. Neither averaging lemma assumes symmetry of the projectors or of an input state. The additional ingredient for the second-order converse is the matrix on fixed-size subsets constructed next.

The matrix $W$ need not be a probability transition matrix or a positive semidefinite operator. It is used only in this inequality. We next construct it so that $\zeta$ is small, its support makes $\eta$ small, and the logarithm of $\ell$ remains negligible compared with $\sqrt n$.

\paragraph{What the averaging construction must achieve.}
Section~\ref{sec:filter} has one concrete objective. For the size-$k$ erased sets, construct a symmetric $W$ and an integer $t=O(\ln n)<k$ such that
\begin{subequations}\label{eq:construction-contract}
\begin{align}
\norm{W-U}&\le2n^{-2} &&\text{(accuracy)},
\label{eq:construction-accuracy}\\
W_{EF}\ne0&\ \Longrightarrow\ |E\cap F|\le t &&\text{(permitted overlaps)},
\label{eq:construction-support}\\
\log_d\ell&=O((\ln n)^2) &&\text{(cost of signed weights)}.
\label{eq:construction-cost}
\end{align}
\end{subequations}
Here $U\coloneqq\binom nk^{-1}\one\one^{\mathsf T}$ and $\ell\coloneqq\max_E\sum_F|W_{EF}|$. The exact degree and row-sum bound will be specified in \eqref{eq:filter-constants} and \eqref{eq:filter-row-budget}. Lemma~\ref{lem:overlap} then allows the choice $\eta\coloneqq d^{n-2k+t}/M$ in Lemma~\ref{lem:average}. Each extra common erased position costs one unit in $\log_dM$, while the signed weights cost $\log_d\ell$ units. Their total cost is $O((\ln n)^2)=o(\sqrt n)$, which explains the quantitative requirements.

\section{Averaging erasure patterns and the fixed-count converse}\label{sec:filter}
Lemma~\ref{lem:average} reduces the converse to choosing scalar weights for pairs of erasure patterns. The weights must approximate uniform averaging, use only pairs with few common erased positions, and have a controlled sum of absolute values. We state the required matrix estimate first and then show explicitly how its three bounds give the fixed-count converse. Appendix~\ref{app:tensor-proof} constructs the matrix and proves every estimate using tensor products of two-dimensional matrices and a classical polynomial identity.

\subsection{An averaging matrix with small intersections}
For integers $n,k$ with $0\le k\le n$, define the pattern space and its uniform averaging matrix by
\begin{equation}\label{eq:subset-space}
\Omega_{n,k}\coloneqq\{E\subseteq[n]:|E|=k\},
\qquad
U\coloneqq\frac1{\binom nk}\one\one^{\mathsf T},
\qquad \one\in\bbR^{\Omega_{n,k}}.
\end{equation}
Here $\one$ is the all-ones vector. A vector $v\in\bbR^{\Omega_{n,k}}$ assigns a real number $v(E)$ to each pattern, and $Uv$ replaces every entry by their average. Thus an operator-norm bound on $W-U$ controls the error of this replacement for every unit vector $v$ at once. The matrices $W,U$ act on pattern labels, whereas the privacy tests $P_E$ act on the code's enlarged Hilbert space.

The second matrix norm that we need is the absolute row-sum norm:
\begin{equation}\label{eq:row-norm}
\norm A_{\row}\coloneqq\max_x\sum_y|A_{xy}|.
\end{equation}
It measures the size of signed coefficients in a row. A row sum equal to one need not have absolute row sum equal to one, because positive and negative entries can cancel.

Fix $p_0\in(0,1/2)$. The constants below depend only on this fixed upper bound on $k/n$:
\begin{align}
r_0&\coloneqq\frac{p_0}{1-p_0},\qquad c_0\coloneqq-\ln r_0,\label{eq:filter-decay-constants}\\
c&\coloneqq r_0^2,\qquad
\Gamma_0\coloneqq1+3(c^{-1}-1),\label{eq:filter-coefficient-constants}\\
t&\coloneqq\left\lceil\frac{4\ln n}{c_0}\right\rceil.
\label{eq:filter-constants}
\end{align}
In particular, $0<r_0,c<1$ and $c_0>0$. Define the row-sum bound
\begin{equation}\label{eq:filter-row-budget}
\ell_n\coloneqq2(n+1)(\Gamma_0n)^t.
\end{equation}
The degree $t$ will bound the number of positions at which the construction can introduce an intersection.

\Needspace{8\baselineskip}
\begin{lemma}[Averaging patterns with small intersections]\label{lem:filter}
Fix $p_0\in(0,1/2)$ and an integer $n\ge3$. Let $r_0,c_0,c,\Gamma_0,t$ be the parameters in \eqref{eq:filter-decay-constants}--\eqref{eq:filter-constants}, and let $\ell_n$ be defined in \eqref{eq:filter-row-budget}. For every integer $k$ with $t<k\le p_0n$, there is a real symmetric matrix $W$ on the pattern space $\Omega_{n,k}$ in \eqref{eq:subset-space} satisfying
\begin{align}
W\one&=\one,\label{eq:filter-row-sum}\\
W_{EF}\ne0&\ \Longrightarrow\ |E\cap F|\le t,\label{eq:filter-support}\\
\norm{W-U}&\le2n^{-2},\label{eq:filter-error}\\
\norm W_{\row}&\le\ell_n.\label{eq:filter-mass}
\end{align}
Here $E,F\in\Omega_{n,k}$, $U,\one$ are defined in \eqref{eq:subset-space}, and the row-sum norm is defined in \eqref{eq:row-norm}. In particular, $W_{EE}=0$. The same constants apply to every allowed $k$.
\end{lemma}

The proof in Appendix~\ref{app:tensor-proof} first works with all subsets, giving each position an independent probability $\alpha\coloneqq k/n$ of being included. A two-dimensional matrix prevents a position from belonging to both patterns. Its two eigenvalues are $1$ and $-r$, where $r\coloneqq\alpha/(1-\alpha)<1$. In the resulting product eigenbasis, a vector with exactly $i$ nonconstant one-position factors has eigenvalue $(-r)^i$. The proof chooses a real polynomial $q_t$ of degree at most $t$ satisfying
\begin{equation}\label{eq:filter-polynomial-target}
q_t(0)=1,\qquad
\max_{1\le i\le n}r^i|q_t(i)|\le2n^{-4}.
\end{equation}
The first condition preserves the constant vector. The second suppresses all other product basis vectors; see \eqref{eq:bt-polynomial-spectral-reduction} and Lemma~\ref{lem:geometric-polynomial}. The geometric factor $r^i$ makes a logarithmic degree sufficient. The required polynomial identity is a classical Meixner orthogonality relation~\cite[Table~18.19.1 and Eq.~18.23.4]{DLMF}; Appendix~\ref{app:tensor-proof} derives it from a geometric series.

Finally, we keep only rows and columns corresponding to sets of size $k$ and normalize their common row sum. The auxiliary product distribution gives this layer probability at least $1/(n+1)$, so the normalization costs only a polynomial factor. A degree-$t$ polynomial permits at most $t$ single-position changes after a disjoint choice, which proves the support restriction. This auxiliary construction does not change the physical erasure probability or add resources to the code.

The signed-averaging approach follows Refs.~\cite{Kondra2026,WildePrivate2026}, and the biased disjointness construction has a classical precedent in Friedgut's method~\cite{Friedgut2008,Filmus2013}. The needed estimates here combine that construction, the Meixner polynomial identity, and restriction to a fixed-size layer. In particular, \eqref{eq:filter-constants}--\eqref{eq:filter-row-budget} give
\begin{equation}\label{eq:filter-cost-orders}
t=O_{p_0}(\ln n),\qquad
\ln\ell_n=O_{p_0}((\ln n)^2).
\end{equation}
The next theorem shows why these two costs are small enough to preserve the Gaussian correction.

\subsection{Performance averaged over a fixed erasure count}
We can now combine Lemma~\ref{lem:overlap} with the matrix constructed in Lemma~\ref{lem:filter}. This is the main converse estimate. It concerns a fixed erasure count and is therefore independent of the erasure probability.

\Needspace{5\baselineskip}
\begin{theorem}[Averaged Singleton bound for privacy tests]\label{thm:fixed}
Fix an integer $d\ge2$ and a number $p_0\in(0,1/2)$. For each integer $n\ge3$, let $t,\ell_n$ be defined in \eqref{eq:filter-constants} and \eqref{eq:filter-row-budget}, and define
\begin{equation}\label{eq:hn}
h_n\coloneqq t+\log_d\ell_n+4\log_dn.
\end{equation}
For every integer $n\ge3$, every integer $M\ge1$, every integer $k$ with $t<k\le p_0n$, and every family of size-$M$ decoded privacy-test projectors $(P_E)_{|E|=k}$ on the common Hilbert space of Lemma~\ref{lem:lift},
\begin{equation}\label{eq:fixed-raw}
\norm{\frac1{\binom nk}\sum_{|E|=k}P_E}
\le2n^{-2}+\ell_n\frac{d^{n-2k+t}}M.
\end{equation}
Here the projectors are those in \eqref{eq:decoder-projector}, with the distinct receiver auxiliary systems specified in Lemma~\ref{lem:lift}. Consequently, \eqref{eq:fixed-raw} also bounds each fixed-count average in \eqref{eq:conditional-fidelity}: the optimized private average $\overline f_{n,k}$, defined using \eqref{eq:conditional-private-definition}, and the entanglement generation average $\overline F_{n,k}$, defined using \eqref{eq:conditional-quantum-definition}. Each bound holds for every code of its indicated type, separately. In particular,
\begin{equation}\label{eq:fixed-threshold}
\log_dM\ge n-2k+h_n
\quad\Longrightarrow\quad
\max\{\overline f_{n,k},\overline F_{n,k}\}\le3n^{-2}.
\end{equation}
The loss satisfies $h_n=O_{p_0,d}((\ln n)^2)$, and every bound is uniform over the code and its size $M$.
\end{theorem}
\begin{proof}
Fix $n,k,M$ as in the statement. We condition on exactly $k$ erasures throughout, so the erasure probability does not enter the proof. The task is to bound one code's average over all size-$k$ erased sets, even when the code treats the physical positions differently. We use Lemma~\ref{lem:lift} to express that average in terms of projectors on one state, Lemma~\ref{lem:overlap} to control selected pairs of projectors, and Lemmas~\ref{lem:filter} and~\ref{lem:average} to turn those pairwise bounds into a bound on the full average.

\emph{The average is an expectation of one operator.}
For convenience, restate the index set and its cardinality:
\begin{equation}\label{eq:fixed-index-set}
\Omega_{n,k}\coloneqq\{E\subseteq[n]:|E|=k\},
\qquad |\Omega_{n,k}|=\binom nk.
\end{equation}
Define the average of the projectors and the uniform averaging matrix by
\begin{align}
\overline P_{n,k}&\coloneqq\frac1{\binom nk}\sum_{E\in\Omega_{n,k}}P_E,
\label{eq:fixed-average-operator}\\
U&\coloneqq\frac1{\binom nk}\one\one^{\mathsf T},
\qquad \one\in\bbR^{\Omega_{n,k}}.
\label{eq:fixed-uniform-matrix}
\end{align}
Here $\one$ is the all-ones vector. The operators in these two lines act on different spaces. The operator $\overline P_{n,k}$ acts on the code's enlarged Hilbert space, whereas $U$ acts only on real vectors indexed by erasure patterns.

For a private code, Lemma~\ref{lem:lift}, specifically \eqref{eq:layer-acceptance}, gives a normalized state $\widetilde\Psi$ that is the same for every $E$, and the projectors satisfy
\begin{equation}\label{eq:fixed-private-acceptance}
f_E\le\Tr(P_E\widetilde\Psi)
\qquad(E\in\Omega_{n,k}).
\end{equation}
The state is the one in \eqref{eq:common-state}; its system labels are suppressed in this proof because every operator acts on the common space \eqref{eq:common-space}. Averaging \eqref{eq:fixed-private-acceptance} gives
\begin{align}
\overline f_{n,k}
&=\frac1{\binom nk}\sum_{E\in\Omega_{n,k}}f_E
\le\Tr(\overline P_{n,k}\widetilde\Psi),
\label{eq:fixed-private-average}\\
\Tr(\overline P_{n,k}\widetilde\Psi)
&\le\norm{\overline P_{n,k}}\Tr\widetilde\Psi
=\norm{\overline P_{n,k}}.
\label{eq:fixed-state-norm}
\end{align}
The last line uses $0\le\overline P_{n,k}\le\norm{\overline P_{n,k}}I$ and $\Tr\widetilde\Psi=1$. For an entanglement generation code, the Bell-test part of Lemma~\ref{lem:lift}, specifically \eqref{eq:quantum-layer-operator}, gives a possibly mixed normalized state $\widetilde\rho$ on the common space and
\begin{align}
F_E&=\Tr(P_E\widetilde\rho),
\label{eq:fixed-quantum-acceptance}\\
\overline F_{n,k}
&=\Tr(\overline P_{n,k}\widetilde\rho)
\le\norm{\overline P_{n,k}}.
\label{eq:fixed-quantum-average}
\end{align}
Thus the operator-norm estimate \eqref{eq:fixed-raw} proves both operational bounds without any assumption that the encoder is invariant under permutations.

\emph{Only pairs with small intersections are needed.}
Because $n\ge3$ and $t<k\le p_0n$, Lemma~\ref{lem:filter} supplies a real symmetric matrix $W$ on $\bbR^{\Omega_{n,k}}$ with
\begin{align}
\norm{W-U}&\le2n^{-2},
\label{eq:fixed-matrix-accuracy}\\
\max_{E\in\Omega_{n,k}}\sum_{F\in\Omega_{n,k}}|W_{EF}|
&\le\ell_n,
\label{eq:fixed-matrix-mass}\\
W_{EF}\ne0&\ \Longrightarrow\ |E\cap F|\le t.
\label{eq:fixed-matrix-support}
\end{align}
These are precisely \eqref{eq:filter-error}, \eqref{eq:filter-mass}, and \eqref{eq:filter-support}. The construction depends on $n,k,p_0$ and not on the code, its size $M$, or the dimensions of its auxiliary systems. Since $|E\cap E|=k>t$, \eqref{eq:fixed-matrix-support} implies
\begin{equation}\label{eq:fixed-zero-diagonal}
W_{EE}=0\qquad(E\in\Omega_{n,k}).
\end{equation}
Thus every nonzero entry $W_{EF}$ couples distinct patterns, as required when we use the erasure-pattern conclusion of Lemma~\ref{lem:overlap}.

For such a pair, decoder $E$ receives $A_{E^c}$ and decoder $F$ receives $A_{F^c}$. Their common physical input is $A_{(E\cup F)^c}$. The receiver auxiliary systems are distinct, so they add no further common input. Since $|E|=|F|=k$,
\begin{align}
\dim A_{(E\cup F)^c}
&=d^{n-|E\cup F|}
=d^{n-2k+|E\cap F|},
\label{eq:fixed-common-dimension}\\
\norm{P_EP_F}
&\le\frac{d^{n-2k+|E\cap F|}}M
\le\frac{d^{n-2k+t}}M.
\label{eq:supported-overlap}
\end{align}
The first inequality in \eqref{eq:supported-overlap} is Lemma~\ref{lem:overlap}; the second is the support property \eqref{eq:fixed-matrix-support}. It is harmless if the final upper bound exceeds one, because Lemma~\ref{lem:average} permits any nonnegative overlap bound.

\emph{Apply the signed-averaging estimate.}
Recall the row-sum bound $\ell_n$ from \eqref{eq:filter-row-budget}, and set
\begin{equation}\label{eq:fixed-averaging-parameters}
\zeta_n\coloneqq2n^{-2},\qquad
\ell_n=2(n+1)(\Gamma_0n)^t,\qquad
\eta_{n,k,M}\coloneqq\frac{d^{n-2k+t}}M.
\end{equation}
Equations~\eqref{eq:fixed-matrix-accuracy}, \eqref{eq:fixed-matrix-mass}, and \eqref{eq:supported-overlap} satisfy all three hypotheses of Lemma~\ref{lem:average}, with index set $\Omega_{n,k}$. Consequently,
\begin{align}
\norm{\overline P_{n,k}}
&\le\zeta_n+\ell_n\eta_{n,k,M},
\label{eq:fixed-averaging-application}\\
&=2n^{-2}+\ell_n\frac{d^{n-2k+t}}M.
\label{eq:fixed-raw-derived}
\end{align}
This proves \eqref{eq:fixed-raw}. Combining it with \eqref{eq:fixed-private-average}--\eqref{eq:fixed-quantum-average} proves the corresponding bounds for private communication and entanglement generation.

\emph{The chosen loss absorbs the signed weights.}
Now assume the code-size condition in \eqref{eq:fixed-threshold}, namely
\begin{equation}\label{eq:fixed-size-condition}
\log_dM\ge n-2k+h_n.
\end{equation}
Because $d>1$, this implies $M\ge d^{n-2k+h_n}$. We can therefore bound the second term in \eqref{eq:fixed-raw-derived} explicitly:
\begin{align}
\ell_n\frac{d^{n-2k+t}}M
&\le\ell_n d^{t-h_n},
\label{eq:threshold-algebra}\\
&=d^{\,\log_d\ell_n+t-h_n},
\label{eq:threshold-one-base}\\
&=d^{-4\log_dn},
\label{eq:threshold-cancellation}\\
&=n^{-4}.
\label{eq:threshold-power}
\end{align}
The equality \eqref{eq:threshold-cancellation} follows by substituting the definition of $h_n$ from \eqref{eq:hn}. The term $t$ pays for the additional common surviving systems in \eqref{eq:supported-overlap}, the term $\log_d\ell_n$ pays for the absolute row sum in \eqref{eq:fixed-matrix-mass}, and $4\log_dn$ leaves the factor $n^{-4}$. Hence
\begin{equation}\label{eq:fixed-final-small-bound}
\max\{\overline f_{n,k},\overline F_{n,k}\}
\le2n^{-2}+n^{-4}\le3n^{-2},
\end{equation}
which proves \eqref{eq:fixed-threshold}.

Finally, \eqref{eq:filter-constants} gives the explicit degree bound
\begin{equation}\label{eq:fixed-degree-growth}
t\le\frac{4\ln n}{c_0}+1.
\end{equation}
Substituting \eqref{eq:filter-row-budget} into \eqref{eq:hn}, and then using \eqref{eq:fixed-degree-growth}, yields
\begin{align}
h_n
&=t\left[1+\frac{\ln\Gamma_0+\ln n}{\ln d}\right]
  +\frac{\ln(2(n+1))+4\ln n}{\ln d},
\label{eq:fixed-loss-expansion}\\
&\le\left(\frac{4\ln n}{c_0}+1\right)
 \left[1+\frac{\ln\Gamma_0+\ln n}{\ln d}\right]
 +\frac{\ln(2(n+1))+4\ln n}{\ln d}.
\label{eq:fixed-loss-growth}
\end{align}
Since $c_0,\Gamma_0$ depend only on $p_0$, the right side is $O_{p_0,d}((\ln n)^2)$. The estimates hold for every permitted $n,k$ and are independent of $M$, the encoder, and the decoder family.
\end{proof}

The linear dependence in Lemma~\ref{lem:average} sharpens the finite-size estimate \eqref{eq:fixed-raw}. At the threshold \eqref{eq:fixed-size-condition}, the earlier square-root estimate in Ref.~\cite[Lemma~4.2]{WildePrivate2026} would also give $3n^{-2}$: \eqref{eq:threshold-power} bounds $\ell\eta$ by $n^{-4}$, so its square root is at most $n^{-2}$. Appendix~\ref{app:tensor-precedents} explains why the biased construction and restriction to a fixed-size layer permit a logarithmic polynomial degree.

No permutation symmetry is assumed of the encoder, the message states, or the decoders. The symmetry belongs to the matrix used to bound their average. The theorem also does not constrain each pattern separately: it says that one code above the threshold cannot maintain appreciable average fidelity on a uniformly chosen size-$k$ pattern.

\section{The Gaussian converses}\label{sec:normal-converse}
The operator estimate in Theorem~\ref{thm:fixed} has reduced both communication tasks to a threshold on the number of erasures. We first state the finite-block bounds, including the different square-root formula for the actual-marginal criterion, and then apply the same binomial quantile calculation to both.

\Needspace{5\baselineskip}
\begin{proposition}[Finite-block converses]\label{prop:finite-converse}
Fix an integer $d\ge2$ and probabilities $0<p<p_0<1/2$. For a blocklength $n$ and code size $M\ge1$, let $t,h_n$ be defined in \eqref{eq:filter-constants} and \eqref{eq:hn}, let $\Ner\sim\Bin(n,p)$ be the erasure count, and set
\begin{align}
a_n&\coloneqq\frac{n-\log_dM+h_n}{2},\label{eq:threshold-a}\\
\tau_n&\coloneqq\Prb\{\Ner\le t\}+\Prb\{\Ner>p_0n\}.
\label{eq:exceptional-count-probability}
\end{align}
There is an integer $n_0$, depending only on $d,p,p_0$, such that, for every integer $n\ge n_0$ and every integer $M\ge1$, every size-$M$ private code for $\cE_{p,d}^{\otimes n}$ obeys the following bounds on $f_{\opt}$ and $f_{\marg}$ from \eqref{eq:private-opt} and \eqref{eq:private-marg}. Every dimension-$M$ entanglement transmission or entanglement generation code for the same channel obeys the corresponding bound on its Bell-target fidelity $F_{\et}$ in \eqref{eq:transmission-fidelity} or $F_{\eg}$ in \eqref{eq:generation-fidelity}:
\begin{align}
f_{\opt},\ F_{\et},\ F_{\eg}
&\le\Prb\{\Ner\le a_n\}+3n^{-2}+\tau_n,\label{eq:finite-converse}\\
\sqrt{f_{\marg}}
&\le\Prb\{\Ner\le a_n\}+\sqrt3\,n^{-1}+\tau_n.
\label{eq:finite-marg-converse}
\end{align}
Each term in $\tau_n$ is exponentially small in $n$.
\end{proposition}
\begin{proof}
We first identify the counts to which Theorem~\ref{thm:fixed} applies. Define the binomial weights and the permitted set of counts by
\begin{align}
\pi_{n,k}&\coloneqq\binom nk p^k(1-p)^{n-k}
\qquad(0\le k\le n),\label{eq:binomial-count-weights}\\
\mathcal I_n&\coloneqq\{k\in\{0,\ldots,n\}:t<k\le p_0n\}.
\label{eq:permitted-counts}
\end{align}
For $k\in\mathcal I_n$ above the threshold in \eqref{eq:threshold-a}, we have
\begin{align}
k>a_n
&\ \Longrightarrow\ 2k>n-\log_dM+h_n,\\
&\ \Longrightarrow\ \log_dM>n-2k+h_n.
\label{eq:count-threshold-rearrangement}
\end{align}
Thus the hypothesis in \eqref{eq:fixed-threshold} is satisfied. Theorem~\ref{thm:fixed} gives both $\overline f_{n,k}\le3n^{-2}$ and $\overline F_{n,k}\le3n^{-2}$ for these counts. At every other count, both averages are at most one.

For the private code, first perform the flag reduction in Lemma~\ref{lem:flags}. It can only increase the fidelities being bounded, so an upper bound for the resulting code also bounds the original code. Write $c_k$ for either the flag-reduced $\overline f_{n,k}$ or the quantum average $\overline F_{n,k}$. Partitioning the binomial sum into counts below the threshold, permitted counts above it, and exceptional counts above it gives
\begin{align}
\sum_{k=0}^n\pi_{n,k}c_k
&=\sum_{k\le a_n}\pi_{n,k}c_k
  +\sum_{\substack{k>a_n\\k\in\mathcal I_n}}\pi_{n,k}c_k
  +\sum_{\substack{k>a_n\\k\notin\mathcal I_n}}\pi_{n,k}c_k,\label{eq:binomial-converse-partition-equality}\\
&\le\Prb\{\Ner\le a_n\}
  +3n^{-2}\Prb\{\Ner>a_n,\ \Ner\in\mathcal I_n\}
  +\Prb\{\Ner\notin\mathcal I_n\},\\
&\le\Prb\{\Ner\le a_n\}+3n^{-2}+\tau_n.
\label{eq:binomial-converse-partition}
\end{align}
All sums here and below are over integers between zero and $n$. The last step uses \eqref{eq:exceptional-count-probability}. The exact mixture identities in \eqref{eq:binomial-mixture} now prove \eqref{eq:finite-converse} for the optimized private fidelity and for both quantum tasks. This argument remains valid when $a_n$ lies outside $[0,n]$.

For the actual-marginal criterion, define its mean square-root fidelity on the size-$k$ layer by
\begin{equation}\label{eq:conditional-root-average}
r_{n,k}\coloneqq\frac1{\binom nk}\sum_{|E|=k}\sqrt{g_E},
\end{equation}
where $g_E$ is the conditional actual-marginal fidelity from Lemma~\ref{lem:flags}. Cauchy--Schwarz and $g_E\le f_E$ imply
\begin{subequations}\label{eq:conditional-jensen}
\begin{align}
r_{n,k}^{2}
&\le\frac1{\binom nk}\sum_{|E|=k}g_E,\\
r_{n,k}
&\le\sqrt{\frac1{\binom nk}\sum_{|E|=k}g_E}
\le\sqrt{\overline f_{n,k}}.
\end{align}
\end{subequations}
In particular, $r_{n,k}\le\sqrt3/n$ on the permitted counts above $a_n$, and $r_{n,k}\le1$ everywhere. The actual-marginal identity \eqref{eq:flag-marginal-identity} and the same partition as in \eqref{eq:binomial-converse-partition-equality} give
\begin{align}
\sqrt{f_{\marg}}
&=\sum_{k=0}^n\pi_{n,k}r_{n,k},\\
&\le\Prb\{\Ner\le a_n\}+\sqrt3\,n^{-1}+\tau_n.
\label{eq:marginal-converse-partition}
\end{align}
This is \eqref{eq:finite-marg-converse}.

Let $\mathsf D$ denote the binary relative entropy with natural logarithms, defined in \eqref{eq:binary-relative-entropy}. Finally, $t=O(\ln n)$ implies $t<np/2$ for all sufficiently large $n$. The exponential estimates in \eqref{eq:binomial-exponential} therefore yield
\begin{align}
\tau_n
&\le\Prb\{\Ner\le np/2\}+\Prb\{\Ner\ge p_0n\},\\
&\le e^{-n\mathsf D(p/2\|p)}+e^{-n\mathsf D(p_0\|p)}.
\label{eq:exceptional-count-tail}
\end{align}
Both relative entropies are positive because $p/2<p<p_0$. This proves the claimed exponential decay, uniformly in the code size $M$.
\end{proof}

\begin{corollary}[Second-order upper bounds]\label{cor:normal-upper}
Fix an integer $d\ge2$, $0<p<1/2$, and $0<\eps<1$. Let $C,V$ be defined in \eqref{eq:qv}, let $\Phi$ be the normal cumulative distribution function in \eqref{eq:normal-cdf}, and set $\eta(\eps)\coloneqq1-\sqrt{1-\eps}$ as in \eqref{eq:one-shot-ordering}. The optimal logarithmic code sizes defined in \eqref{eq:operational-Q} and \eqref{eq:operational-P} satisfy
\begin{align}
Q_{n,\et}^{\eps},\ Q_{n,\eg}^{\eps},\ P_{n,\opt}^{\eps},\ P_{n,\max}^{\eps}
&\le nC+\sqrt{nV}\,\Phi^{-1}(\eps)+O((\ln n)^2),\label{eq:normal-upper}\\
P_{n,\marg}^{\eps}
&\le nC+\sqrt{nV}\,\Phi^{-1}(\eta(\eps))+O((\ln n)^2).
\label{eq:normal-marg-upper}
\end{align}
\end{corollary}
\begin{proof}
Choose $p<p_0<1/2$. For the quantum or optimized private criterion, a code with error at most $\eps$ requires fidelity at least $1-\eps$. For the actual-marginal criterion, it requires square-root fidelity at least $\sqrt{1-\eps}$. The corresponding required binomial probabilities are
\begin{align}
\alpha_{\opt}&\coloneqq1-\eps
&&\text{for either quantum task or optimized private infidelity},\\
\alpha_{\marg}&\coloneqq\sqrt{1-\eps}
&&\text{for actual-marginal private infidelity}.
\label{eq:required-binomial-probabilities}
\end{align}
Write $\alpha$ for the appropriate one of these two numbers.
Proposition~\ref{prop:finite-converse} implies in either case
\begin{equation}\label{eq:quantile-start}
\Prb\{\Ner\le a_n\}\ge\alpha-O(n^{-1}).
\end{equation}
The uniform Berry--Esseen bound in Lemma~\ref{lem:binomial-tools}, with $\sigma_n\coloneqq\sqrt{np(1-p)}$, gives
\begin{equation}\label{eq:cdf-inversion}
\Phi\!\left(\frac{a_n-np}{\sigma_n}\right)
\ge\alpha-O(n^{-1/2}).
\end{equation}
There is no prior restriction on the threshold $a_n$ in this step.

Because $0<\alpha<1$ is fixed, the normal density is positive at $\Phi^{-1}(\alpha)$, and the inverse distribution function has bounded derivative in a neighborhood of $\alpha$. Inverting \eqref{eq:cdf-inversion} and multiplying by $\sigma_n$ therefore gives
\begin{equation}\label{eq:quantile-upper-proof}
a_n\ge np+\sigma_n\Phi^{-1}(\alpha)-O(1).
\end{equation}
Lemma~\ref{lem:quantile} records this argument explicitly. To make the resulting rate bound explicit, substitute \eqref{eq:threshold-a} into \eqref{eq:quantile-upper-proof} and rearrange:
\begin{align}
\frac{n-\log_dM+h_n}{2}
&\ge np+\sigma_n\Phi^{-1}(\alpha)-O(1),\\
\log_dM
&\le n(1-2p)-2\sigma_n\Phi^{-1}(\alpha)+h_n+O(1),\\
\log_2M
&\le nC+\sqrt{nV}\,\Phi^{-1}(1-\alpha)
+h_n\log_2d+O(1).
\label{eq:upper-with-h}
\end{align}
The last step uses $-\Phi^{-1}(\alpha)=\Phi^{-1}(1-\alpha)$, the definitions of $C,V$ in \eqref{eq:qv}, and the fact that $d$ is fixed.
The two choices of $\alpha$ prove the claimed quantum, optimized private, and actual-marginal bounds. The maximal-message upper bound follows from $\eps_{\opt}\le\eps_{\max}$ in \eqref{eq:private-ordering}.
\end{proof}

The optimized private converse retains the original error parameter $\eps$, so it subsumes the entanglement transmission converse through \eqref{eq:one-shot-ordering}. The entanglement generation bound follows directly from the Bell-test part of Theorem~\ref{thm:fixed}. The changed quantile for the actual marginal comes from the exact fidelity identity \eqref{eq:flag-identities}, not from an estimate converting one distance into another.

\subsection{Completion of the coding theorems}\label{subsec:completion}
We now combine the converse with the achievable construction and explain the role of each error criterion. Throughout this subsection, $d\ge2$, $0<p<1/2$, and $0<\eps<1$ are fixed, and $C,V$ are defined in \eqref{eq:qv}. The complete achievable proofs remain in Appendix~\ref{sec:ach}; the following argument shows how their conclusions give Theorems~\ref{thm:main} and~\ref{thm:other-criteria}.

For entanglement transmission, Corollary~\ref{cor:normal-lower} constructs codes with fidelity $F_e\ge1-\eps$ and logarithmic dimension at least $nC+\sqrt{nV}\,\Phi^{-1}(\eps)-O(1)$. Preparing the encoded half of a maximally entangled state gives an entanglement generation code of the same dimension and fidelity. Using basis states as messages and measuring the decoded output instead gives the private code in Lemma~\ref{lem:q-to-p}, for which
\begin{equation}\label{eq:completion-optimized-error}
f_{\opt}\ge F_e\ge1-\eps.
\end{equation}
Thus all three tasks in \eqref{eq:main-tasks} attain the same lower bound at the original infidelity $\eps$. Corollary~\ref{cor:normal-upper} gives their common upper bound. This proves \eqref{eq:main-bounds}; both remainders vanish after division by $\sqrt n$, giving the limit in \eqref{eq:second-order-limit}.

For the actual-marginal private criterion, Lemma~\ref{lem:q-to-p} gives $f_{\marg}\ge F_e^2$. We therefore start with a quantum code at the smaller infidelity $\eta(\eps)\coloneqq1-\sqrt{1-\eps}$ from \eqref{eq:one-shot-ordering}. The resulting code obeys
\begin{equation}\label{eq:completion-marginal-error}
f_{\marg}\ge F_e^2\ge\bigl(1-\eta(\eps)\bigr)^2=1-\eps.
\end{equation}
Corollary~\ref{cor:normal-lower} at error $\eta(\eps)$ therefore supplies the lower bound in \eqref{eq:marg-main}. It matches the converse \eqref{eq:normal-marg-upper}, including its Gaussian quantile. The square in \eqref{eq:completion-marginal-error} is essential: omitting it would change the second-order term.

For the maximal-message criterion, good performance on the uniform message alone does not guarantee good performance for every message. Lemma~\ref{lem:expurgation} resolves this by retaining messages that all have sufficiently large fidelity with one common environment state. Start with an optimized private code at error $\eps_n\coloneqq\eps-n^{-1/2}$, which lies in $(0,\eps)$ for all sufficiently large $n$. Its logarithmic size has the same Gaussian term at $\eps$, up to a bounded loss: the inverse normal distribution has bounded derivative near the fixed number $\eps$, so multiplying its $O(n^{-1/2})$ change by $\sqrt{nV}$ gives $O(1)$, as quantified in \eqref{eq:expurgation-quantile-cost}. The achievable remainder is uniform in this neighborhood; the common density bound in \eqref{eq:expurgation-uniform-density} makes this explicit in Appendix~\ref{sec:ach}.

Let $M_n$ be the size of this preliminary code, and let $M_n'$ be the number of messages retained by Lemma~\ref{lem:expurgation}. Its retained fraction $r_n$, given by \eqref{eq:expurgation-fraction} with $\eps_0=\eps_n$, has $\log_2r_n=-\tfrac12\log_2n+O(1)$ by \eqref{eq:expurgation-loss}--\eqref{eq:expurgation-log-fraction}. Consequently,
\begin{align}
\log_2M_n'&\ge\log_2M_n+\log_2r_n,
\label{eq:completion-maximal-size}\\
&\ge nC+\sqrt{nV}\,\Phi^{-1}(\eps)-\tfrac12\log_2n-O(1).
\label{eq:completion-maximal-rate}
\end{align}
Every retained message satisfies the required infidelity $\eps$. The converse follows from $\eps_{\opt}\le\eps_{\max}$ in \eqref{eq:private-ordering}. This proves \eqref{eq:max-main}. The logarithmic loss in \eqref{eq:completion-maximal-rate} is smaller than $\sqrt n$, so it leaves the second-order coefficient unchanged; its necessity is not established by this construction.

\section{Conclusion}\label{sec:conclusion}
Theorems~\ref{thm:main} and~\ref{thm:other-criteria} determine the second-order correction to unassisted quantum and private communication over the erasure channel. For every fixed erasure probability strictly between zero and one half and every fixed infidelity strictly between zero and one, the correction is governed by the fluctuations in the number of lost systems. Each additional erasure both removes a system from the receiver and transfers one to the environment. The balance $n-2\Ner$ in \eqref{eq:dimension-balance} explains the factor of two in the standard deviation $\sqrt{nV}$, with $V$ given in \eqref{eq:qv}.

The converse makes this balance precise for one encoder facing all erasure patterns. A bound for one chosen pattern would be insufficient: the encoder could simply place its information in the surviving positions. Theorem~\ref{thm:fixed} instead bounds the average fidelity over every pattern with the same number of erasures. Its proof combines the overlap of two recovery tests with a matrix that approximates uniform averaging while comparing only patterns with few common erased positions. Tensor products reduce the construction to a two-dimensional spectral calculation. A classical polynomial identity then suppresses the nonconstant components, and conditioning on the number of selected positions gives the required uniform average. Their cost is $h_n=O((\ln n)^2)$ in logarithmic dimension, as defined in \eqref{eq:hn}. Because that loss is smaller than the $\sqrt n$ fluctuation scale, the binomial distribution determines the optimal second-order coefficient.

Private communication adds a further distinction. Both the optimized joint private criterion and the maximal-message criterion have the same Gaussian coefficient as quantum communication. Fixing the comparison state to the actual environment marginal gives the different quantile in \eqref{eq:marg-main}. The exact flag identities in Lemma~\ref{lem:flags} explain this difference. Thus a specification of the error criterion is needed to interpret a finite-block private coding rate, even when the vanishing-error capacities agree.

Several questions remain. The achievable and converse remainders do not determine a third-order term, and the present asymptotic converse is not a tight numerical estimate at moderate blocklengths. Sharper control of the averaging matrix's absolute row sums may reduce its loss further. Efficient explicit codes attaining the Gaussian term are another open direction. Extending the argument to other channels would require replacements for both the erasure-pattern overlap estimate and the corresponding averaging construction; degradability alone does not supply these ingredients. Section~\ref{app:outlook} discusses these questions further. The endpoints in the erasure probability require separate arguments, recorded in Appendix~\ref{app:boundaries}.

\phantomsection
\section*{Use of artificial intelligence}
\addcontentsline{toc}{section}{Use of artificial intelligence}
OpenAI's ChatGPT was used to assist with developing and examining mathematical arguments, checking literature, drafting and revising the text, preparing figures and computational checks, and providing internal critiques of correctness and exposition. These AI-assisted critiques were part of manuscript preparation and do not constitute external peer review. Responsibility for the manuscript's mathematical claims, citations, and presentation rests with the author.\label{sec:end-main-text}

\begingroup
\footnotesize
\setlength{\bibsep}{0.5pt plus 0.5pt minus 0.5pt}
\renewcommand{\bibsection}{%
  \section*{\refname}%
  \addcontentsline{toc}{section}{References}%
}
\bibliographystyle{unsrtnat}
\bibliography{references}
\endgroup
\clearpage
\appendix
\section{Operational comparisons, attainment, and privacy tests}\label{app:operational}
This appendix gives the arguments deferred from Section~\ref{sec:setting}. Sections~\ref{app:quantum-dimension} and~\ref{app:private-dimension} show that the operational maxima in \eqref{eq:operational-M} and \eqref{eq:operational-P} are finite and attained. Section~\ref{app:q-to-p} proves the two fidelity comparisons in Lemma~\ref{lem:q-to-p}. Finally, Section~\ref{app:overlap-proof} gives the general overlap proof used in Lemma~\ref{lem:overlap}. These arguments use only finite-dimensional quantum mechanics and apply before any construction of the subset matrix.

\subsection{Quantum dimension bounds and attainment}\label{app:quantum-dimension}
Fix integers $n\ge1$ and $d\ge2$, an erasure probability $0\le p\le1$, and an infidelity $0\le\eps<1$. For a code of dimension $M$, the reference $R$ and decoded system $\widehat S$ have dimension $M$, and the Bell target $\Phi_{R\widehat S}^{M}$ is defined in \eqref{eq:bell}. The state $\rho_{RA^n}$ prepared for entanglement generation may be arbitrary, as in \eqref{eq:generation-fidelity}. We prove that the maxima $M_\star^*(n,\eps;p,d)$ in \eqref{eq:operational-M}, for $\star\in\{\et,\eg\}$, are finite and attained. To see finiteness even for entanglement generation, let $D\coloneqq d^n$ and let $\cT\colon A^n\to\widehat S$ be the channel followed by the decoder. Its adjoint $\cT^\dagger$, characterized by $\Tr[Y\cT(X)]=\Tr[\cT^\dagger(Y)X]$, is unital: $\cT^\dagger(I_{\widehat S})=I_{A^n}$. The positive semidefinite operator
\begin{equation}\label{eq:dimension-test}
B\coloneqq(\id_R\otimes\cT^\dagger)(\PhiM)
\end{equation}
has the following partial trace, using $\Tr_R\Phi_{R\widehat S}^{M}=I_{\widehat S}/M$ and the unitality of $\cT^\dagger$:
\begin{align}
\Tr_RB
&=\cT^\dagger(\Tr_R\Phi_{R\widehat S}^{M})
=\frac1M\cT^\dagger(I_{\widehat S})
=\frac1M I_{A^n},\label{eq:dimension-test-partial-trace}\\
\Tr B&=\frac1M\Tr I_{A^n}=\frac DM.\label{eq:dimension-test-trace}
\end{align}
For an arbitrary prepared state, the adjoint identity and positivity of $B$ give
\begin{align}
F_{\eg}&=\Tr(B\rho_{RA^n}),\label{eq:dimension-bound-expectation}\\
\Tr(B\rho_{RA^n})
&\le\norm B\Tr\rho_{RA^n}
=\norm B\le\Tr B=\frac DM.\label{eq:dimension-bound}
\end{align}
The first inequality uses $B\le\norm B\,I$, and the last uses the fact that the largest eigenvalue of a positive semidefinite operator is no larger than the sum of its eigenvalues.
The fidelity constraint therefore implies $M\le D/(1-\eps)$. There are finitely many candidate integers, and $M=1$ is feasible. For each candidate, the relevant state set and the sets of channels represented by their Choi operators are compact. Fidelity is continuous, so its maximum is attained. This also justifies later choices of optimal decoders in individual flag sectors.

\subsection{Private dimension bounds and attainment}\label{app:private-dimension}
Fix $n,d,p,\eps$ in the same ranges as in Section~\ref{app:quantum-dimension}. Consider a private code of size $M$, with input states $\rho_{A^n}^m$, receiver POVM effects $\Lambda_{B^n}^j$, and normalized message--estimate--environment state $\omega_{K\widehat K Z^n}$ from \eqref{eq:private-output}. Its ideal message correlation is $\overline\Phi_{K\widehat K}^{M}$ from \eqref{eq:ideal-key}. We prove finiteness and attainment for each criterion in \eqref{eq:operational-P}. To see the dimension bound explicitly, let $p_{\mathrm{cor}}$ be the probability that the decoded message is correct. The two-outcome measurement that checks $K=\widehat K$ sends the actual state to $(p_{\mathrm{cor}},1-p_{\mathrm{cor}})$ and every ideal state to $(1,0)$. Fidelity monotonicity in \eqref{eq:fidelity-monotonicity} therefore gives, for every normalized $\sigma_{Z^n}$,
\begin{align}
F(\omega_{K\widehat K Z^n},\overline\Phi_{K\widehat K}^{M}\otimes\sigma_{Z^n})
&\le F\bigl((p_{\mathrm{cor}},1-p_{\mathrm{cor}}),(1,0)\bigr),\label{eq:correct-test-monotonicity}\\
F\bigl((p_{\mathrm{cor}},1-p_{\mathrm{cor}}),(1,0)\bigr)
&=(\sqrt{p_{\mathrm{cor}}})^2=p_{\mathrm{cor}}.\label{eq:correct-test-fidelity}
\end{align}
Define the pulled-back POVM by $\Gamma_{A^n}^{m}\coloneqq(\cE_{p,d}^{\otimes n})^\dagger(\Lambda_{B^n}^{m})$. The adjoint channel is positive and unital, so $\Gamma^m\ge0$ and $\sum_m\Gamma^m=I_{A^n}$. Taking the maximum over $\sigma_{Z^n}$ and using $0\le\rho_{A^n}^{m}\le I_{A^n}$ yields
\begin{align}
f_{\opt}&\le p_{\mathrm{cor}}
=\frac1M\sum_m\Tr(\Gamma_{A^n}^{m}\rho_{A^n}^{m}),\label{eq:private-dimension-decoding}\\
\frac1M\sum_m\Tr(\Gamma_{A^n}^{m}\rho_{A^n}^{m})
&\le\frac1M\sum_m\Tr\Gamma_{A^n}^{m}
=\frac1M\Tr I_{A^n}=\frac{d^n}{M}.\label{eq:private-dimension-bound}
\end{align}
For any criterion label $\bullet\in\{\opt,\marg,\max\}$, an admissible code has $f_{\opt}\ge1-\eps$ by \eqref{eq:private-ordering}. Equation~\eqref{eq:private-dimension-bound} therefore implies $M\le d^n/(1-\eps)$. For each of the finitely many candidate $M$, the state and POVM sets are compact, as is the set of ideal environment states. Continuity then gives attainment. The other two criteria are bounded by the optimized one through \eqref{eq:private-ordering}.

\subsection{Quantum-to-private comparisons}\label{app:q-to-p}
We prove Lemma~\ref{lem:q-to-p} for a finite-dimensional channel $\mathcal N\colon A\to B$ and integers $n,M\ge1$. The input is a dimension-$M$ entanglement transmission code of fidelity $F_e$, defined as in \eqref{eq:transmission-fidelity}. The desired private code has the optimized joint fidelity $f_{\opt}$ in \eqref{eq:private-opt} and actual-marginal joint fidelity $f_{\marg}$ in \eqref{eq:private-marg}. The claims are $f_{\opt}\ge F_e$ and $f_{\marg}\ge F_e^2$, followed by the code-size comparisons in \eqref{eq:one-shot-ordering}. In both private criteria, the ideal environment state is common to every message.

\begin{proof}[Proof of Lemma~\ref{lem:q-to-p}]
The private code uses the same encoder and decoder as the quantum code, with basis states as inputs and a basis measurement at the output. We first verify that this produces uniform messages, and then compare its fidelity with the entanglement fidelity.

Let $\cC\colon S\to A^n$ and $\cD\colon B^n\to\widehat S$ be the quantum encoder and decoder, with $\dim S=\dim\widehat S=M$. Retain a complementary output $Z^n$ of $\mathcal N^{\otimes n}$, and let $\mathcal T\colon S\to\widehat S Z^n$ denote the composite channel consisting of the encoder, the channel isometry, and the decoder. The final state and its reference marginal are
\begin{equation}\label{eq:q-to-p-final-state}
\rho_{R\widehat S Z^n}\coloneqq(\id_R\otimes\mathcal T)(\Phi_{RS}^{M}),
\qquad \rho_R=\pi_R.
\end{equation}
Define the basis-measurement channels
\begin{align}
\mathcal M_{R\to K}(X_R)
&\coloneqq\sum_{m=1}^M\bra m X_R\ket m\,\ket m\bra m_K,\\
\mathcal M_{\widehat S\to\widehat K}(Y_{\widehat S})
&\coloneqq\sum_{j=1}^M\bra j Y_{\widehat S}\ket j\,\ket j\bra j_{\widehat K}.
\label{eq:basis-measurements}
\end{align}
Measuring $R$ in the maximally entangled input from \eqref{eq:bell} gives
\begin{equation}\label{eq:measured-bell-input}
(\mathcal M_{R\to K}\otimes\id_S)(\Phi_{RS}^{M})
=\frac1M\sum_{m=1}^M\ket m\bra m_K\otimes\ket m\bra m_S.
\end{equation}
This measurement commutes with all operations on $S$. Thus the state
\begin{equation}\label{eq:measured-quantum-output}
\omega_{K\widehat K Z^n}\coloneqq
(\mathcal M_{R\to K}\otimes\mathcal M_{\widehat S\to\widehat K}\otimes\id_{Z^n})(\rho_{R\widehat S Z^n})
\end{equation}
is the output of a valid private code with uniform messages, input states
$\rho_{A^n}^m\coloneqq\cC(\ket m\bra m_S)$, and receiver POVM
$\Lambda_{B^n}^j\coloneqq\cD^\dagger(\ket j\bra j_{\widehat S})$.
The adjoint decoder is positive and unital, so these effects are positive semidefinite and sum to $I_{B^n}$. The same measurements send the ideal entangled state to the ideal key state in \eqref{eq:ideal-key}:
\begin{equation}\label{eq:measured-bell-target}
(\mathcal M_{R\to K}\otimes\mathcal M_{\widehat S\to\widehat K})(\Phi_{R\widehat S}^{M})
=\overline\Phi_{K\widehat K}^{M}.
\end{equation}

To identify a suitable environment state, define the positive semidefinite operator
\begin{equation}\label{eq:bell-environment-block}
A_{Z^n}\coloneqq\bra{\Phi^M}_{R\widehat S}\rho_{R\widehat S Z^n}\ket{\Phi^M}_{R\widehat S}.
\end{equation}
Its trace is the entanglement fidelity:
\begin{equation}\label{eq:bell-block-trace}
\Tr A_{Z^n}=\Tr\!\left[(\Phi_{R\widehat S}^{M}\otimes I_{Z^n})\rho_{R\widehat S Z^n}\right]=F_e.
\end{equation}
Complete $\ket{\Phi^M}_{R\widehat S}$ to an orthonormal basis $(\ket{e_a})_{a=1}^{M^2}$ with $\ket{e_1}=\ket{\Phi^M}$. The partial trace is then
\begin{equation}\label{eq:environment-block-order}
\rho_{Z^n}=A_{Z^n}+\sum_{a=2}^{M^2}\bra{e_a}\rho_{R\widehat S Z^n}\ket{e_a}
\ge A_{Z^n}\ge0,
\end{equation}
because every summand is positive semidefinite.

We extend $F$ to positive operators by the same trace-norm expression in \eqref{eq:fidelity-definition}; their traces need not both equal one. For a normalized $\sigma_{Z^n}$, define the positive semidefinite operator
\begin{equation}\label{eq:rank-one-sandwich-definition}
Y_{R\widehat S Z^n}\coloneqq
(\Phi_{R\widehat S}^{M}\otimes\sqrt{\sigma_{Z^n}})
\rho_{R\widehat S Z^n}
(\Phi_{R\widehat S}^{M}\otimes\sqrt{\sigma_{Z^n}}).
\end{equation}
The rank-one projector on $R\widehat S$ selects exactly the block $A_{Z^n}$ in \eqref{eq:bell-environment-block}. Thus
\begin{align}
Y_{R\widehat S Z^n}
&=\Phi_{R\widehat S}^{M}\otimes\sqrt{\sigma_{Z^n}}A_{Z^n}\sqrt{\sigma_{Z^n}},\label{eq:rank-one-sandwich-factorization}\\
F(\rho_{R\widehat S Z^n},\Phi_{R\widehat S}^{M}\otimes\sigma_{Z^n})
&=\left(\Tr\sqrt{Y_{R\widehat S Z^n}}\right)^2,\label{eq:rank-one-fidelity-trace}\\
\left(\Tr\sqrt{Y_{R\widehat S Z^n}}\right)^2
&=\left(\Tr\sqrt{\sqrt{\sigma_{Z^n}}A_{Z^n}\sqrt{\sigma_{Z^n}}}\right)^2
=F(A_{Z^n},\sigma_{Z^n}).\label{eq:rank-one-fidelity}
\end{align}
The last equality uses $\Tr\Phi_{R\widehat S}^{M}=1$ and the definition of fidelity.
If $F_e>0$, choose $\sigma_{Z^n}\coloneqq A_{Z^n}/F_e$. Then
\begin{equation}\label{eq:optimized-block-fidelity}
F(A_{Z^n},A_{Z^n}/F_e)
=\left(\frac{\Tr A_{Z^n}}{\sqrt{F_e}}\right)^2=F_e.
\end{equation}
Using \eqref{eq:fidelity-monotonicity} for the two measurements in \eqref{eq:measured-quantum-output}, together with \eqref{eq:measured-bell-target}, yields
\begin{equation}\label{eq:optimized-conversion-chain}
f_{\opt}\ge
F\!\left(\omega_{K\widehat K Z^n},\overline\Phi_{K\widehat K}^{M}\otimes A_{Z^n}/F_e\right)
\ge F_e.
\end{equation}
If $F_e=0$, both claims in \eqref{eq:q-to-p-fidelity} follow from nonnegativity.

For the actual-marginal criterion, \eqref{eq:environment-block-order} implies
\begin{equation}\label{eq:marginal-block-order}
\sqrt{A_{Z^n}}\rho_{Z^n}\sqrt{A_{Z^n}}\ge A_{Z^n}^2.
\end{equation}
Operator monotonicity of the square root and positivity of the trace therefore give
\begin{equation}\label{eq:marginal-conversion}
\sqrt{F(A_{Z^n},\rho_{Z^n})}
=\Tr\sqrt{\sqrt{A_{Z^n}}\rho_{Z^n}\sqrt{A_{Z^n}}}
\ge\Tr A_{Z^n}=F_e.
\end{equation}
The local measurements do not change the environment marginal, so $\omega_{Z^n}=\rho_{Z^n}$. Applying \eqref{eq:fidelity-monotonicity} and \eqref{eq:rank-one-fidelity} once more gives
\begin{equation}\label{eq:actual-marginal-conversion-chain}
f_{\marg}
=F\!\left(\omega_{K\widehat K Z^n},\overline\Phi_{K\widehat K}^{M}\otimes\rho_{Z^n}\right)
\ge F(A_{Z^n},\rho_{Z^n})\ge F_e^2.
\end{equation}
Finally, $F_e\ge1-\eps$ implies $f_{\opt}\ge1-\eps$, while
$F_e\ge1-\eta(\eps)=\sqrt{1-\eps}$ implies $f_{\marg}\ge1-\eps$.
Taking the largest admissible code size proves \eqref{eq:one-shot-ordering}.
\end{proof}

\subsection{Proof of the privacy-test overlap bound}\label{app:overlap-proof}
We prove the general inequality \eqref{eq:overlap-general} in Lemma~\ref{lem:overlap}, following Ref.~\cite[Lemma~3.2, Eqs.~(3.19)--(3.36)]{WildePrivate2026}. Recall its setting: the size-$M$ privacy-test projectors $P,Q$ share sender systems $K,L$, with $\dim K=M$ and the same basis on $K$. Their receiver inputs are $XC$ and $CY$, where $X,C,Y$ are distinct tensor factors; the decoder auxiliary inputs belong separately to $X,Y$. Both tests have the form \eqref{eq:privacy-test}, pulled back through receiver unitaries as in \eqref{eq:decoder-projector} and extended by identities on unused factors. No restriction is imposed on $\dim L$. The claim is $\norm{PQ}\le\min\{1,(\dim C)/M\}$. The full calculation is included so that the dimension factor can be checked in the notation of this paper.

\begin{proof}[Proof of \eqref{eq:overlap-general}]
We first prove the exact value $1/M$ when the receiver input systems are disjoint. We then introduce a separate copy of the common input and reconnect it by a partial trace. The cost of that partial trace is the dimension appearing in \eqref{eq:overlap-general}.

\emph{Disjoint receiver inputs.}
When $C$ is trivial, the two decoder unitaries act on disjoint tensor factors. Conjugating by their product leaves the norm unchanged and places both privacy tests on their output spaces, which we call $K_1T_1$ and $K_2T_2$. Let $\mathcal S\coloneqq LT_1T_2$ be the full shield system. Write $V_1,\ldots,V_M$ for the first test's unitaries on $LT_1$ and $W_1,\ldots,W_M$ for the second test's unitaries on $LT_2$, extending each by the identity on the unused receiver shield. The projector ranges are the images of the isometries
\begin{align}
J_1(\ket j_{K_2}\ket\xi_{\mathcal S})
&\coloneqq\frac1{\sqrt M}\sum_i\ket i_K\ket i_{K_1}\ket j_{K_2}V_i\ket\xi,
\label{eq:first-overlap-isometry}\\
J_2(\ket j_{K_1}\ket\xi_{\mathcal S})
&\coloneqq\frac1{\sqrt M}\sum_i\ket i_K\ket j_{K_1}\ket i_{K_2}W_i\ket\xi.
\label{eq:second-overlap-isometry}
\end{align}
To calculate their overlap, apply $J_1^\dagger$ to the second line. The inner products on $K$ and $K_1$ force both summation labels to equal $j$. Consequently,
\begin{equation}\label{eq:disjoint-overlap-on-vector}
J_1^\dagger J_2(\ket j_{K_1}\ket\xi_{\mathcal S})
=\frac1M\ket j_{K_2}V_j^\dagger W_j\ket\xi_{\mathcal S}.
\end{equation}
Equivalently,
\begin{equation}\label{eq:disjoint-privacy-overlap}
J_1^\dagger J_2
=\frac1M\sum_j\ket j_{K_2}\bra j_{K_1}\otimes V_j^\dagger W_j.
\end{equation}
Each product $V_j^\dagger W_j$ is unitary, even though its factors need not commute on $L$. Thus
\begin{equation}\label{eq:disjoint-overlap-norm-square}
(J_1^\dagger J_2)^\dagger(J_1^\dagger J_2)
=\frac1{M^2}\sum_j\ket j\bra j_{K_1}\otimes I_{\mathcal S}
=\frac1{M^2}I_{K_1\mathcal S}.
\end{equation}
It follows that $\norm{J_1^\dagger J_2}=1/M$. In these coordinates $P=J_1J_1^\dagger$ and $Q=J_2J_2^\dagger$. Submultiplicativity and the unit norms of the isometries give both inequalities
\begin{align}
\norm{PQ}
&=\norm{J_1(J_1^\dagger J_2)J_2^\dagger}
\le\norm{J_1^\dagger J_2},
\label{eq:disjoint-projector-upper}\\
\norm{J_1^\dagger J_2}
&=\norm{J_1^\dagger PQJ_2}
\le\norm{PQ}.
\label{eq:disjoint-projector-lower}
\end{align}
Therefore $\norm{PQ}=1/M$ for disjoint receiver inputs.

\emph{A common receiver input.}
Now let $C$ be nontrivial. Introduce a formal tensor factor $C'\cong C$, and let $Q'$ be $Q$ with its action on $C$ relabeled as an action on $C'$. This constructs a new operator; it does not copy a quantum state. Extend $P$ by the identity on $C'$ and $Q'$ by the identity on $C$. Their receiver inputs are disjoint, so the result just proved gives
\begin{equation}\label{eq:separated-projector-overlap}
\norm{PQ'}=\frac1M.
\end{equation}
Let $F_{CC'}$ be the swap operator between $C$ and $C'$. We claim that
\begin{equation}\label{eq:swap-link}
PQ=\Tr_{C'}[(PQ')F_{CC'}].
\end{equation}
To verify the order of all operator factors, fix bases of $C$ and $C'$ with the same labels. Define the operator blocks $P_{ij}\coloneqq\bra i_CP\ket j_C$ and $Q_{ab}\coloneqq\bra a_CQ\ket b_C$, which act on the remaining systems. Then
\begin{align}
P&=\sum_{i,j}P_{ij}\otimes\ket i\bra j_C,
\label{eq:overlap-first-blocks}\\
Q'&=\sum_{a,b}Q_{ab}\otimes\ket a\bra b_{C'},
\label{eq:overlap-second-blocks}\\
F_{CC'}&=\sum_{r,s}\ket r\bra s_C\otimes\ket s\bra r_{C'}.
\label{eq:overlap-swap-expansion}
\end{align}
Multiplying these sums, the $C$ and $C'$ matrix units impose $r=j$ and $s=b$. Thus
\begin{align}
(PQ')F_{CC'}
&=\sum_{i,j,a,b}P_{ij}Q_{ab}\otimes
\ket i\bra b_C\otimes\ket a\bra j_{C'},
\label{eq:swap-link-product}\\
\Tr_{C'}[(PQ')F_{CC'}]
&=\sum_{i,j,b}P_{ij}Q_{jb}\otimes\ket i\bra b_C
=PQ.
\label{eq:swap-link-expanded}
\end{align}
No two blocks have been interchanged, so this calculation also applies when they fail to commute on the sender systems.

For any operator $Z$ on the enlarged space, the partial trace is a sum of diagonal compressions. Since each compression has operator norm at most $\norm Z$,
\begin{align}
\norm{\Tr_{C'}Z}
&=\norm{\sum_{r=1}^{\dim C}(I\otimes\bra r_{C'})Z(I\otimes\ket r_{C'})}
\label{eq:partial-trace-norm-expansion}\\
&\le\sum_{r=1}^{\dim C}
\norm{(I\otimes\bra r_{C'})Z(I\otimes\ket r_{C'})}
\label{eq:partial-trace-norm-triangle}\\
&\le(\dim C)\norm Z.
\label{eq:partial-trace-operator-bound}
\end{align}
Apply \eqref{eq:partial-trace-operator-bound} to $Z=(PQ')F_{CC'}$. The swap is unitary, so \eqref{eq:swap-link} and \eqref{eq:separated-projector-overlap} imply
\begin{equation}\label{eq:shared-input-overlap-bound}
\norm{PQ}\le(\dim C)\norm{(PQ')F_{CC'}}
=(\dim C)\norm{PQ'}=\frac{\dim C}{M}.
\end{equation}
Also, $\norm{PQ}\le\norm P\norm Q\le1$, because $P$ and $Q$ are projectors. These two bounds prove \eqref{eq:overlap-general}.

\end{proof}

\section{Proof of the averaging-matrix lemma}\label{app:tensor-proof}

This appendix proves Lemma~\ref{lem:filter}. The matrix in that lemma compares erased sets of one fixed size, but its construction begins with all subsets of $[n]$. On this larger space, independent one-position choices give a tensor-product eigenbasis. We first calculate the two one-position eigenvalues. We then choose a polynomial that suppresses every nonconstant tensor-product eigenvector. Finally, we restrict the matrix to sets of size $k$ and normalize its row sums. The probability of that size-$k$ layer is at least $1/(n+1)$, so the restriction costs only a polynomial factor.

The construction uses classical ingredients. Biased disjointness operators and polynomial modifications of their spectra appear in Friedgut's method~\cite{Friedgut2008}, as developed explicitly in Ref.~\cite[Section~3.3, Lemmas~3.10--3.12]{Filmus2013}. The polynomial used below is built from the classical Meixner polynomials~\cite[Table~18.19.1 and Eqs.~18.20.7, 18.23.4]{DLMF}. We derive every property needed here from two-dimensional linear algebra, the binomial theorem, and a geometric series.

For a reader beginning with this appendix, we repeat all parameters. Fix $p_0\in(0,1/2)$ and an integer $n\ge3$, and define
\begin{align}
r_0&\coloneqq\frac{p_0}{1-p_0},\qquad
c_0\coloneqq-\ln r_0,\qquad c\coloneqq r_0^2,\label{eq:bt-constants}\\
t&\coloneqq\left\lceil\frac{4\ln n}{c_0}\right\rceil,\qquad
\Gamma_0\coloneqq1+3(c^{-1}-1),\qquad
\ell_n\coloneqq2(n+1)(\Gamma_0n)^t.\label{eq:bt-degree-budget}
\end{align}
These are the parameters in \eqref{eq:filter-decay-constants}, \eqref{eq:filter-coefficient-constants}, \eqref{eq:filter-constants}, and \eqref{eq:filter-row-budget}. Let $k$ be an integer satisfying $t<k\le p_0n$. Write $[n]\coloneqq\{1,\ldots,n\}$, and define
\begin{equation}\label{eq:bt-layer}
\Omega_{n,k}\coloneqq\{E\subseteq[n]:|E|=k\},
\qquad U\coloneqq\binom nk^{-1}\one\one^{\mathsf T},
\end{equation}
where $\one$ is the vector of ones indexed by $\Omega_{n,k}$. The goal is to construct a real symmetric matrix $W$ on this space such that
\begin{align}
W\one&=\one,\label{eq:bt-goal-rows}\\
W_{EF}\ne0&\ \Longrightarrow\ |E\cap F|\le t,\label{eq:bt-goal-support}\\
\norm{W-U}&\le2n^{-2},\label{eq:bt-goal-error}\\
\max_E\sum_F|W_{EF}|&\le\ell_n.\label{eq:bt-goal-mass}
\end{align}
The norm in \eqref{eq:bt-goal-error} is the ordinary Euclidean operator norm. In particular, \eqref{eq:bt-goal-support} implies $W_{EE}=0$, since $|E|=k>t$.

\subsection{Two one-position matrices and an orthonormal product basis}\label{app:bt-tensor-basis}

We first choose a probability distribution on all subsets. Each position will belong to a subset independently with probability $\alpha=k/n$. This choice makes sets of size $k$ collectively have sufficiently large probability; the exact bound is proved in \eqref{eq:bt-layer-probability}. Define
\begin{equation}\label{eq:bt-bias}
\alpha\coloneqq\frac kn,\qquad
r\coloneqq\frac{\alpha}{1-\alpha}\le r_0<1.
\end{equation}
Rows and columns of the following matrices are indexed, in order, by $0$ and $1$:
\begin{equation}\label{eq:bt-one-position}
\mathsf P\coloneqq
\begin{pmatrix}1-\alpha&\alpha\\1-\alpha&\alpha\end{pmatrix},
\qquad
\mathsf R\coloneqq
\begin{pmatrix}1-r&r\\1&0\end{pmatrix}
=\mathsf P-r(I-\mathsf P).
\end{equation}
Here $I$ is the identity matrix. The first matrix averages a function against the probabilities $(1-\alpha,\alpha)$. The second matrix has the useful zero entry $\mathsf R_{11}=0$: an input position and an output position cannot both be present in a nonzero matrix entry. Both matrices have nonnegative entries and row sums one. They are auxiliary matrices on subset labels, not operations added to the communication protocol.

For real-valued functions $f,g$ on $\{0,1\}$, define
\begin{equation}\label{eq:bt-one-position-inner}
\langle f,g\rangle_\alpha
\coloneqq(1-\alpha)f(0)g(0)+\alpha f(1)g(1).
\end{equation}
This inner product weights the two possible values by their probabilities. It makes $\mathsf P$ an orthogonal projection onto constant functions:
\begin{equation}\label{eq:bt-averaging-projection}
\mathsf Pf=\langle\one,f\rangle_\alpha\one,
\qquad \langle\one,\one\rangle_\alpha=1.
\end{equation}
The function
\begin{equation}\label{eq:bt-centered-function}
h(x)\coloneqq\frac{x-\alpha}{\sqrt{\alpha(1-\alpha)}},
\qquad x\in\{0,1\},
\end{equation}
is the centered, normalized one-position indicator. Indeed,
\begin{align}
\langle\one,h\rangle_\alpha
&=\frac{-(1-\alpha)\alpha+\alpha(1-\alpha)}
{\sqrt{\alpha(1-\alpha)}}=0,\label{eq:bt-centered-mean}\\
\langle h,h\rangle_\alpha
&=\frac{(1-\alpha)\alpha^2+\alpha(1-\alpha)^2}
{\alpha(1-\alpha)}=1.\label{eq:bt-centered-norm}
\end{align}
Thus $\one,h$ form an orthonormal basis, and \eqref{eq:bt-one-position} gives
\begin{align}
\mathsf P\one&=\one,\qquad \mathsf Ph=0,\label{eq:bt-local-projection-spectrum}\\
\mathsf R\one&=\one,\qquad \mathsf Rh=-rh.\label{eq:bt-local-disjointness-spectrum}
\end{align}
The strict inequality $r<1$ in \eqref{eq:bt-bias} is the source of spectral decay.

Identify a subset $E\subseteq[n]$ with its indicator string $(e_1,\ldots,e_n)\in\{0,1\}^n$. The product probability distribution and its inner product are
\begin{align}
\pi(E)&\coloneqq\alpha^{|E|}(1-\alpha)^{n-|E|},\label{eq:bt-product-probability}\\
\langle f,g\rangle_\pi
&\coloneqq\sum_{E\subseteq[n]}\pi(E)f(E)g(E).\label{eq:bt-product-inner}
\end{align}
For an operator $\mathsf A$ on this function space, let $\norm{\mathsf A}_{\pi\to\pi}$ denote the operator norm induced by \eqref{eq:bt-product-inner}. The corresponding vector norm is denoted by $\norm{f}_\pi$.

Let $\mathsf P_j$ act as $\mathsf P$ on position $j$ and as the identity on all other positions. Define
\begin{equation}\label{eq:bt-product-operators}
\mathsf K\coloneqq\mathsf R^{\otimes n},
\qquad
\mathsf H\coloneqq\sum_{j=1}^n(I-\mathsf P_j),
\qquad
\mathsf V\coloneqq\mathsf P^{\otimes n}.
\end{equation}
The matrix $\mathsf V$ averages against the product distribution:
\begin{equation}\label{eq:bt-product-average}
\mathsf V_{EF}=\pi(F),\qquad
(\mathsf Vf)(E)=\sum_{F\subseteq[n]}\pi(F)f(F).
\end{equation}
The disjointness property of $\mathsf K$ follows directly from its entries:
\begin{equation}\label{eq:bt-product-disjointness}
\mathsf K_{EG}=\prod_{j=1}^n\mathsf R_{e_jg_j}=0
\quad\text{if }E\cap G\ne\varnothing.
\end{equation}
A common element supplies a factor $\mathsf R_{11}=0$.

For $S\subseteq[n]$, define the product function
\begin{equation}\label{eq:bt-product-basis}
h_S(E)\coloneqq\prod_{j\in S}h(e_j),
\qquad h_\varnothing\coloneqq\one.
\end{equation}
We write $\one\{\text{condition}\}$ for the scalar indicator, equal to one when the condition holds and zero otherwise. Independence in \eqref{eq:bt-product-probability} and \eqref{eq:bt-centered-mean}--\eqref{eq:bt-centered-norm} imply
\begin{equation}\label{eq:bt-product-orthogonality}
\langle h_S,h_T\rangle_\pi
=\prod_{j=1}^n
\left\langle h^{\one\{j\in S\}},h^{\one\{j\in T\}}\right\rangle_\alpha
=\one\{S=T\}.
\end{equation}
Here $h^0$ is the constant function one, and $h^1=h$. There are $2^n$ such functions in a $2^n$-dimensional space, so they form an orthonormal basis.

The tensor-product eigenvalues now follow by multiplying or adding the one-position eigenvalues in \eqref{eq:bt-local-projection-spectrum}--\eqref{eq:bt-local-disjointness-spectrum}:
\begin{align}
\mathsf K h_S&=(-r)^{|S|}h_S,\label{eq:bt-tensor-disjointness-spectrum}\\
\mathsf H h_S&=|S|h_S,\label{eq:bt-tensor-counting-spectrum}\\
\mathsf V h_S&=\one\{S=\varnothing\}h_S.\label{eq:bt-tensor-average-spectrum}
\end{align}
In particular, $\mathsf H$ counts how many factors of the product basis vector are nonconstant. Its eigenvalues are the integers $0,1,\ldots,n$.

Suppose that $q$ is a real polynomial with $q(0)=1$. The three matrices in \eqref{eq:bt-product-operators} are diagonal in the same orthonormal basis. Thus
\begin{equation}\label{eq:bt-polynomial-spectral-reduction}
\norm{\mathsf Kq(\mathsf H)-\mathsf V}_{\pi\to\pi}
=\max_{1\le i\le n}r^i|q(i)|.
\end{equation}
The constant basis vector gives eigenvalue zero in this difference. Equation~\eqref{eq:bt-polynomial-spectral-reduction} reduces the spectral problem to finding a polynomial that equals one at zero and is small at positive integers after multiplication by the geometric factor $r^i$.

\subsection{A polynomial for the geometric weight}\label{app:bt-polynomial}

We give an explicit polynomial and prove its estimate. For a real variable $x$ and an integer $\ell\ge0$, define
\begin{equation}\label{eq:bt-polynomial-binomial}
\binom x0\coloneqq1,\qquad
\binom x\ell\coloneqq
\frac{x(x-1)\cdots(x-\ell+1)}{\ell!}\quad(\ell\ge1).
\end{equation}
For $0<c<1$ and integer $j\ge0$, define
\begin{equation}\label{eq:bt-meixner}
M_j(x)\coloneqq
\sum_{\ell=0}^j\binom j\ell\binom x\ell(1-c^{-1})^\ell.
\end{equation}
These are the Meixner polynomials with parameter $\beta=1$~\cite[Eq.~18.20.7]{DLMF}. The following proof uses their explicit finite sum, so no facts about orthogonal polynomials are assumed. Our aim is to choose a degree-at-most-$t$ polynomial that equals one at zero and minimizes the geometrically weighted sum of its squared values. The orthogonality proved below makes the coefficients of this polynomial explicit. The argument following the lemma verifies this minimization; no optimization result is assumed in the proof.

\begin{lemma}[Polynomial estimate for a geometric weight]\label{lem:geometric-polynomial}
Let $0<c<1$ and let $t$ be a nonnegative integer. With the polynomials $M_j$ defined in \eqref{eq:bt-meixner}, set
\begin{equation}\label{eq:bt-polynomial}
q_t(x)\coloneqq
\frac{1-c}{1-c^{t+1}}\sum_{j=0}^t c^jM_j(x).
\end{equation}
Then $q_t$ has degree at most $t$ and satisfies
\begin{align}
q_t(0)&=1,\label{eq:bt-polynomial-zero}\\
\sum_{i=1}^\infty c^iq_t(i)^2
&=\frac{c^{t+1}}{1-c^{t+1}},\label{eq:bt-polynomial-energy}\\
\sup_{i\in\{1,2,\ldots\}}c^{i/2}|q_t(i)|
&\le\frac{c^{(t+1)/2}}{\sqrt{1-c^{t+1}}}.\label{eq:bt-polynomial-sup}
\end{align}
The supremum in \eqref{eq:bt-polynomial-sup} is over positive integers.
\end{lemma}

\begin{proof}
Each summand in \eqref{eq:bt-meixner} has degree at most $j$, so $q_t$ has degree at most $t$. Since $M_j(0)=1$, the finite geometric series gives
\begin{equation}\label{eq:bt-polynomial-normalization-detail}
q_t(0)=\frac{1-c}{1-c^{t+1}}\sum_{j=0}^tc^j
=\frac{1-c}{1-c^{t+1}}\frac{1-c^{t+1}}{1-c}=1.
\end{equation}

We next establish the identity that will remove all cross terms in the squared norm of $q_t$. Fix a nonnegative integer $i$. For $z$ sufficiently close to zero, define the generating function $F_i(z)\coloneqq\sum_{j=0}^\infty M_j(i)z^j$. Substituting \eqref{eq:bt-meixner} and using $\binom i\ell=0$ for $\ell>i$ gives
\begin{align}
F_i(z)
&=\sum_{\ell=0}^i
\binom i\ell(1-c^{-1})^\ell
\sum_{j=\ell}^\infty\binom j\ell z^j\label{eq:bt-generating-first}\\
&=\frac1{1-z}\sum_{\ell=0}^i
\binom i\ell
\left(\frac{(1-c^{-1})z}{1-z}\right)^\ell\label{eq:bt-generating-second}\\
&=\frac{(1-z/c)^i}{(1-z)^{i+1}}.\label{eq:bt-generating-final}
\end{align}
The second line uses $\sum_{j=\ell}^\infty\binom j\ell z^j=z^\ell/(1-z)^{\ell+1}$, obtained by differentiating the geometric series $\ell$ times and multiplying by $z^\ell/\ell!$. The last line is the binomial theorem. Equation~\eqref{eq:bt-generating-final} is the $\beta=1$ specialization of the classical generating function~\cite[Eq.~18.23.4]{DLMF}.

For complex variables $z,w$ sufficiently close to zero, \eqref{eq:bt-generating-final} yields another geometric series:
\begin{align}
\sum_{i=0}^\infty c^iF_i(z)F_i(w)
&=\frac1{(1-z)(1-w)}
\sum_{i=0}^\infty
\left(\frac{c(1-z/c)(1-w/c)}{(1-z)(1-w)}\right)^i
\label{eq:bt-orthogonal-series}\\
&=\frac1{(1-z)(1-w)-c(1-z/c)(1-w/c)}
\label{eq:bt-orthogonal-rational}\\
&=\frac1{(1-c)(1-zw/c)}
\label{eq:bt-orthogonal-factor}\\
&=\frac1{1-c}\sum_{j=0}^\infty c^{-j}z^jw^j.
\label{eq:bt-orthogonal-diagonal}
\end{align}
We justify the comparison of coefficients by an absolute-convergence estimate. Choose $0<\rho<1$ sufficiently small that
\begin{equation}\label{eq:bt-convergence-radius}
a_\rho\coloneqq1+\frac{(c^{-1}-1)\rho}{1-\rho},
\qquad ca_\rho^2<1.
\end{equation}
Such a choice exists because $a_\rho\to1$ as $\rho\to0$ and $c<1$. Taking absolute values in the finite-sum calculation of \eqref{eq:bt-generating-first} gives
\begin{align}
\sum_{j=0}^\infty |M_j(i)|\rho^j
&\le\frac1{1-\rho}\sum_{s=0}^i\binom is
\left(\frac{(c^{-1}-1)\rho}{1-\rho}\right)^s
\label{eq:bt-generating-absolute-first}\\
&=\frac{a_\rho^i}{1-\rho}.
\label{eq:bt-generating-absolute-final}
\end{align}
Applying this bound to both polynomial factors yields
\begin{equation}\label{eq:bt-triple-absolute-convergence}
\sum_{i,j,\ell=0}^\infty c^i|M_j(i)M_\ell(i)|\rho^{j+\ell}
\le\frac1{(1-\rho)^2}\sum_{i=0}^\infty(ca_\rho^2)^i
=\frac1{(1-\rho)^2(1-ca_\rho^2)}<\infty.
\end{equation}
We may therefore rearrange the power series for $|z|,|w|\le\rho$ and compare coefficients. The coefficient of $z^jw^\ell$ in \eqref{eq:bt-orthogonal-diagonal} proves
\begin{equation}\label{eq:bt-meixner-orthogonality}
\sum_{i=0}^\infty c^iM_j(i)M_\ell(i)
=\frac{c^{-j}}{1-c}\,\one\{j=\ell\}.
\end{equation}
This is the Meixner orthogonality relation for the geometric weight~\cite[Table~18.19.1]{DLMF}, obtained here directly from the finite sum.

Substituting \eqref{eq:bt-polynomial} and applying \eqref{eq:bt-meixner-orthogonality}, we obtain
\begin{align}
\sum_{i=0}^\infty c^iq_t(i)^2
&=\frac{(1-c)^2}{(1-c^{t+1})^2}
\sum_{j,\ell=0}^t c^{j+\ell}
\sum_{i=0}^\infty c^iM_j(i)M_\ell(i)
\label{eq:bt-energy-first}\\
&=\frac{1-c}{(1-c^{t+1})^2}\sum_{j=0}^t c^j
\label{eq:bt-energy-second}\\
&=\frac1{1-c^{t+1}}.\label{eq:bt-energy-final}
\end{align}
The $i=0$ term is one by \eqref{eq:bt-polynomial-normalization-detail}. Subtracting it proves \eqref{eq:bt-polynomial-energy}. Each summand in \eqref{eq:bt-polynomial-energy} is nonnegative, so each is bounded by the entire sum. Taking square roots proves \eqref{eq:bt-polynomial-sup}.
\end{proof}

The coefficients in \eqref{eq:bt-polynomial} have a simple motivation. We want to preserve the value at zero while making the weighted squared values at positive integers small. The leading coefficient of $M_j$ is $(1-c^{-1})^j/j!\ne0$, so the polynomials $M_0,\ldots,M_t$ form a basis for the polynomials of degree at most $t$. Write any such polynomial as $q=\sum_{j=0}^t a_jM_j$. The condition $q(0)=1$ becomes $\sum_j a_j=1$. By \eqref{eq:bt-meixner-orthogonality} and Cauchy--Schwarz,
\begin{align}
1=\left(\sum_{j=0}^ta_j\right)^2
&\le
\left(\sum_{j=0}^t\frac{a_j^2c^{-j}}{1-c}\right)
\left((1-c)\sum_{j=0}^tc^j\right)
\label{eq:bt-least-squares-first}\\
&=(1-c^{t+1})\sum_{i=0}^\infty c^iq(i)^2.
\label{eq:bt-least-squares-second}
\end{align}
The coefficients in \eqref{eq:bt-polynomial} attain equality. Since $q(0)=1$, removing the $i=0$ term subtracts the fixed constant one. Thus $q_t$ also minimizes $\sum_{i=1}^\infty c^iq(i)^2$ among degree-at-most-$t$ polynomials with value one at zero. This optimization explains the choice of coefficients.

We now use Lemma~\ref{lem:geometric-polynomial} with the fixed value $c=r_0^2$ in \eqref{eq:bt-constants}, independently of $k$. Define
\begin{equation}\label{eq:bt-filtered-operator}
\mathsf T\coloneqq\mathsf Kq_t(\mathsf H),
\qquad \delta_n\coloneqq2n^{-4},
\end{equation}
with $\mathsf K,\mathsf H$ from \eqref{eq:bt-product-operators}. Its entries may have either sign. Equations~\eqref{eq:bt-polynomial-spectral-reduction} and \eqref{eq:bt-polynomial-sup}, together with $r\le r_0$, give
\begin{align}
\norm{\mathsf T-\mathsf V}_{\pi\to\pi}
&=\max_{1\le i\le n}r^i|q_t(i)|
\label{eq:bt-full-error-first}\\
&\le\frac{r_0^{t+1}}{\sqrt{1-r_0^{2(t+1)}}}
\label{eq:bt-full-error-second}\\
&\le\frac{n^{-4}}{\sqrt{1-n^{-8}}}
\label{eq:bt-full-error-third}\\
&\le\delta_n.\label{eq:bt-full-error-final}
\end{align}
For the third line, $r_0=e^{-c_0}$ and $t\ge4\ln n/c_0$ imply $r_0^{t+1}\le n^{-4}$. The last line holds for $n\ge2$. The scalar sum in \eqref{eq:bt-polynomial-energy} is over the integer values $i$, not over all product basis vectors. Their multiplicities do not enter \eqref{eq:bt-full-error-first}, because an operator norm takes the largest absolute eigenvalue.

\subsection{Restriction to the fixed-size layer}\label{app:bt-restriction}

We next transfer the approximation in \eqref{eq:bt-full-error-final} from all subsets to $\Omega_{n,k}$. Two details need to be checked: the weighted operator norm must become the ordinary norm on the layer, and the restricted matrix must have row sums that can be normalized without a large loss.

For the first detail, let $\mathsf D$ be the diagonal matrix with entries $\mathsf D_{EE}\coloneqq\pi(E)$, and define
\begin{equation}\label{eq:bt-symmetric-representation}
\mathsf S\coloneqq\mathsf D^{1/2}\mathsf T\mathsf D^{-1/2},
\qquad u_E\coloneqq\sqrt{\pi(E)}.
\end{equation}
Every $\pi(E)$ is positive, because $0<\alpha<1$. The coordinate change $f\mapsto\mathsf D^{1/2}f$ preserves norms:
\begin{equation}\label{eq:bt-isometric-change}
\norm{f}_\pi^2=f^{\mathsf T}\mathsf Df
=\norm{\mathsf D^{1/2}f}_2^2.
\end{equation}
The product basis calculation shows that $\mathsf T$ is self-adjoint for \eqref{eq:bt-product-inner}. Equivalently, $\mathsf D\mathsf T=\mathsf T^{\mathsf T}\mathsf D$, which implies that $\mathsf S$ is real symmetric. Furthermore, \eqref{eq:bt-product-average} gives
\begin{equation}\label{eq:bt-average-similarity}
(\mathsf D^{1/2}\mathsf V\mathsf D^{-1/2})_{EF}
=\sqrt{\pi(E)}\,\pi(F)\,\frac1{\sqrt{\pi(F)}}
=u_Eu_F.
\end{equation}
Equations~\eqref{eq:bt-isometric-change}, \eqref{eq:bt-average-similarity}, and \eqref{eq:bt-full-error-final} therefore yield the ordinary Euclidean bound
\begin{equation}\label{eq:bt-euclidean-bound}
\norm{\mathsf S-uu^{\mathsf T}}
=\norm{\mathsf T-\mathsf V}_{\pi\to\pi}
\le\delta_n.
\end{equation}

Define the restricted matrix and the probability of its layer by
\begin{align}
\mathsf B&\coloneqq(\mathsf T_{EF})_{E,F\in\Omega_{n,k}},
\label{eq:bt-restricted-matrix}\\
\mu&\coloneqq\sum_{E\in\Omega_{n,k}}\pi(E)
=\binom nk\alpha^k(1-\alpha)^{n-k}.
\label{eq:bt-layer-mass-definition}
\end{align}
The probabilities $\pi(E)$ are equal for all $E\in\Omega_{n,k}$. Thus the size-$k$ principal block of $\mathsf S$ is exactly $\mathsf B$, and the corresponding principal block of $uu^{\mathsf T}$ is $\mu U$. In particular, $\mathsf B$ is real symmetric.

For completeness, let $z$ be a vector indexed by $\Omega_{n,k}$, and extend it to a vector $\widetilde z$ on all subsets by assigning zero outside $\Omega_{n,k}$. Then
\begin{align}
z^{\mathsf T}(\mathsf B-\mu U)z
&=\widetilde z^{\mathsf T}(\mathsf S-uu^{\mathsf T})\widetilde z,
\label{eq:bt-compression-form}\\
\left|z^{\mathsf T}(\mathsf B-\mu U)z\right|
&\le\delta_n\norm{\widetilde z}_2^2
=\delta_n\norm z_2^2.
\label{eq:bt-compression-bound}
\end{align}
Taking the supremum over unit vectors, using symmetry, proves
\begin{equation}\label{eq:bt-compressed-error}
\norm{\mathsf B-\mu U}\le\delta_n.
\end{equation}
This is the precise norm estimate inherited by restriction.

We now normalize the row sums. Every permutation $\tau$ of $[n]$ acts on the subset labels. The tensor product $\mathsf K$ and the sum $\mathsf H$ are unchanged by this simultaneous permutation of their rows and columns, so $\mathsf T_{\tau E,\tau F}=\mathsf T_{EF}$. Therefore, for $E\in\Omega_{n,k}$,
\begin{equation}\label{eq:bt-row-invariance}
\sum_{F\in\Omega_{n,k}}\mathsf B_{\tau E,F}
=\sum_{F\in\Omega_{n,k}}\mathsf B_{\tau E,\tau F}
=\sum_{F\in\Omega_{n,k}}\mathsf B_{EF}.
\end{equation}
Any two size-$k$ subsets are related by such a permutation. All row sums are consequently equal to their average
\begin{equation}\label{eq:bt-normalizer-definition}
a\coloneqq\binom nk^{-1}\one^{\mathsf T}\mathsf B\one,
\qquad \mathsf B\one=a\one.
\end{equation}
Applying \eqref{eq:bt-compressed-error} to the unit vector $\binom nk^{-1/2}\one$ gives
\begin{equation}\label{eq:bt-normalizer-comparison}
|a-\mu|
=\left\|(\mathsf B-\mu U)\binom nk^{-1/2}\one\right\|_2
\le\delta_n.
\end{equation}

The probability $\mu$ is not exponentially small. Indeed, $\alpha=k/n$ makes $k$ a most likely value of a binomial random variable with parameters $(n,\alpha)$. To verify this without an asymptotic approximation, define
\begin{equation}\label{eq:bt-binomial-probabilities}
b_j\coloneqq\binom nj\alpha^j(1-\alpha)^{n-j},
\qquad j\in\{0,\ldots,n\}.
\end{equation}
For $0\le j<n$, the ratio
\begin{equation}\label{eq:bt-binomial-ratio}
\frac{b_{j+1}}{b_j}
=\frac{n-j}{j+1}\frac{k}{n-k}
\end{equation}
decreases with $j$, and its two values around $k$ are
\begin{equation}\label{eq:bt-binomial-mode}
\frac{b_k}{b_{k-1}}=\frac{n-k+1}{n-k}>1,
\qquad
\frac{b_{k+1}}{b_k}=\frac{k}{k+1}<1.
\end{equation}
Thus $b_k$ is the largest of $n+1$ nonnegative numbers whose sum is one. Since $\mu=b_k$,
\begin{equation}\label{eq:bt-layer-probability}
\mu\ge\frac1{n+1}.
\end{equation}
For $n\ge3$, $2n^{-4}\le1/(2(n+1))$. Combining \eqref{eq:bt-normalizer-comparison} and \eqref{eq:bt-layer-probability} gives
\begin{equation}\label{eq:bt-normalizer-positive}
a\ge\mu-\delta_n\ge\frac1{2(n+1)}>0.
\end{equation}
The desired matrix is therefore well defined by
\begin{equation}\label{eq:bt-final-matrix}
W\coloneqq a^{-1}\mathsf B.
\end{equation}
It is real symmetric and satisfies $W\one=\one$, proving \eqref{eq:bt-goal-rows}.

To finish the operator-norm estimate, decompose the layer space into the span of $\one$ and its orthogonal complement. Both $\mathsf B$ and $U$ preserve this decomposition, because $\mathsf B$ is symmetric and $\mathsf B\one=a\one$. The difference $W-U$ vanishes on the first subspace. For a vector $z$ in the second subspace, $Uz=0$ and \eqref{eq:bt-compressed-error} implies
\begin{equation}\label{eq:bt-mean-zero-bound}
\norm{(W-U)z}_2
=a^{-1}\norm{\mathsf Bz}_2
=a^{-1}\norm{(\mathsf B-\mu U)z}_2
\le a^{-1}\delta_n\norm z_2.
\end{equation}
Taking the maximum on the two orthogonal subspaces and using \eqref{eq:bt-normalizer-positive}, we obtain
\begin{align}
\norm{W-U}
&\le\frac{\delta_n}{a}\label{eq:bt-slice-error-first}\\
&\le2(n+1)\delta_n=4(n+1)n^{-4}\label{eq:bt-slice-error-second}\\
&\le2n^{-2}\qquad(n\ge3).\label{eq:bt-slice-error-final}
\end{align}
This proves \eqref{eq:bt-goal-error}. The key point is that the product distribution assigns probability at least $1/(n+1)$ to the selected layer; this keeps the normalization loss polynomial in $n$.

\subsection{Support and absolute row sums}\label{app:bt-support-rows}

The spectral estimate used an orthonormal product basis. The two remaining estimates are proved in the coordinate basis, where entries are indexed by subsets. This is where the degree of the polynomial has a direct combinatorial meaning.

\paragraph{At most \(t\) common positions.}
For any subsets $G,F$, a nonzero entry of $I-\mathsf P_j$ requires that $G,F$ agree outside position $j$. Thus $\mathsf H_{GF}\ne0$ can hold only when $G,F$ differ in at most one position. Write $G\mathbin{\triangle}F\coloneqq(G\setminus F)\cup(F\setminus G)$ for their symmetric difference. Expanding matrix multiplication shows that
\begin{equation}\label{eq:bt-power-support}
(\mathsf H^\ell)_{GF}\ne0
\ \Longrightarrow\
|G\mathbin{\triangle}F|\le\ell,
\qquad \ell\in\{0,1,\ldots\},
\end{equation}
Indeed, a nonzero product contributing to the entry specifies a sequence of $\ell$ steps, each changing at most one position.

\begin{figure}[tbp]
\centering
\resizebox{0.96\linewidth}{!}{
\begin{tikzpicture}[
 font=\small,
 patharrow/.style={-{Stealth[length=2mm]},line width=.8pt,draw=black!65},
 card/.style={draw=erasureGray!60,rounded corners=2pt,line width=.6pt},
 bit/.style={draw=black!35,minimum size=3.2mm,inner sep=0pt,font=\footnotesize},
 onebit/.style={bit,fill=erasureTeal!18},
 sharedbit/.style={bit,fill=erasureOrange!25}
]
\foreach \xc/\name/\size in {0/E/3,3.8/G_0/3,7.6/G_1/4,11.4/F/3}{
 \draw[card] (\xc-1.5,-1.15) rectangle (\xc+1.5,1.15);
 \node at (\xc,.79) {$\name$};
 \node[font=\footnotesize] at (\xc,-.77) {$|\name|=\size$};
}
\foreach \j/\value in {1/1,2/1,3/1,4/0,5/0,6/0,7/0,8/0}{
 \ifnum\value=1
  \node[sharedbit] at ({(\j-4.5)*.32},0) {\value};
 \else
  \node[bit] at ({(\j-4.5)*.32},0) {\value};
 \fi
 \node[font=\scriptsize,text=black!60] at ({(\j-4.5)*.32},.34) {\j};
}
\foreach \j/\value in {1/0,2/0,3/0,4/1,5/1,6/1,7/0,8/0}{
 \ifnum\value=1
  \node[onebit] at ({3.8+(\j-4.5)*.32},0) {\value};
 \else
  \node[bit] at ({3.8+(\j-4.5)*.32},0) {\value};
 \fi
}
\foreach \j/\value in {1/1,2/0,3/0,4/1,5/1,6/1,7/0,8/0}{
 \ifnum\j=1
  \node[sharedbit] at ({7.6+(\j-4.5)*.32},0) {\value};
 \else
  \ifnum\value=1
   \node[onebit] at ({7.6+(\j-4.5)*.32},0) {\value};
  \else
   \node[bit] at ({7.6+(\j-4.5)*.32},0) {\value};
  \fi
 \fi
}
\foreach \j/\value in {1/1,2/0,3/0,4/1,5/1,6/0,7/0,8/0}{
 \ifnum\j=1
  \node[sharedbit] at ({11.4+(\j-4.5)*.32},0) {\value};
 \else
  \ifnum\value=1
   \node[onebit] at ({11.4+(\j-4.5)*.32},0) {\value};
  \else
   \node[bit] at ({11.4+(\j-4.5)*.32},0) {\value};
  \fi
 \fi
}
\draw[patharrow] (1.55,0)--(2.25,0);
\draw[patharrow] (5.35,0)--(6.05,0);
\draw[patharrow] (9.15,0)--(9.85,0);
\node[align=center,font=\footnotesize,text width=3.4cm] at (1.9,-1.8)
 {Choose a set\\disjoint from $E$};
\node[align=center,font=\footnotesize,text width=3.4cm] at (5.7,-1.8)
 {Add\\position $1$};
\node[align=center,font=\footnotesize,text width=3.4cm] at (9.5,-1.8)
 {Remove\\position $6$};
\node[align=center,text width=13.6cm,font=\small] at (5.7,-3.15)
 {Two single-position changes; $E\cap F=\{1\}$, so $|E\cap F|=1\le2$.\\[3pt]
 Intermediate sets may change size. Only the initial and final sets must have size $k$.};
\end{tikzpicture}}
\caption{Why a polynomial of degree $t$ permits at most $t$ common erased positions. The example uses $n=8$ and two single-position changes: first choose $G_0$ disjoint from $E$, then add position $1$ and remove position $6$. The final set $F$ has only one position in common with $E$. The intermediate set $G_1$ has size four, although the initial and final sets have size three. This illustrates the support argument in \eqref{eq:bt-power-support}--\eqref{eq:bt-intersection-bound}; it is independent of the asymptotic choice of $t$ in \eqref{eq:bt-degree-budget}.}
\label{fig:tensor-support}
\end{figure}
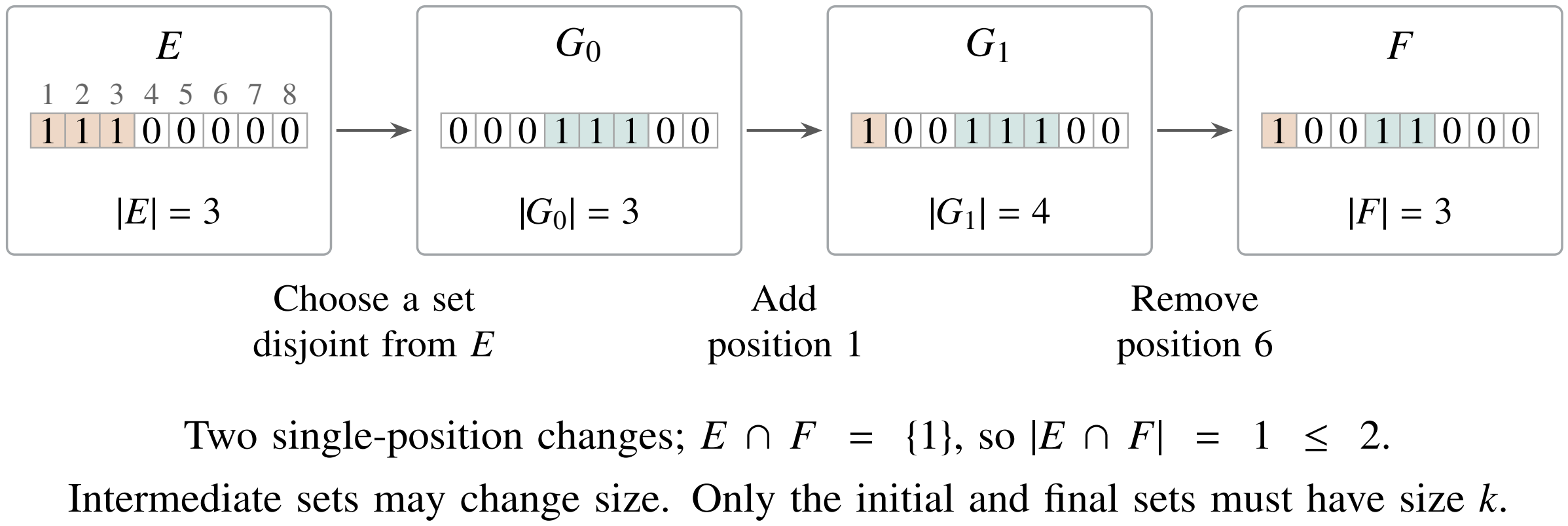

Since $q_t$ has degree at most $t$, a nonzero entry of $\mathsf T=\mathsf Kq_t(\mathsf H)$ requires at least one $\ell\in\{0,\ldots,t\}$ and an intermediate set $G$ for which
\begin{equation}\label{eq:bt-nonzero-path}
\mathsf K_{EG}\ne0,\qquad
(\mathsf H^\ell)_{GF}\ne0.
\end{equation}
Equations~\eqref{eq:bt-product-disjointness} and \eqref{eq:bt-power-support} then give
\begin{equation}\label{eq:bt-intersection-bound}
E\cap G=\varnothing,\qquad
|E\cap F|\le|F\setminus G|
\le|G\mathbin{\triangle}F|\le\ell\le t.
\end{equation}
Thus $\mathsf T_{EF}=0$ whenever $|E\cap F|>t$. Restricting to $\Omega_{n,k}$ and dividing by the nonzero number $a$ cannot create a nonzero entry. This proves \eqref{eq:bt-goal-support}. Intermediate sets need not have size $k$; only the initial and final labels of $W$ are restricted to that size. Figure~\ref{fig:tensor-support} illustrates this support argument.

\paragraph{Controlling signed coefficients.}
For a real matrix $\mathsf A$, recall the absolute row-sum norm
\begin{equation}\label{eq:bt-row-norm}
\norm{\mathsf A}_{\row}
\coloneqq\max_E\sum_F|\mathsf A_{EF}|.
\end{equation}
Its submultiplicativity follows from the triangle inequality:
\begin{equation}\label{eq:bt-row-submultiplicativity}
\sum_G|(\mathsf A\mathsf B)_{EG}|
\le\sum_F|\mathsf A_{EF}|\sum_G|\mathsf B_{FG}|
\le\norm{\mathsf B}_{\row}\sum_F|\mathsf A_{EF}|.
\end{equation}
Taking the maximum over $E$ proves $\norm{\mathsf A\mathsf B}_{\row}\le\norm{\mathsf A}_{\row}\norm{\mathsf B}_{\row}$.

The matrices $\mathsf K$ and $\mathsf P_j$ have nonnegative entries and row sums one. Consequently,
\begin{align}
\norm{\mathsf K}_{\row}&=1,\qquad
\norm{\mathsf P_j}_{\row}=1,\label{eq:bt-stochastic-row-norm}\\
\norm{\mathsf H}_{\row}
&\le\sum_{j=1}^n\norm{I-\mathsf P_j}_{\row}
\le\sum_{j=1}^n\bigl(\norm I_{\row}+\norm{\mathsf P_j}_{\row}\bigr)
=2n.\label{eq:bt-counting-row-norm}
\end{align}
The assumption $t<k\le p_0n<n$ implies $t\le n$. Evaluating the polynomial \eqref{eq:bt-polynomial-binomial} at $\mathsf H$, for $1\le\ell\le t$ we obtain
\begin{align}
\left\|\binom{\mathsf H}{\ell}\right\|_{\row}
&\le\frac1{\ell!}
\prod_{s=0}^{\ell-1}\norm{\mathsf H-sI}_{\row}
\label{eq:bt-binomial-row-first}\\
&\le\frac1{\ell!}\prod_{s=0}^{\ell-1}(2n+s)
\label{eq:bt-binomial-row-second}\\
&\le\frac{(3n)^\ell}{\ell!}.
\label{eq:bt-binomial-row-final}
\end{align}
The same final bound holds for $\ell=0$, because $\binom{\mathsf H}{0}=I$.

Let $v\coloneqq c^{-1}-1>0$. Using \eqref{eq:bt-meixner} and \eqref{eq:bt-binomial-row-final}, and then $1/\ell!\le1$, gives
\begin{align}
\norm{M_j(\mathsf H)}_{\row}
&\le\sum_{\ell=0}^j
\binom j\ell v^\ell\frac{(3n)^\ell}{\ell!}
\label{eq:bt-meixner-row-first}\\
&\le(1+3nv)^j
\le(\Gamma_0n)^t,\qquad 0\le j\le t.
\label{eq:bt-meixner-row-final}
\end{align}
The last inequality uses $\Gamma_0=1+3v$ from \eqref{eq:bt-degree-budget}, $n\ge1$, and $j\le t$.

The coefficients in the sum \eqref{eq:bt-polynomial} are nonnegative and sum to one:
\begin{equation}\label{eq:bt-positive-coefficients}
\frac{(1-c)c^j}{1-c^{t+1}}\ge0,\qquad
\sum_{j=0}^t\frac{(1-c)c^j}{1-c^{t+1}}=1.
\end{equation}
Applying the triangle inequality to that sum, and then \eqref{eq:bt-stochastic-row-norm}, proves
\begin{align}
\norm{q_t(\mathsf H)}_{\row}
&\le(\Gamma_0n)^t,\label{eq:bt-polynomial-row-bound}\\
\norm{\mathsf T}_{\row}
&\le\norm{\mathsf K}_{\row}\norm{q_t(\mathsf H)}_{\row}
\le(\Gamma_0n)^t.\label{eq:bt-full-row-bound}
\end{align}
Taking a principal block cannot increase the absolute row-sum norm, because it only removes terms from each retained row. Therefore, \eqref{eq:bt-final-matrix} and \eqref{eq:bt-normalizer-positive} give
\begin{equation}\label{eq:bt-final-row-bound}
\norm W_{\row}
=a^{-1}\norm{\mathsf B}_{\row}
\le a^{-1}\norm{\mathsf T}_{\row}
\le2(n+1)(\Gamma_0n)^t=\ell_n.
\end{equation}
This proves \eqref{eq:bt-goal-mass} and completes the proof of Lemma~\ref{lem:filter}. Every estimate holds for all $n\ge3$ and every $t<k\le p_0n$; the constants $r_0,c_0,c,\Gamma_0$ depend only on the fixed upper bound $p_0$.

\subsection{Relation to the earlier Boolean construction}\label{app:tensor-precedents}

The signed-averaging arguments of Refs.~\cite[Lemma~S3]{Kondra2026} and~\cite[Lemma~4.1]{WildePrivate2026} also use a polynomial and a tensor-product eigenbasis. Their matrices are supported on pairs of binary strings with few agreements, where both $00$ and $11$ count as agreements. Our support condition counts only common erased positions, corresponding to $11$.

The distinction prevents a direct restriction of those earlier matrices. If $E,F$ each have size $k$, their number of agreements is
\begin{equation}\label{eq:bt-agreement-obstruction}
|E\cap F|+|E^c\cap F^c|
=n-2k+2|E\cap F|\ge(1-2p_0)n.
\end{equation}
For a support parameter of order $\ln n$, their entire size-$k$ principal block would therefore be zero for sufficiently large $n$. Here we change the one-position matrix before restricting: $\mathsf R_{11}=0$ enforces disjointness, while $\mathsf R_{00}=1-r>0$ permits a position to be absent from both sets.

This biased one-position matrix gives the additional spectral factor $r^i$ in \eqref{eq:bt-polynomial-spectral-reduction}. At $\alpha=1/2$, $\mathsf R$ would be complementation and $r=1$, so that decay would disappear. The assumption $p_0<1/2$ keeps it uniformly present. The product-space operator with eigenvalues $(-r)^iq_t(i)$ belongs to the polynomial family in Ref.~\cite[Section~3.3, Lemmas~3.10--3.12]{Filmus2013}; their intersection threshold is $t+1$ when the polynomial degree is at most $t$. The classical matrix family and Meixner identities supply the ingredients, while \eqref{eq:bt-compressed-error}--\eqref{eq:bt-slice-error-final} and \eqref{eq:bt-final-row-bound} give the quantitative approximation and absolute-row-sum estimates needed for the erasure converse.

\section{A direct achievable bound}\label{sec:ach}
This appendix proves the achievable parts of Theorems~\ref{thm:main} and~\ref{thm:other-criteria}. We derive a direct finite-block bound, extract its Gaussian rate, and then treat the private criteria. Section~\ref{app:haar} supplies the Haar second-moment calculation used in the first step. The quantum achievable expansion is known from the work discussed in the introduction. The purpose of the argument here is to make the erasure threshold explicit, to cover every integer input dimension, and to retain the original infidelity parameter throughout. The calculation also gives the finite-block bound used in the numerical illustration.

\subsection{Haar-isometry encoding and recovery}
Fix a target dimension $M\le d^n$. A Haar-random isometry embeds the logical system into a uniformly random $M$-dimensional subspace of the physical input. For a specified erased set, recovery is possible when the reference is nearly independent of the erased systems. Uhlmann's theorem \eqref{eq:uhlmann-form} converts a quantitative comparison with a product state into a decoder.

The finite-block expression in Proposition~\ref{prop:haar} is already given in infidelity form in Ref.~\cite[Eq.~(65)]{ChengDiscrimination2026}. That reference writes root fidelity as $F$ and infidelity as $1-F^2$; these correspond to $\sqrt F$ and $1-F$ under our squared-fidelity convention \eqref{eq:fidelity-definition}. The proof below gives a complete derivation in the present notation.

\begin{proposition}[Haar-isometry achievable bound]\label{prop:haar}
Let $n\ge1$ and $d\ge2$ be integers, let $0\le p\le1$, set $D\coloneqq d^n$, and let $M$ be an integer with $1\le M\le D$. There exists an isometric entanglement transmission encoder and a decoder for $\cE_{p,d}^{\otimes n}$ whose Bell-target fidelity $F_{\et}$, defined in \eqref{eq:transmission-fidelity}, satisfies
\begin{equation}\label{eq:haar-bound}
F_{\et}\ge\E_{\Ner}\!\left[\frac1{1+a_{\Ner}}\right]
\ge\E_{\Ner}\!\left[\frac1{1+Md^{2\Ner-n}}\right],
\end{equation}
where $\Ner\sim\Bin(n,p)$ and
\begin{equation}\label{eq:ak}
a_k\coloneqq\frac{(d^{2k}-1)(MD-1)}{D^2-1},
\qquad 0\le k\le n.
\end{equation}
\end{proposition}
\begin{proof}
We first bound the optimal recovery fidelity for one isometry and one erased set in terms of the purity of the reference--erased-system state. Averaging that purity is a two-copy Haar calculation. We then average over the erased sets before selecting a single encoder.

Choose an isometry $V\colon\bbC^M\to\bbC^D$ according to Haar measure, equivalently the first $M$ columns of a Haar-random unitary on $\bbC^D$. Haar distribution means that multiplying the random unitary on either side by any fixed unitary leaves its distribution unchanged. Encode one share of $\Phi_{RS}^{M}$ with $V$. For a fixed erased set $E\subseteq[n]$ with $|E|=k$, denote the surviving system by $B\coloneqq A_{E^c}$ and the erased system by $Z\coloneqq A_E$. The set label $E$ and the system label $Z$ are thus distinct. Set
\begin{equation}\label{eq:haar-dimensions}
b\coloneqq d^{n-k},\qquad e\coloneqq d^k,\qquad D=be.
\end{equation}
Define $\ket\omega_{RBZ}\coloneqq(I_R\otimes V)\ket{\Phi^M}_{RS}$ after identifying $A^n$ with $BZ$, and let $\omega_{RZ}\coloneqq\Tr_B\ket\omega\bra\omega_{RBZ}$. Because $V$ is an isometry, $\omega_R=\pi_R$.

Define the best recovery fidelity from $B$ by
\begin{equation}\label{eq:conditional-optimal-recovery}
F_{\mathrm{opt}}(V,E)\coloneqq
\max_{\cD_E:B\to\widehat S}
\Tr\!\left[\Phi^M_{R\widehat S}
(\id_R\otimes\cD_E)(\Tr_Z\ket\omega\bra\omega_{RBZ})\right],
\end{equation}
where the maximum is over channels. Uhlmann's theorem~\cite{Uhlmann1976} applied to $\omega_{RZ}$ and $\pi_R\otimes\pi_Z$ gives
\begin{equation}\label{eq:uhlmann-lower}
F_{\mathrm{opt}}(V,E)
\ge F(\omega_{RZ},\pi_R\otimes\pi_Z)
=\frac{(\Tr\sqrt{\omega_{RZ}})^2}{Me}.
\end{equation}
Here is the recovery step in operator form. Choose an auxiliary system $T$ large enough to purify $\pi_Z$ and to accommodate an isometry from $B$ to $\widehat S T$. Let $\ket\xi_{TZ}$ purify $\pi_Z$. Then $\ket{\Phi^M}_{R\widehat S}\otimes\ket\xi_{TZ}$ purifies $\pi_R\otimes\pi_Z$. Uhlmann's theorem supplies an isometry $U_E\colon B\to\widehat S T$ such that, with
\begin{align}
\ket{\omega'}_{R\widehat S TZ}
&\coloneqq(I_R\otimes U_E\otimes I_Z)\ket\omega_{RBZ},\\
\cD_E(X_B)&\coloneqq\Tr_T(U_EX_BU_E^\dagger),
\label{eq:recovery-isometry-channel}
\end{align}
we have
\begin{align}
F(\omega_{RZ},\pi_R\otimes\pi_Z)
&=\left|\bigl(\bra{\Phi^M}_{R\widehat S}\otimes\bra\xi_{TZ}\bigr)
\ket{\omega'}_{R\widehat S TZ}\right|^2,\\
&=\Tr\!\left[(\Phi^M_{R\widehat S}\otimes\ket\xi\bra\xi_{TZ})
\ket{\omega'}\bra{\omega'}_{R\widehat S TZ}\right],\\
&\le\Tr\!\left[(\Phi^M_{R\widehat S}\otimes I_{TZ})
\ket{\omega'}\bra{\omega'}_{R\widehat S TZ}\right],\\
&\le F_{\mathrm{opt}}(V,E).
\label{eq:recovery-bell-acceptance}
\end{align}
The first inequality uses $\ket\xi\bra\xi_{TZ}\le I_{TZ}$; the last uses the optimization in \eqref{eq:conditional-optimal-recovery}. This proves the first inequality in \eqref{eq:uhlmann-lower} without assuming that $\omega_{RZ}$ has full rank.

We next lower-bound the trace-square-root expression using purity. If $\lambda_j$ are the eigenvalues of a normalized state $\omega$, H\"older's inequality with exponents $3/2$ and $3$ gives
\begin{equation}\label{eq:holder}
1=\sum_j(\lambda_j^{1/2})^{2/3}(\lambda_j^2)^{1/3}
\le\left(\sum_j\sqrt{\lambda_j}\right)^{2/3}
\left(\sum_j\lambda_j^2\right)^{1/3}.
\end{equation}
Cubing the nonnegative quantities in \eqref{eq:holder} and then dividing by the positive purity gives
\begin{align}
1&\le(\Tr\sqrt\omega)^2\Tr(\omega^2),\\
(\Tr\sqrt\omega)^2&\ge\frac1{\Tr(\omega^2)}.
\label{eq:holder-purity-rearrangement}
\end{align}
Zero eigenvalues cause no difficulty. Combining this inequality with \eqref{eq:uhlmann-lower} yields
\begin{equation}\label{eq:purity-recovery}
F_{\mathrm{opt}}(V,E)\ge\frac1{Me\Tr(\omega_{RZ}^2)}.
\end{equation}
The second-moment identity \eqref{eq:purity-final}, derived in Section~\ref{app:haar}, is
\begin{equation}\label{eq:haar-purity}
\E_V\Tr(\omega_{RZ}^2)
=\frac{e(b^2-1)/M+b(e^2-1)}{D^2-1}.
\end{equation}
This expectation depends only on the size of $E$.

The function $x\mapsto1/x$ is convex on the strictly positive purities of normalized states. Jensen's inequality applied to \eqref{eq:purity-recovery} therefore gives
\begin{align}
\E_VF_{\mathrm{opt}}(V,E)
&\ge\frac1{Me}\E_V\!\left[\frac1{\Tr(\omega_{RZ}^2)}\right],\\
&\ge\frac1{Me\E_V\Tr(\omega_{RZ}^2)}.
\label{eq:haar-jensen-step}
\end{align}
For the denominator, \eqref{eq:haar-purity} and $D=be$ give
\begin{align}
Me\E_V\Tr(\omega_{RZ}^2)
&=\frac{e^2(b^2-1)+Mbe(e^2-1)}{D^2-1},\\
&=\frac{D^2-e^2+MD(e^2-1)}{D^2-1},\\
&=1+\frac{(e^2-1)(MD-1)}{D^2-1}
=1+a_k.
\label{eq:haar-denominator-algebra}
\end{align}
Consequently,
\begin{equation}\label{eq:haar-average}
\E_VF_{\mathrm{opt}}(V,E)\ge\frac1{1+a_k}.
\end{equation} Also,
\begin{equation}\label{eq:ak-simplification}
\frac{e^2-1}{D^2-1}\le\frac{e^2}{D^2},
\qquad MD-1\le MD,
\end{equation}
because $1\le e\le D$. Hence $a_k\le Me^2/D=Md^{2k-n}$.

For each fixed isometry, the orthogonal flag sectors permit the receiver to use a separate optimal recovery channel for each $E$. There are finitely many patterns, and each optimum is attained by compactness. The resulting fidelity is
\begin{equation}\label{eq:chosen-decoder-fidelity}
F(V)\coloneqq\sum_{E\subseteq[n]}w_p(E)F_{\mathrm{opt}}(V,E).
\end{equation}
This is a continuous function of $V$: it is the maximum of a continuous fidelity over a fixed compact set of decoders, summed over finitely many patterns. We now average over the encoder \emph{after} including every erased set:
\begin{align}
\E_VF(V)
&=\sum_{E\subseteq[n]}w_p(E)\E_VF_{\mathrm{opt}}(V,E),\\
&\ge\sum_{E\subseteq[n]}\frac{w_p(E)}{1+a_{|E|}},\\
&=\sum_{k=0}^n\binom nk p^k(1-p)^{n-k}\frac1{1+a_k}
=\E_{\Ner}\!\left[\frac1{1+a_{\Ner}}\right].
\label{eq:haar-all-pattern-average}
\end{align}
The first equality interchanges an expectation with a finite sum. The inequality uses \eqref{eq:haar-average}; the final equality groups equal-size erased sets, all of which have the same probability. The space of isometries is compact and $F(V)$ is continuous, so a maximizing isometry $V_*$ exists and satisfies
\begin{equation}\label{eq:haar-encoder-selection}
F(V_*)\ge\E_VF(V)
\ge\E_{\Ner}\!\left[\frac1{1+a_{\Ner}}\right].
\end{equation}
This selects one encoder for the complete channel. The bound on $a_k$ proves the second inequality in \eqref{eq:haar-bound}.
\end{proof}

The order of the averages in this argument matters. Equation~\eqref{eq:haar-average} is an ensemble statement at a specified erased set. It does not assert that one isometry attains that lower bound at every set or every value of $k$. A single code is selected only after the channel's erasure probabilities have been included in \eqref{eq:chosen-decoder-fidelity}.

\subsection{Extracting the Gaussian achievable rate}
The simplified bound in \eqref{eq:haar-bound} changes from high fidelity to low fidelity around $\log_dM=n-2k$. Its transition has a width independent of $n$ in the integer variable $k$. We show that replacing it by a sharp threshold costs only $O(n^{-1/2})$ in probability and therefore only $O(1)$ in the logarithmic dimension.

\begin{corollary}[Second-order lower bound]\label{cor:normal-lower}
Fix an integer $d\ge2$, $0<p<1/2$, and $0<\eps<1$. Let $Q_{n,\et}^{\eps}$ denote the optimal entanglement transmission logarithmic code size in \eqref{eq:operational-Q}, let $C,V$ be given by \eqref{eq:qv}, and let $\Phi$ be defined in \eqref{eq:normal-cdf}. There is a finite constant $C_-$, depending only on $d,p,\eps$, such that, for all sufficiently large integers $n$,
\begin{equation}\label{eq:normal-lower}
Q_{n,\et}^{\eps}
\ge nC+\sqrt{nV}\,\Phi^{-1}(\eps)-C_-.
\end{equation}
\end{corollary}
\begin{proof}
For an integer $u$ with $0\le u<n/2$, choose the integer code dimension $M\coloneqq d^{n-2u}$. From Proposition~\ref{prop:haar} and $x/(1+x)\le\min\{1,x\}$ for $x\ge0$, the achievable infidelity satisfies
\begin{align}
1-F_{\et}
&\le\E\min\{1,d^{2(\Ner-u)}\}\\
&\le\Prb\{\Ner>u\}
+\sum_{j=0}^{u}\Prb\{\Ner=u-j\}d^{-2j}\\
&\le\Prb\{\Ner>u\}+\frac{b_n}{1-d^{-2}},
\label{eq:ach-tail}
\end{align}
where $b_n\coloneqq\max_j\Prb\{\Ner=j\}$. The last line extends a finite geometric sum to an infinite one.

Lemma~\ref{lem:binomial-tools} gives constants $B_p,C_p$ for which the uniform normal-approximation error is at most $B_p/\sqrt n$ and $b_n\le C_p/\sqrt n$. Set $z_\eps\coloneqq\Phi^{-1}(1-\eps)$, $\sigma_n\coloneqq\sqrt{np(1-p)}$, and
\begin{equation}\label{eq:positive-density}
\mu_\eps\coloneqq\min_{|x-z_\eps|\le1}\phi(x)>0.
\end{equation}
Choose a fixed positive constant $c_{\rm ach}$ large enough that
\begin{equation}\label{eq:ach-C}
\frac{\mu_\eps c_{\rm ach}}{\sqrt{p(1-p)}}
>B_p+\frac{C_p}{1-d^{-2}},
\end{equation}
and define
\begin{equation}\label{eq:ach-threshold}
u\coloneqq\left\lceil np+\sigma_nz_\eps+c_{\rm ach}\right\rceil.
\end{equation}
For large $n$, $c_{\rm ach}/\sigma_n\le1$, so integration of the normal density gives
\begin{equation}\label{eq:normal-gain}
\Phi\!\left(\frac{u-np}{\sigma_n}\right)
\ge\Phi\!\left(z_\eps+\frac{c_{\rm ach}}{\sigma_n}\right)
\ge1-\eps+\frac{\mu_\eps c_{\rm ach}}{\sigma_n}.
\end{equation}
The Berry--Esseen estimate \eqref{eq:berry-esseen} and \eqref{eq:normal-gain} imply
\begin{align}
\Prb\{\Ner>u\}
&\le1-\Phi\!\left(\frac{u-np}{\sigma_n}\right)+\frac{B_p}{\sqrt n},\\
&\le\eps-\frac{\mu_\eps c_{\rm ach}}{\sigma_n}
+\frac{B_p}{\sqrt n}.
\label{eq:ach-tail-with-slack}
\end{align}
Using $b_n\le C_p/\sqrt n$ in \eqref{eq:ach-tail} now gives
\begin{align}
1-F_{\et}
&\le\eps+\frac1{\sqrt n}
\left[B_p+\frac{C_p}{1-d^{-2}}
-\frac{\mu_\eps c_{\rm ach}}{\sqrt{p(1-p)}}\right],\\
&\le\eps.
\label{eq:ach-error-bound}
\end{align}
The bracket is negative by the explicit choice \eqref{eq:ach-C}.

Because $0<p<1/2$ is fixed, the choice in \eqref{eq:ach-threshold} satisfies $0\le u<n/2$ for all sufficiently large $n$. Thus it gives an admissible code dimension. The ceiling in \eqref{eq:ach-threshold} gives
\begin{equation}\label{eq:ach-rounding}
u<np+\sigma_nz_\eps+c_{\rm ach}+1.
\end{equation}
Thus rounding upward costs at most $2\log_2d$ in $\log_2M$. Substituting \eqref{eq:ach-rounding} and using $z_\eps=\Phi^{-1}(1-\eps)=-\Phi^{-1}(\eps)$ yields
\begin{align}
\log_2M
&=(n-2u)\log_2d\\
&\ge n(1-2p)\log_2d
+2\sigma_n\log_2d\,\Phi^{-1}(\eps)
-2(c_{\rm ach}+1)\log_2d.
\label{eq:ach-final}
\end{align}
This proves \eqref{eq:normal-lower}.
\end{proof}

\subsection{Private achievability and the maximal-message criterion}
Lemma~\ref{lem:q-to-p} immediately transfers the quantum lower bound to optimized private infidelity at the same error. Applying it at quantum error $\eta(\eps)$ from \eqref{eq:one-shot-ordering} transfers the lower bound to the actual-marginal criterion. For the maximal-message criterion, we keep a subset of messages that all work with one common ideal environment state.

\begin{lemma}[Expurgation with a common environment state]\label{lem:expurgation}
Let $M\ge1$ be an integer, let $0<\eps_0<\eps<1$, and consider a size-$M$ private code for a finite-dimensional quantum channel with environment output $Z$. Suppose its optimized joint fidelity \eqref{eq:private-opt} satisfies $f_{\opt}\ge1-\eps_0$. There is an unassisted private code for the same channel with at least $rM$ messages and maximal-message joint infidelity \eqref{eq:private-max} at most $\eps$, where
\begin{equation}\label{eq:expurgation-fraction}
r\coloneqq
\frac{\sqrt{1-\eps_0}-\sqrt{1-\eps}}
{1-\sqrt{1-\eps}}.
\end{equation}
\end{lemma}
\begin{proof}
Choose a common environment state $\sigma_Z$ that attains the optimization in \eqref{eq:private-opt}. Such a state exists because the finite-dimensional state space is compact and fidelity is continuous. For each message $m$, define
\begin{equation}\label{eq:expurgation-message-root}
a_m\coloneqq\sqrt{F(\omega^m_{\widehat K Z},
\ket m\bra m_{\widehat K}\otimes\sigma_Z)}.
\end{equation}
The direct-sum identity \eqref{eq:message-fidelity} and the assumed joint fidelity give
\begin{equation}\label{eq:expurgation-root-mean}
\frac1M\sum_{m=1}^M a_m=\sqrt{f_{\opt}}\ge\sqrt{1-\eps_0}.
\end{equation}
Set $a\coloneqq\sqrt{1-\eps}$, define the retained set $\mathcal G\coloneqq\{m:a_m\ge a\}$, and let $q\coloneqq|\mathcal G|/M$ be its fraction. On $\mathcal G$ we use $a_m\le1$, and on its complement we use $a_m<a$. Consequently,
\begin{align}
\sqrt{1-\eps_0}
&\le\frac1M\sum_{m=1}^M a_m,\\
&\le\frac{|\mathcal G|}{M}
  +\left(1-\frac{|\mathcal G|}{M}\right)a
=q+(1-q)a,\\
q&\ge\frac{\sqrt{1-\eps_0}-a}{1-a}=r>0.
\label{eq:expurgation-count}
\end{align}
Thus $\mathcal G$ is nonempty and contains at least $rM$ messages.

To construct a decoder with exactly these retained labels, fix $m_0\in\mathcal G$. Let $r_0(j)$ equal $j$ for $j\in\mathcal G$ and equal $m_0$ otherwise. Define the classical postprocessing channel
\begin{equation}\label{eq:expurgation-postprocess}
\mathcal R(X_{\widehat K})\coloneqq
\sum_{j=1}^M\bra jX_{\widehat K}\ket j\,
\ket{r_0(j)}\bra{r_0(j)}_{\widehat K'},
\end{equation}
where $\widehat K'$ has an orthonormal basis indexed by $\mathcal G$. Use the original input states for $m\in\mathcal G$ with a uniform prior, and apply $\mathcal R$ after the original decoder. For every retained message, its new conditional output is
\begin{equation}\label{eq:expurgation-new-output}
\omega^{\prime,m}_{\widehat K'Z}\coloneqq
(\mathcal R\otimes\id_Z)(\omega^m_{\widehat K Z}).
\end{equation}
Because $r_0(m)=m$ for retained labels, the same channel maps each original ideal state to the corresponding new ideal state. Fidelity monotonicity therefore yields
\begin{align}
F(\omega^{\prime,m}_{\widehat K'Z},
\ket m\bra m_{\widehat K'}\otimes\sigma_Z)
&\ge F(\omega^m_{\widehat K Z},
\ket m\bra m_{\widehat K}\otimes\sigma_Z),\\
&=a_m^2\ge1-\eps
\qquad(m\in\mathcal G).
\label{eq:expurgation-message-bound}
\end{align}
The normalized state $\sigma_Z$ is common to all retained messages, so it is an admissible choice in the optimization defining \eqref{eq:private-max}. The new code consequently has maximal-message joint infidelity at most $\eps$. Its actual environment marginal may change when the prior is restricted; that marginal is not the comparison state used in this criterion.
\end{proof}

The following proof supplies the detailed estimates used in Section~\ref{subsec:completion}, including the uniformity in the error parameter needed for the maximal-message bound. We retain the other task comparisons here so that the achievable argument can also be read within this appendix.

\begin{proof}[Proofs of Theorems~\ref{thm:main} and~\ref{thm:other-criteria}]
Corollary~\ref{cor:normal-upper} gives all upper bounds. The entanglement transmission lower bound in Corollary~\ref{cor:normal-lower} also applies to entanglement generation, because applying an entanglement transmission encoder to one share of a maximally entangled state prepares an admissible entanglement generation state with the same fidelity. Lemma~\ref{lem:q-to-p} gives the optimized private lower bound at the same error, and the actual-marginal lower bound from an entanglement transmission code at the error $\eta(\eps)$. To distinguish the two error parameters explicitly, let $F_e$ be the fidelity of this quantum code. Corollary~\ref{cor:normal-lower} gives $F_e\ge1-\eta(\eps)$, and \eqref{eq:one-shot-ordering} defines $\eta(\eps)=1-\sqrt{1-\eps}$. The comparison in \eqref{eq:q-to-p-fidelity} therefore yields
\begin{equation}\label{eq:actual-marginal-achievable-error}
f_{\marg}\ge F_e^2\ge\bigl(1-\eta(\eps)\bigr)^2=1-\eps.
\end{equation}
Thus the resulting private infidelity is at most $\eps$. This proves \eqref{eq:main-bounds} and \eqref{eq:marg-main}.

For the maximal-message criterion, use the optimized lower bound at $\eps_n\coloneqq\eps-n^{-1/2}$, which is positive for sufficiently large $n$. The constant in Corollary~\ref{cor:normal-lower} can be chosen uniformly for error values in a fixed compact neighborhood of $\eps$: the proof uses only bounded normal quantiles and a common positive lower bound on their nearby normal densities. Choose the compact interval $J\coloneqq[\eps/2,(1+\eps)/2]\subset(0,1)$, which contains $\eps_n$ and $\eps$ for all sufficiently large $n$. Define
\begin{equation}\label{eq:expurgation-uniform-density}
\mu_J\coloneqq\min_{\substack{u\in J\\v\in[-1,1]}}
\phi\!\left(\Phi^{-1}(1-u)+v\right)>0.
\end{equation}
The minimum is positive because the function being minimized is continuous and positive on the compact set $J\times[-1,1]$. For every error $u\in J$, the density minimum in \eqref{eq:positive-density}, with $\eps=u$, is at least $\mu_J$. Thus replacing $\mu_\eps$ by $\mu_J$ in \eqref{eq:ach-C} gives one admissible constant $c_{\rm ach}$, and hence one constant loss in \eqref{eq:ach-final}, for every $u\in J$. The normal quantiles are bounded on $J$, so one blocklength cutoff also ensures that the integer threshold in \eqref{eq:ach-threshold} is nonnegative and strictly smaller than $n/2$ for every error in $J$. To quantify the effect on the Gaussian term, the derivative bound
\begin{equation}\label{eq:expurgation-quantile-lipschitz}
L_J\coloneqq\max_{u\in J}\frac1{\phi(\Phi^{-1}(u))}<\infty
\end{equation}
gives, by the mean value theorem,
\begin{equation}\label{eq:expurgation-quantile-cost}
\sqrt{nV}\left|\Phi^{-1}(\eps_n)-\Phi^{-1}(\eps)\right|
\le\sqrt{nV}\,L_Jn^{-1/2}=L_J\sqrt V.
\end{equation}
Thus the preliminary reduction of the error changes the achievable logarithmic code size by only a bounded amount.

In \eqref{eq:expurgation-fraction}, with $\eps_0=\eps_n$, rationalizing the numerator gives
\begin{equation}\label{eq:expurgation-loss}
r=\frac{n^{-1/2}}
{\bigl(\sqrt{1-\eps+n^{-1/2}}+\sqrt{1-\eps}\bigr)
\bigl(1-\sqrt{1-\eps}\bigr)}.
\end{equation}
The denominator in \eqref{eq:expurgation-loss} converges to the positive constant $2\sqrt{1-\eps}(1-\sqrt{1-\eps})$, so
\begin{equation}\label{eq:expurgation-log-fraction}
\log_2r=-\tfrac12\log_2n+O(1).
\end{equation}
Let $M$ be the size of the optimized private code at error $\eps_n$, and let $M'$ be the retained size. Lemma~\ref{lem:expurgation}, \eqref{eq:expurgation-quantile-cost}, and \eqref{eq:expurgation-log-fraction} imply
\begin{align}
\log_2M'&\ge\log_2M+\log_2r,\\
&\ge nC+\sqrt{nV}\,\Phi^{-1}(\eps)
-\tfrac12\log_2n-O(1).
\label{eq:expurgation-final-rate}
\end{align}
This proves the maximal-message lower bound in \eqref{eq:max-main}. Finally, dividing the upper and lower remainders by $\sqrt n$ and using $(\ln n)^2/\sqrt n\to0$ proves the second-order limits.
\end{proof}

\subsection{The Haar second-moment calculation}\label{app:haar}
We prove \eqref{eq:haar-purity} using only the second moments of one random unitary column and of two distinct columns. This avoids a general integration formula. Let $D\ge2$, and let $\ket{u_1},\ldots,\ket{u_D}$ be the columns of a Haar-distributed unitary on $\bbC^D$. Write $F_D$ for the swap on two copies of this space.

\subsubsection{One column and two distinct columns}\label{app:haar-columns}
For a uniformly distributed unit vector $\ket u$, write $u_a\coloneqq\braket{a}{u}$ for its $a$th coordinate. Phase invariance makes a fourth moment vanish unless its unbarred and conjugated coordinate indices match as multisets. Permutation invariance leaves only two scalar moments to determine:
\begin{equation}\label{eq:haar-coordinate-moments}
\alpha\coloneqq\E|u_1|^4,
\qquad \beta\coloneqq\E|u_1|^2|u_2|^2.
\end{equation}
Unitary invariance permits replacing the first coordinate by $(u_1+u_2)/\sqrt2$. Expanding its fourth absolute power and discarding the phase-dependent terms gives
\begin{align}
\alpha
&=\frac14\E|u_1+u_2|^4,\\
&=\frac14\E\left[|u_1|^4+|u_2|^4
+4|u_1|^2|u_2|^2\right]
=\frac14(2\alpha+4\beta),\\
\alpha&=2\beta.
\label{eq:haar-moment-ratio}
\end{align}
The terms omitted in the second line contain a nonzero net phase power of $u_1$ or $u_2$, and hence have zero expectation. Normalization supplies the other relation:
\begin{align}
1&=\E\left(\sum_{a=1}^D|u_a|^2\right)^2,\\
&=D\alpha+D(D-1)\beta
=D(D+1)\beta.
\label{eq:haar-moment-normalization}
\end{align}
Thus $\beta=1/[D(D+1)]$ and $\alpha=2/[D(D+1)]$. For arbitrary coordinate indices $a,b,c,d$, phase invariance and these two values give
\begin{equation}\label{eq:haar-coordinate-fourth-moment}
\E[u_a\overline u_bu_c\overline u_d]
=\frac{\delta_{ab}\delta_{cd}+\delta_{ad}\delta_{cb}}{D(D+1)}.
\end{equation}
The first product of Kronecker deltas is the matrix element of the identity, and the second is the matrix element of the swap. Hence
\begin{equation}\label{eq:haar-one-column}
A\coloneqq\E\left[\ket{u_i}\bra{u_i}^{\otimes2}\right]
=\frac{I+F_D}{D(D+1)}.
\end{equation}

For $i\ne j$, invariance under column permutations makes
\begin{equation}\label{eq:haar-distinct-moment-definition}
B\coloneqq\E\left[\ket{u_i}\bra{u_j}\otimes\ket{u_j}\bra{u_i}\right]
\qquad(i\ne j)
\end{equation}
independent of the chosen ordered pair. Completeness of the columns gives the exact identity
\begin{equation}\label{eq:haar-column-completeness}
\sum_{j=1}^{D}\ket{u_i}\bra{u_j}\otimes\ket{u_j}\bra{u_i}
=(\ket{u_i}\bra{u_i}\otimes I)F_D.
\end{equation}
Indeed, applying either side to a product vector $\ket x\otimes\ket y$ gives $\braket{u_i}{y}\ket{u_i}\otimes\ket x$. On averaging, the term $j=i$ becomes $A$, and each of the other $D-1$ terms becomes $B$. Unitary invariance gives $\E\ket{u_i}\bra{u_i}=I/D$, so
\begin{align}
A+(D-1)B&=F_D/D,\\
B&=\frac1{D-1}\left[\frac{F_D}{D}
-\frac{I+F_D}{D(D+1)}\right].
\label{eq:haar-two-column-solution}
\end{align}
Combining the two fractions proves
\begin{equation}\label{eq:haar-two-columns}
B=\frac{F_D-I/D}{D^2-1}.
\end{equation}
Thus the distinct-column moment follows from orthogonality and completeness, without a second integration calculation.

\subsubsection{Purity of the reference and erased systems}\label{app:haar-purity}
Let $b,e\ge1$ be integers with $D\coloneqq be\ge2$, where $B$ is the surviving system of dimension $b$ and $Z$ is the erased system of dimension $e$. Let $1\le M\le D$. The first $M$ random columns define the isometric encoder, giving
\begin{equation}\label{eq:encoded-haar-state}
\ket\psi_{RBZ}\coloneqq\frac1{\sqrt M}\sum_{i=1}^{M}\ket i_R\ket{u_i}_{BZ},
\qquad \omega_{RZ}\coloneqq\Tr_B\ket\psi\bra\psi_{RBZ}.
\end{equation}
The reduced state has the block expansion
\begin{equation}\label{eq:haar-reduced-blocks}
\omega_{RZ}=\frac1M\sum_{i,j=1}^M\ket i\bra j_R\otimes X_{ij},
\qquad X_{ij}\coloneqq\Tr_B(\ket{u_i}\bra{u_j}_{BZ}).
\end{equation}
Since $\Tr(\ket i\bra j\ket l\bra m)=\delta_{jl}\delta_{mi}$, squaring \eqref{eq:haar-reduced-blocks} and taking its trace gives
\begin{equation}\label{eq:purity-expanded}
\Tr\omega_{RZ}^2
=\frac1{M^2}\sum_{i,j=1}^{M}
\Tr\!\left[
\Tr_B(\ket{u_i}\bra{u_j}_{BZ})\,
\Tr_B(\ket{u_j}\bra{u_i}_{BZ})\right].
\end{equation}
Let $B'$ and $Z'$ be copies of $B$ and $Z$, respectively, and let $F_{ZZ'}$ be the operator that swaps $Z$ and $Z'$. For arbitrary operators $X,Y$ on $BZ$, fix product bases and write
$X_{az,az'}\coloneqq\bra a_B\bra z_ZX\ket a_B\ket{z'}_Z$, with analogous notation for $Y$. Then
\begin{align}
\Tr[(\Tr_BX)(\Tr_BY)]
&=\sum_{z,z'}\left(\sum_aX_{az,az'}\right)
                 \left(\sum_cY_{cz',cz}\right),\\
&=\sum_{a,c,z,z'}X_{az,az'}Y_{cz',cz},\\
&=\Tr[(X\otimes Y)(I_{BB'}\otimes F_{ZZ'})].
\label{eq:swap-purity}
\end{align}
The last equality follows by inserting the basis expansion
$F_{ZZ'}=\sum_{z,z'}\ket z\bra{z'}_Z\otimes\ket{z'}\bra z_{Z'}$. Tensor factors in this expression are ordered as $BB'ZZ'$, by the natural identification with $BZB'Z'$.

Define $S\coloneqq I_{BB'}\otimes F_{ZZ'}$. The two contractions needed for \eqref{eq:haar-one-column} and \eqref{eq:haar-two-columns} are
\begin{subequations}\label{eq:haar-contractions}
\begin{align}
\Tr(IS)&=\Tr(I_{BB'})\Tr(F_{ZZ'})=b^2e,\\
\Tr(F_DS)&=\Tr(F_{BB'})\Tr(F_{ZZ'}^2)=be^2.
\end{align}
\end{subequations}
Here $F_D=F_{BB'}\otimes F_{ZZ'}$, the trace of the swap on two copies of an $r$-dimensional space is $r$, and $F_{ZZ'}^2=I_{ZZ'}$. Thus the equal-index and distinct-index expectations in \eqref{eq:purity-expanded} are, respectively,
\begin{align}
\Tr(AS)&=\frac{b^2e+be^2}{D(D+1)}
=\frac{b+e}{D+1},\\
\Tr(BS)&=\frac{be^2-b^2e/D}{D^2-1}
=\frac{b(e^2-1)}{D^2-1}.
\label{eq:haar-two-purity-contributions}
\end{align}

There are $M$ equal-index terms and $M(M-1)$ distinct-index terms in \eqref{eq:purity-expanded}. Applying \eqref{eq:haar-one-column}--\eqref{eq:haar-contractions} yields
\begin{align}
\E\Tr\omega_{RZ}^2
&=\frac1{M^2}\left[M\Tr(AS)+M(M-1)\Tr(BS)\right],\\
&=\frac1M\frac{b+e}{D+1}
+\frac{M-1}{M}\frac{b(e^2-1)}{D^2-1},\\
&=\frac{(b+e)(D-1)+(M-1)b(e^2-1)}{M(D^2-1)},\\
&=\frac{e(b^2-1)/M+b(e^2-1)}{D^2-1}.
\label{eq:purity-final}
\end{align}
For the final simplification, use $D=be$ and
$(b+e)(be-1)-b(e^2-1)=e(b^2-1)$.
This is \eqref{eq:haar-purity}. For $e=1$ it gives $1/M$, as required for the maximally mixed reference. For $b=1$ it gives one, because the reference and erased systems together are pure. Both checks include arbitrary $1\le M\le D$.

\section{Interpretation and finite-block examples}\label{sec:interpretation}
This appendix explains the main results for fixed integer input dimension $d\ge2$ and erasure probability $0<p<1/2$. We use the erasure channel in \eqref{eq:channel}, the capacity $C$ and variance $V$ in \eqref{eq:qv}, the normal distribution function $\Phi$ in \eqref{eq:normal-cdf}, and the binomial erasure count $\Ner\sim\Bin(n,p)$. The proof separates the physical fluctuations from the choice of error criterion. We first explain the common variance, then show why the actual-marginal convention changes the quantile. The finite-block illustration concerns a rigorous achievable bound. Parameter boundaries are treated separately in Appendix~\ref{app:boundaries}.

\subsection{The two values of the coherent-information density}
The dispersion in \eqref{eq:qv} has a direct interpretation at the maximally entangled input. Let
\begin{equation}\label{eq:choi-state}
\omega_{RB}\coloneqq(\id_R\otimes\cE_{p,d})(\Phi_{RA}^{d})
=(1-p)\Phi_{RB}^{d}\oplus p\pi_R\otimes\ket e\bra e_B.
\end{equation}
Set $\omega_B\coloneqq\Tr_R\omega_{RB}$. On the support of $\omega_{RB}$, the log-likelihood operator
\begin{equation}\label{eq:information-density}
L_{RB}\coloneqq\log_2\omega_{RB}-\log_2(I_R\otimes\omega_B)
\end{equation}
takes the value $+\log_2d$ in the transmitted sector and $-\log_2d$ in the erased sector. Indeed, in the transmitted sector the relevant eigenvalues of $\omega_{RB}$ and $I_R\otimes\omega_B$ are $1-p$ and $(1-p)/d$. In the erased sector they are $p/d$ and $p$. Thus the two values are explicitly
\begin{align}
\log_2(1-p)-\log_2((1-p)/d)&=\log_2d,\\
\log_2(p/d)-\log_2p&=-\log_2d.
\label{eq:density-sector-values}
\end{align} All logarithms in this calculation are restricted to the support of $\omega_{RB}$; no logarithm of a zero eigenvalue is required.

The mean and variance are consequently
\begin{align}
\Tr(\omega_{RB}L_{RB})
&=(1-p)\log_2d-p\log_2d
=(1-2p)\log_2d=C,\label{eq:density-mean}\\
\Tr\!\left[\omega_{RB}(L_{RB}-CI)^2\right]
&=(1-p)(\log_2d-C)^2+p(-\log_2d-C)^2,\\
&=(1-p)(2p\log_2d)^2+p(2(1-p)\log_2d)^2,\\
&=4p(1-p)(\log_2d)^2=V.
\label{eq:density-variance}
\end{align}
The mean is the coherent information $H(B)_\omega-H(RB)_\omega$, where $H(B)_\omega\coloneqq H(\omega_B)$, $H(RB)_\omega\coloneqq H(\omega_{RB})$, and $H(\rho)\coloneqq-\Tr(\rho\log_2\rho)$ is the von Neumann entropy. These two values also show why the sum over $n$ uses is $(n-2\Ner)\log_2d$ at this product input. Theorem~\ref{thm:main} establishes that arbitrary encoders cannot improve the coefficient obtained from these fluctuations. Merely computing \eqref{eq:density-variance} at a candidate input would not establish that converse.

\Needspace{5\baselineskip}
\subsection{An example with one channel use}\label{app:one-use}
The two private criteria already differ for a simple physical code. Take one use with $M=d$ orthogonal basis inputs, decode perfectly when the input arrives, and guess uniformly if it is erased. In the delivered sector, both conditional fidelities are one. In the erased sector, the environment holds an orthogonal copy of the message and the correct guess has probability $1/M$; both conditional fidelities are $1/M^2$. The direct-sum calculation in Lemma~\ref{lem:flags} then gives
\begin{equation}\label{eq:one-use-metric-example}
f_{\opt}=1-p+\frac p{M^2},
\qquad
f_{\marg}=\left(1-p+\frac pM\right)^2.
\end{equation}
To verify the erased-sector optimum, let $Z$ denote the environment's $M$-dimensional system in this sector. For a comparison state $\sigma_Z$, define $r_m\coloneqq\bra m_Z\sigma_Z\ket m_Z$. The messagewise root fidelities are $\sqrt{r_m/M}$. Since $\sum_mr_m=\Tr\sigma_Z=1$, Cauchy--Schwarz gives
\begin{equation}\label{eq:one-use-erased-optimum}
\frac1M\sum_{m=1}^{M}\sqrt{\frac{r_m}{M}}
\le\frac1{M\sqrt M}\sqrt{M\sum_{m=1}^{M}r_m}
=\frac1M.
\end{equation}
The choice $\sigma_Z=I_Z/M$ attains equality. Squaring \eqref{eq:one-use-erased-optimum} gives the erased-sector fidelity $1/M^2$. The maximally mixed state is also the actual erased-sector marginal. Thus the difference in \eqref{eq:one-use-metric-example} comes entirely from the ideal pattern weights.

\subsection{Why the private error convention changes the quantile}\label{subsec:criteria-meaning}
\begin{figure}[!t]
\centering
\resizebox{\textwidth}{!}{
\begin{tikzpicture}[
  font=\small,
  box/.style={draw,rounded corners=2pt,line width=.65pt,align=center,
    inner xsep=6pt,inner ysep=7pt},
  optbox/.style={box,draw=erasureTeal,fill=erasureTeal!5},
  margbox/.style={box,draw=erasureBlue,fill=erasureBlue!5},
  arrow/.style={-{Stealth[length=2mm]},line width=.75pt,draw=black!65}
]
\node[box,fill=black!3,text width=10.8cm] (flags) at (0,6.4)
  {After the decoder reads the erasure flags\\[2pt]
   $\omega'_{K\widehat K Z^n}=\bigoplus_E w_p(E)\,\omega^E_{K\widehat K A_E}$};

\node[optbox,text width=5.65cm,minimum height=2.9cm] (optimized)
  at (-3.55,3.8)
  {Optimize the ideal flag weights $q_E$\\[3pt]
   $\textstyle
     \max_q\left(\sum_E\sqrt{w_p(E)q_Ef_E}\right)^2$\\[4pt]
   $\displaystyle f_{\mathrm{opt}}=\sum_Ew_p(E)f_E$};
\node[margbox,text width=5.65cm,minimum height=2.9cm] (marginal)
  at (3.55,3.8)
  {Use the actual marginal\\as the ideal environment state\\[3pt]
   Ideal flag weights remain $w_p(E)$\\[4pt]
   $\displaystyle \sqrt{f_{\mathrm{marg}}}=\sum_Ew_p(E)\sqrt{g_E}$};
\draw[arrow] ([xshift=-35.5mm]flags.south)--(optimized.north);
\draw[arrow] ([xshift=35.5mm]flags.south)--(marginal.north);

\node[align=center,text=black!75] at (0,1.5)
  {$f_E$: optimized conditional fidelity;\quad
   $g_E$: fidelity to the actual conditional marginal.\\[2pt]
   $0\le g_E\le f_E\le1$};

\node[align=center] at (0,.45)
  {\textbf{Schematic model:} perfect good patterns and zero-fidelity bad patterns};
\filldraw[fill=erasureTeal!16,draw=erasureTeal,line width=.65pt]
  (-6.45,-1.2) rectangle (1.25,-.15);
\filldraw[fill=black!9,draw=black!50,line width=.65pt]
  (1.25,-1.2) rectangle (6.45,-.15);
\node[align=center] at (-2.6,-.675)
  {Good patterns: probability $s$\\$f_E=g_E=1$};
\node[align=center] at (3.85,-.675)
  {Bad patterns: probability $1-s$\\$f_E=g_E=0$};

\node[optbox,text width=5.65cm] at (-3.55,-2.25)
  {$\displaystyle f_{\mathrm{opt}}=s$\\[3pt]
   $f_{\mathrm{opt}}\ge1-\varepsilon
     \ \Longleftrightarrow\ 1-s\le\varepsilon$};
\node[margbox,text width=5.65cm] at (3.55,-2.25)
  {$\displaystyle f_{\mathrm{marg}}=s^2$\\[3pt]
   $f_{\mathrm{marg}}\ge1-\varepsilon
     \ \Longleftrightarrow\ 1-s\le1-\sqrt{1-\varepsilon}$};
\node[align=center,text=black!75] at (0,-3.45)
  {General codes may have intermediate conditional fidelities.};
\end{tikzpicture}}
\caption{The role of the ideal environment's erasure-pattern weights. The upper identities are \eqref{eq:flag-optimized-identity} and \eqref{eq:flag-marginal-identity}, obtained after the flag reduction in Lemma~\ref{lem:flags}. The lower bars illustrate a schematic zero-or-one conditional-fidelity model: good patterns have total probability $s$. Optimizing the ideal weights gives fidelity $s$, while retaining the actual weights gives fidelity $s^2$. This example explains the two required binomial probabilities; it is not an assertion that every code has a perfect step transition.}
\label{fig:privacy-criteria}
\end{figure}
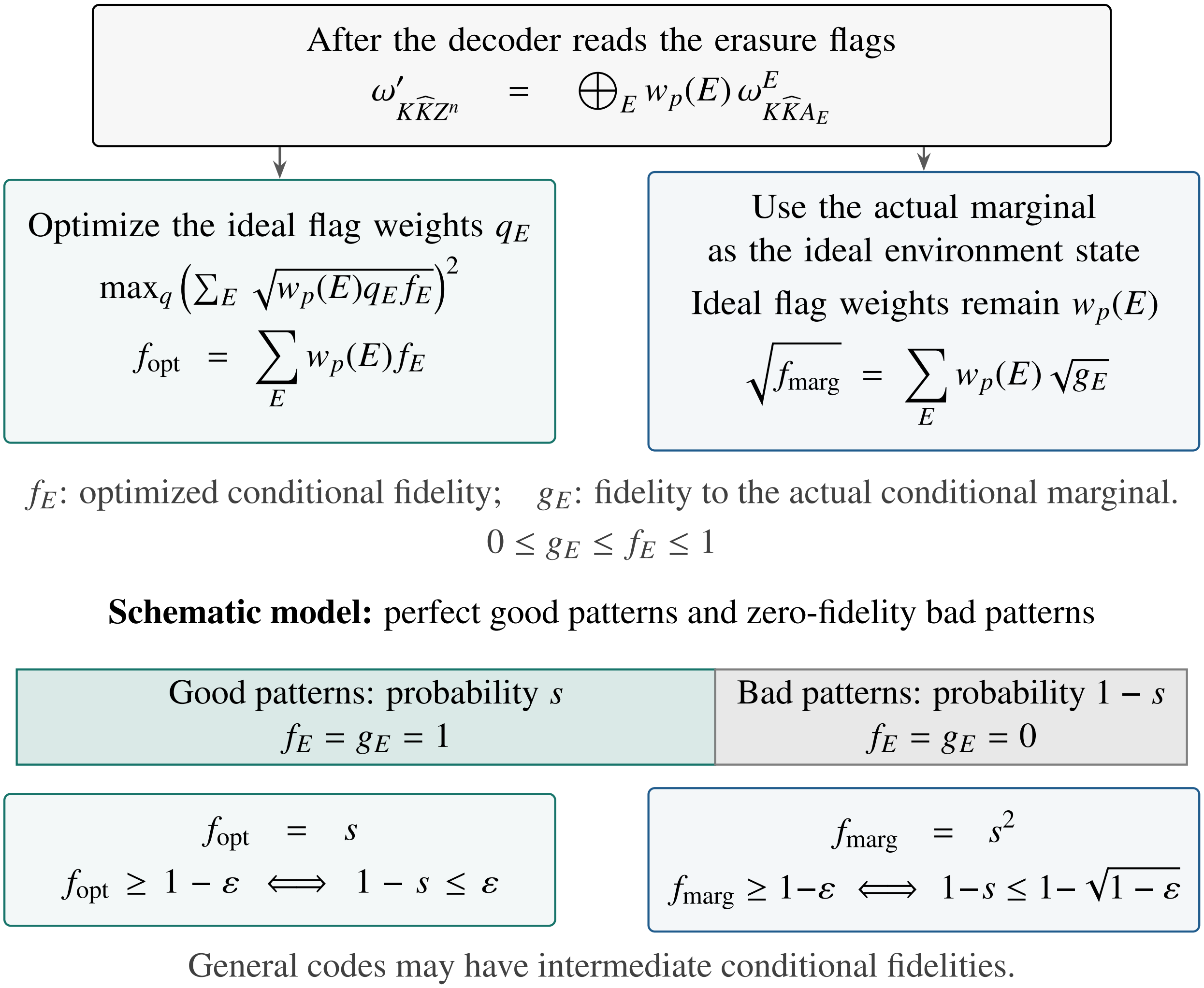

The one-use code in \eqref{eq:one-use-metric-example} and the schematic $s$ versus $s^2$ example following Theorem~\ref{thm:other-criteria} isolate the effect of the ideal flag weights. Figure~\ref{fig:privacy-criteria} places the exact identities beside that example. The optimizer can reweight the ideal environment's erasure sectors, but must still use one environment state common to all messages. The actual-marginal convention fixes both the sector weights and the conditional states.

For arbitrary block codes, the fixed-count converse and ensemble-averaged achievable bound sharpen this distinction into a difference at order $\sqrt n$. Optimized infidelity $\eps$ requires good-pattern probability approximately $1-\eps$, whereas actual-marginal infidelity $\eps$ requires approximately $\sqrt{1-\eps}$.

\begin{samepage}
Table~\ref{tab:criteria} collects the second-order terms. The maximal-message result uses the optimized common environment state and an expurgation argument; it should not be confused with the actual-marginal condition.\par
\end{samepage}

\begin{table}[!t]
\centering\small
\renewcommand{\arraystretch}{1.22}
\begin{tabular}{@{}p{.43\textwidth}p{.32\textwidth}p{.17\textwidth}@{}}
\toprule
Task and error convention & Coefficient of $\sqrt{nV}$ & Achievable loss\\
\midrule
Entanglement transmission or entanglement generation & $\Phi^{-1}(\eps)$ & $O(1)$\\
Private, optimized joint infidelity & $\Phi^{-1}(\eps)$ & $O(1)$\\
Private, maximal-message joint infidelity & $\Phi^{-1}(\eps)$ & $\tfrac12\log_2n+O(1)$\\
Private, actual environment marginal & $\Phi^{-1}(1-\sqrt{1-\eps})$ & $O(1)$\\
\bottomrule
\end{tabular}
\caption{Second-order results for fixed $d\ge2$, $0<p<1/2$, and $0<\eps<1$. Every row has leading term $nC$ and converse remainder $O((\ln n)^2)$. The last column is the loss in the proved achievable lower bound, not an optimal third-order term.}
\label{tab:criteria}
\end{table}

For instance, the optimized correction vanishes at $\eps=1/2$. The actual-marginal correction vanishes only at $\eps=3/4$, where $\eta(\eps)=1/2$; it is negative at $\eps=1/2$. For $\eta(\eps)$ defined in \eqref{eq:one-shot-ordering}, every fixed $0<\eps<1$ satisfies $\eta(\eps)<\eps$, so the actual-marginal requirement reduces the second-order code size. This is consistent with \eqref{eq:private-ordering}. The distinction concerns the specified fidelity criteria; the paper makes no same-parameter claim for separately bounded decoding and trace-distance secrecy errors.

\subsection{A finite-block illustration}
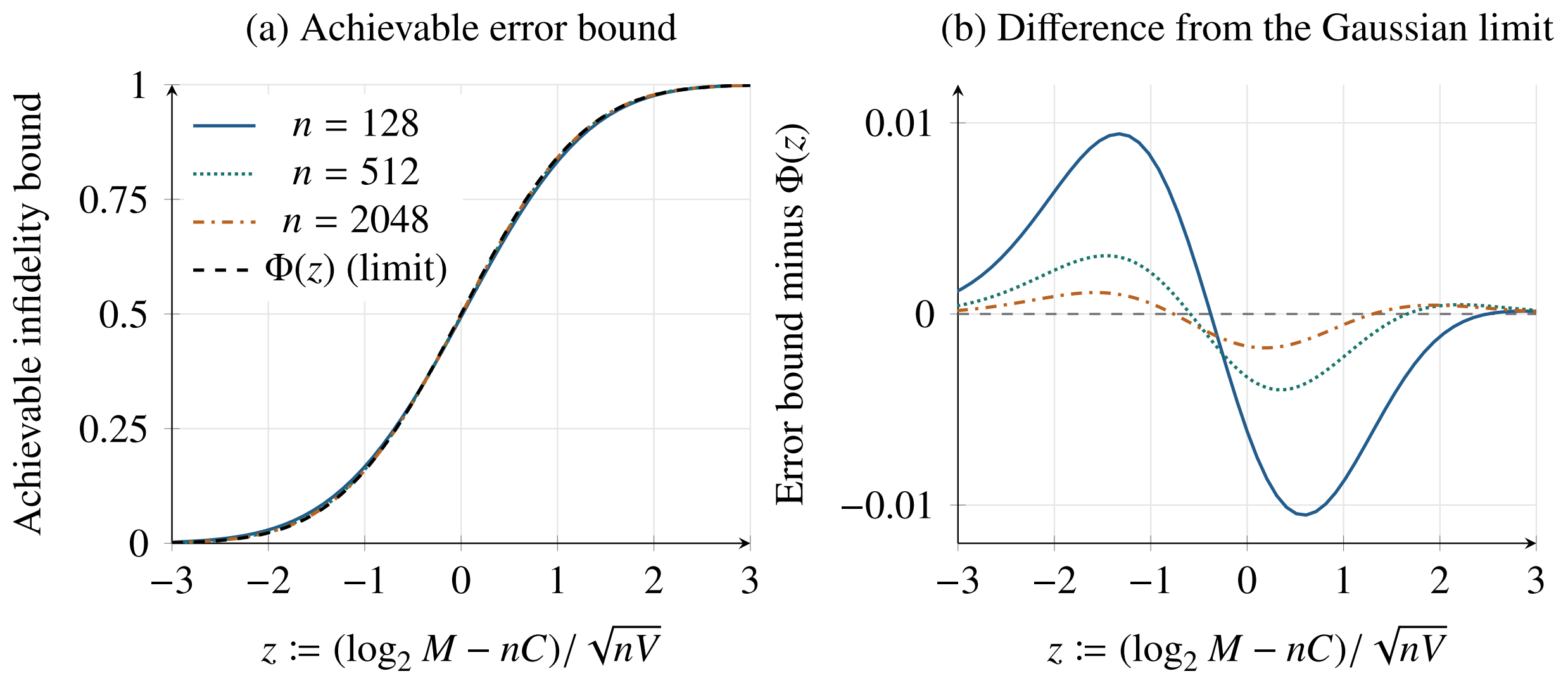
\begin{figure}[t]
\centering
\resizebox{\textwidth}{!}{
\begin{tikzpicture}
\begin{groupplot}[
  group style={group size=2 by 1,horizontal sep=2.0cm},
  width=7.15cm,height=6.0cm,
  tick label style={font=\small},label style={font=\small},
  title style={font=\small},
  xlabel={$z\coloneqq(\log_2M-nC)/\sqrt{nV}$},
  xmin=-3,xmax=3,xtick={-3,-2,-1,0,1,2,3},
  axis lines=left,
  grid=major,grid style={black!10},
  every axis plot/.append style={line width=.9pt},
]
\nextgroupplot[
  title={(a) Achievable error bound},
  ylabel={Achievable infidelity bound},
  ymin=0,ymax=1,ytick={0,.25,.5,.75,1},
  legend style={font=\small,draw=none,fill=white,at={(.02,.98)},anchor=north west,
    inner sep=2pt,row sep=0pt},
]
\addplot[erasureBlue] table[x=z,y=error,col sep=comma]{figures/gaussian_n128.csv};
\addlegendentry{$n=128$}
\addplot[erasureTeal,densely dotted] table[x=z,y=error,col sep=comma]{figures/gaussian_n512.csv};
\addlegendentry{$n=512$}
\addplot[erasureOrange,dash dot] table[x=z,y=error,col sep=comma]{figures/gaussian_n2048.csv};
\addlegendentry{$n=2048$}
\addplot[black,dashed] table[x=z,y=normal,col sep=comma]{figures/gaussian_normal.csv};
\addlegendentry{$\Phi(z)$ (limit)}
\nextgroupplot[
  title={(b) Difference from the Gaussian limit},
  ylabel={Error bound minus $\Phi(z)$},
  ymin=-.012,ymax=.012,ytick={-.01,0,.01},
  scaled y ticks=false,
  yticklabel style={/pgf/number format/fixed,/pgf/number format/precision=2},
]
\addplot[black!55,dashed,line width=.6pt] coordinates {(-3,0)(3,0)};
\addplot[erasureBlue] table[x=z,y=difference,col sep=comma]{figures/gaussian_n128.csv};
\addplot[erasureTeal,densely dotted] table[x=z,y=difference,col sep=comma]{figures/gaussian_n512.csv};
\addplot[erasureOrange,dash dot] table[x=z,y=difference,col sep=comma]{figures/gaussian_n2048.csv};
\end{groupplot}
\end{tikzpicture}}
\caption{A rigorous achievable bound and its Gaussian limit for $d=2$, $p=1/4$, and $n\in\{128,512,2048\}$. Panel (a) plots \eqref{eq:plot-bound}, evaluated at integer code exponents $m=\log_2M$, and the normal cumulative distribution function. Panel (b) shows their difference, which is difficult to see on the scale of panel (a). For each plotted $M$, a code exists whose infidelity is no larger than the corresponding finite-block curve; its distance from the optimal value is not determined here. The constants $C,V$ are defined in \eqref{eq:qv}. The plotted data, parameters, and a log-domain reproduction script are included with the source.}
\label{fig:gaussian}
\end{figure}

Figure~\ref{fig:gaussian} compares the rigorous achievable infidelity bound
\begin{equation}\label{eq:plot-bound}
\eps_{\mathrm{Haar}}(n,M)
\coloneqq1-\sum_{k=0}^{n}\binom nk p^k(1-p)^{n-k}\frac1{1+a_k}
\end{equation}
with its Gaussian limit. Here $a_k$ is defined in \eqref{eq:ak}, and the parameter values are $d=2$ and $p=1/4$. For each plotted integer exponent $m$, define the code dimension $M\coloneqq2^m$ and horizontal coordinate $z\coloneqq(m-nC)/\sqrt{nV}$, with $C,V$ from \eqref{eq:qv}. The curve $\Phi(z)$ is the asymptotic error prediction. It is not a finite-block converse or an assertion that the displayed code bound is optimal at those blocklengths.

The limit of the plotted expression can also be justified in both directions. For general fixed $0<p<1/2$ and integer $d\ge2$, set $D\coloneqq d^n$ and define $z\coloneqq(\log_2M-nC)/\sqrt{nV}$, $\kappa\coloneqq(n-\log_dM)/2$, and $x_k\coloneqq Md^{2k-n}=d^{2(k-\kappa)}$. Equation~\eqref{eq:ak} gives the exact ratio
\begin{equation}\label{eq:plot-ratio}
\frac{a_k}{x_k}
=\frac{(1-d^{-2k})(1-(MD)^{-1})}{1-D^{-2}}.
\end{equation}
For fixed $0<p<1/2$ and $d\ge2$, this ratio is $1+O(e^{-cn})$ uniformly on $k\ge np/2$ and $1\le M\le D$, for some $c>0$. The complementary event has exponentially small binomial probability. Replacing $a_k/(1+a_k)$ by $x_k/(1+x_k)$ therefore changes the expectation by an exponentially small amount. To see the last statement, write $a_k=(1+\theta_k)x_k$ with $|\theta_k|\le1/2$. Direct subtraction gives
\begin{align}
\left|\frac{a_k}{1+a_k}-\frac{x_k}{1+x_k}\right|
&=\frac{|\theta_k|x_k}{(1+a_k)(1+x_k)},\\
&\le\frac{2|\theta_k|x_k}{(1+x_k)^2}
\le2|\theta_k|.
\label{eq:plot-logistic-perturbation}
\end{align}
Here $1+a_k\ge(1+x_k)/2$ and $x_k/(1+x_k)^2\le1$. On the complementary event $k<np/2$, the difference is at most one, so its contribution is bounded by the exponentially small tail in \eqref{eq:binomial-exponential}.

The remaining comparison is with a step at $\kappa$:
\begin{subequations}\label{eq:plot-step}
\begin{align}
\left|\frac{x_k}{1+x_k}-\one\{k>\kappa\}\right|
&\le d^{-2|k-\kappa|},\\
\sum_{k\in\bbZ}d^{-2|k-\kappa|}
&\le\frac2{1-d^{-2}}.
\end{align}
\end{subequations}
For the first inequality, if $k\le\kappa$ use $x_k/(1+x_k)\le x_k$, while if $k>\kappa$ use $1/(1+x_k)\le1/x_k$. The second inequality sums a geometric series on each side of the real threshold. Multiplying by the peak binomial mass from \eqref{eq:binomial-peak} yields
\begin{align}
\left|\E\!\left[\frac{x_{\Ner}}{1+x_{\Ner}}\right]
-\Prb\{\Ner>\kappa\}\right|
&\le\sum_{k=0}^n\pi_{n,k}d^{-2|k-\kappa|},\\
&\le\frac{2C_p}{(1-d^{-2})\sqrt n}.
\label{eq:plot-step-average}
\end{align}
Here $\pi_{n,k}$ is the binomial weight defined in \eqref{eq:binomial-count-weights}. Combining this estimate with \eqref{eq:plot-logistic-perturbation} gives
\begin{equation}\label{eq:plot-limit}
\eps_{\mathrm{Haar}}(n,M)
=\Prb\{\Ner>\kappa\}+O(n^{-1/2})
=\Phi(z)+O(n^{-1/2}).
\end{equation}
For the last equality, $(\kappa-np)/\sqrt{np(1-p)}=-z$ and the uniform Berry--Esseen estimate \eqref{eq:berry-esseen} in Lemma~\ref{lem:binomial-tools} applies. In particular, \eqref{eq:plot-limit} justifies the Gaussian comparison throughout every fixed bounded interval of the plotted coordinate.

\FloatBarrier

The comparison also distinguishes the two remainders in Theorem~\ref{thm:main}. The achievable transition is already narrow enough for a bounded remainder in the transmitted logarithmic dimension. The converse loss $h_n$ in \eqref{eq:hn} comes from controlling arbitrary codes through the signed subset matrix. Its constants were chosen for a transparent asymptotic proof and do not give a tight numerical benchmark at the blocklengths shown in the figure.

\subsection{Perspective and open questions}\label{app:outlook}
An averaged Singleton bound for privacy tests gives the erasure channel's second-order quantum and private coding rates for fixed $0<p<1/2$. Its loss is smaller than the $\sqrt n$ fluctuation scale. The converse allows arbitrary mixed encoders and collective decoders, covers every fixed $0<\eps<1$, and sharpens Cheng's $o(\sqrt n)$ quantum remainder for $0<\eps<1/2$~\cite[Eq.~(78)]{ChengDiscrimination2026}. Quantum achievability and the exact flag-fidelity identities give the matching Gaussian terms. The changed quantile for the actual environment marginal shows that the private error convention remains consequential beyond vanishing-error capacity.

The third-order term remains open. The polynomial has degree $O(\ln n)$, and its absolute row sum has logarithm $O((\ln n)^2)$. Improving the balance between approximation, support, and row sums could reduce the converse loss; an $O(\log n)$ remainder and the necessity of the maximal-message expurgation loss are unresolved. At $p=1/2$, the auxiliary one-position eigenvalue approaches $-1$ for the typical subset sizes, while elementary quantum bounds distinguish errors below and above one half. Appendix~\ref{app:boundaries} gives the separate boundary results.

Further questions concern efficient explicit codes attaining the Gaussian term, sharper finite-block converses at moderate $n$, and extensions to other flagged channels. Such extensions require an overlap estimate and an averaging matrix adapted to the pattern structure. Second-order rates under separate reliability and secrecy constraints also remain distinct from the joint-fidelity criteria of Section~\ref{subsec:private-definition}.

\section{Binomial estimates and fixed-error quantiles}\label{app:probability}
The coding proof uses three properties of the binomial distribution: a uniform normal approximation, an upper bound on its largest point probability, and exponential tails away from its mean. We state all three with their parameter dependence and prove the two elementary estimates. We then record the quantile inversion used in the converse.

\Needspace{5\baselineskip}
\begin{lemma}[Binomial estimates]\label{lem:binomial-tools}
Fix $0<p<1$, set $q\coloneqq1-p$, and let $\Ner\sim\Bin(n,p)$ for an integer $n\ge1$. Let $\Phi$ be the standard normal cumulative distribution function in \eqref{eq:normal-cdf}. There are finite constants $B_p,C_p$, depending only on $p$, such that for every $n\ge1$,
\begin{equation}\label{eq:berry-esseen}
\sup_{x\in\bbR}\left|
\Prb\!\left\{\frac{\Ner-np}{\sqrt{npq}}\le x\right\}-\Phi(x)
\right|\le\frac{B_p}{\sqrt n},
\end{equation}
and
\begin{equation}\label{eq:binomial-peak}
\max_{j\in\bbZ}\Prb\{\Ner=j\}\le\frac{C_p}{\sqrt n}.
\end{equation}
One can take $C_p\coloneqq\sqrt\pi/(2\sqrt{2pq})$. Define the binary relative entropy by
\begin{equation}\label{eq:binary-relative-entropy}
\mathsf D(x\|p)\coloneqq x\ln\frac xp+(1-x)\ln\frac{1-x}{1-p}
\qquad(0<x<1).
\end{equation}
It is positive for $x\ne p$. For every $0<a<p<b<1$,
\begin{equation}\label{eq:binomial-exponential}
\Prb\{\Ner\le na\}\le e^{-n\mathsf D(a\|p)},
\qquad
\Prb\{\Ner\ge nb\}\le e^{-n\mathsf D(b\|p)}.
\end{equation}
The thresholds in these inequalities may be real; the events have their usual integer-valued meaning.
\end{lemma}
\begin{proof}
For the first statement we use the Berry--Esseen theorem~\cite{Berry1941,Esseen1942}. A Bernoulli random variable with mean $p$ has variance $pq$ and centered third absolute moment $pq(p^2+q^2)$. The theorem therefore gives \eqref{eq:berry-esseen} with $B_p\coloneqq C_{\mathrm{BE}}(p^2+q^2)/\sqrt{pq}$ for a universal finite constant $C_{\mathrm{BE}}$. Its uniformity over real $x$ includes thresholds between adjacent integers.

For the point-probability bound, Fourier inversion of the binomial characteristic function gives
\begin{equation}\label{eq:fourier-inversion}
\Prb\{\Ner=j\}=\frac1{2\pi}\int_{-\pi}^{\pi}(q+pe^{iu})^n e^{-iju}\,du.
\end{equation}
To verify the inversion directly, expand $(q+pe^{iu})^n$ by the binomial theorem. The integral of $e^{i(k-j)u}$ over $[-\pi,\pi]$ vanishes unless $k=j$, when it equals $2\pi$.
Since $|q+pe^{iu}|^2=1-4pq\sin^2(u/2)$ and $1-v\le e^{-v}$ for $v\ge0$,
\begin{equation}\label{eq:fourier-modulus}
|q+pe^{iu}|^n\le e^{-2npq\sin^2(u/2)}.
\end{equation}
On $|u|\le\pi$, concavity of sine on $[0,\pi/2]$ gives $\sin(|u|/2)\ge|u|/\pi$. Taking absolute values in \eqref{eq:fourier-inversion} and extending a Gaussian integral to the real line yields
\begin{align}
\Prb\{\Ner=j\}
&\le\frac1{2\pi}\int_{-\pi}^{\pi}e^{-2npq u^2/\pi^2}\,du\\
&\le\frac1{2\pi}\sqrt{\frac{\pi^3}{2npq}}
=\frac{\sqrt\pi}{2\sqrt{2npq}}.
\label{eq:peak-proof}
\end{align}
This proves \eqref{eq:binomial-peak}.

Finally, $\E e^{s\Ner}=(q+pe^s)^n$. For $s<0$, Markov's inequality on the event $\Ner\le na$ gives
\begin{equation}\label{eq:chernoff-step}
\Prb\{\Ner\le na\}\le\exp\{n[\ln(q+pe^s)-sa]\}.
\end{equation}
To minimize the exponent, define $g_a(s)\coloneqq\ln(q+pe^s)-as$. Its derivatives are
\begin{align}
g_a'(s)&=\frac{pe^s}{q+pe^s}-a,\\
g_a''(s)&=\frac{pqe^s}{(q+pe^s)^2}>0.
\label{eq:chernoff-convexity}
\end{align}
The unique minimizer is therefore
\begin{equation}\label{eq:chernoff-optimizer}
s_a\coloneqq\ln\frac{aq}{p(1-a)}<0.
\end{equation}
At this value $q+pe^{s_a}=q/(1-a)$, and direct substitution gives
\begin{align}
g_a(s_a)
&=\ln\frac q{1-a}-a\ln\frac{aq}{p(1-a)},\\
&=-a\ln\frac ap-(1-a)\ln\frac{1-a}{q}
=-\mathsf D(a\|p).
\label{eq:chernoff-optimized-exponent}
\end{align}
For the upper tail, Markov's inequality with $s>0$ gives
\begin{equation}\label{eq:chernoff-upper-step}
\Prb\{\Ner\ge nb\}\le\exp\{n[\ln(q+pe^s)-sb]\}.
\end{equation}
The same derivative calculation has minimizer $s_b\coloneqq\ln[bq/(p(1-b))]>0$ and minimum $-\mathsf D(b\|p)$. This proves both bounds in \eqref{eq:binomial-exponential}. Finally,
\begin{align}
\frac{d}{dx}\mathsf D(x\|p)
&=\ln\frac{xq}{p(1-x)},\\
\frac{d^2}{dx^2}\mathsf D(x\|p)
&=\frac1{x(1-x)}>0.
\label{eq:binary-relative-entropy-convexity}
\end{align}
Together with $\mathsf D(p\|p)=0$, these identities show that the relative entropy has its unique minimum at $p$, proving its strict positivity for $x\ne p$.
\end{proof}

\begin{lemma}[Quantile inversion with a bounded remainder]\label{lem:quantile}
Fix $0<p<1$ and $0<\alpha<1$. For each integer $n\ge1$, let $\Ner\sim\Bin(n,p)$. Let $(a_n)_n$ be a real sequence, let $\Phi$ be defined in \eqref{eq:normal-cdf}, and suppose
\begin{equation}\label{eq:quantile-assumption}
\Prb\{\Ner\le a_n\}\ge\alpha-\delta_n,
\qquad 0\le\delta_n=O(n^{-1/2}).
\end{equation}
Then
\begin{equation}\label{eq:quantile-result}
a_n\ge np+\sqrt{np(1-p)}\,\Phi^{-1}(\alpha)-O(1),
\end{equation}
where the constant may depend on $p,\alpha$ and the bound on $\delta_n$.
\end{lemma}
\begin{proof}
Apply Lemma~\ref{lem:binomial-tools}, specifically \eqref{eq:berry-esseen}, at the real value $(a_n-np)/\sqrt{np(1-p)}$. It gives
\begin{equation}\label{eq:quantile-normal}
\Phi\!\left(\frac{a_n-np}{\sqrt{np(1-p)}}\right)
\ge\alpha-\delta_n-B_pn^{-1/2}.
\end{equation}
Choose a closed interval $I\subset(0,1)$ whose interior contains $\alpha$. For sufficiently large $n$, the right side of \eqref{eq:quantile-normal} lies in $I$. The derivative of $\Phi^{-1}$ is $1/\phi(\Phi^{-1}(u))$, which is continuous and bounded on $I$. Let $L_I$ be such a bound. Monotonicity and the mean value theorem imply
\begin{equation}\label{eq:inverse-perturbation}
\frac{a_n-np}{\sqrt{np(1-p)}}
\ge\Phi^{-1}(\alpha)-L_I(\delta_n+B_pn^{-1/2}).
\end{equation}
After multiplication by $\sqrt{np(1-p)}$, the last term is bounded uniformly in $n$. This proves \eqref{eq:quantile-result}.
\end{proof}

The achievable bound uses the other direction of this local quantile reasoning: increasing an integer threshold by a fixed constant changes its cumulative probability by order $n^{-1/2}$ near an interior normal quantile. Equations~\eqref{eq:ach-C}--\eqref{eq:normal-gain} give that direction with explicit inequalities, including the rounding needed to make the code dimension an integer.

\section{Parameter boundaries}\label{app:boundaries}
The main theorems concern fixed $0<p<1/2$. Throughout this appendix, $n\ge1$ is the number of channel uses, $d\ge2$ is the integer input dimension, and the optimal logarithmic code sizes are those in \eqref{eq:operational-Q} and \eqref{eq:operational-P}. We use $\star\in\{\et,\eg\}$ for the two quantum tasks. We record what the same tools show outside this range. The bound for $p>1/2$ covers entanglement transmission, entanglement generation, and all three private criteria. At $p=1/2$ and at the endpoints, the statements below concern entanglement transmission and entanglement generation.

\subsection{Erasure probabilities above one half}
When $p>1/2$, the set of surviving systems, rather than the erased set, typically has size below $n/2$. Applying the same subset construction to survivor sets gives a useful extension of the converse.

\begin{proposition}[Zero second-order coefficient above one half]\label{prop:above-half}
Fix an integer $d\ge2$, $1/2<p<1$, and $0<\eps<1$. For the optimal logarithmic code sizes in \eqref{eq:operational-Q} and \eqref{eq:operational-P},
\begin{equation}\label{eq:above-half}
0\le L_n^\eps=O((\ln n)^2),\qquad
L_n^\eps\in\{Q_{n,\et}^{\eps},Q_{n,\eg}^{\eps},
P_{n,\opt}^{\eps},P_{n,\marg}^{\eps},P_{n,\max}^{\eps}\}.
\end{equation}
The constants depend only on $d,p,\eps$. Consequently, every listed quantity satisfies $L_n^\eps/\sqrt n\to0$.
\end{proposition}
\begin{proof}
Set $q\coloneqq1-p<1/2$, choose $q<p_0<1/2$, and define $t,\ell_n,h_n$ by \eqref{eq:filter-constants}, \eqref{eq:filter-row-budget}, and \eqref{eq:hn}. Condition on a survivor set $S\subseteq[n]$ of size $s$. Its decoder acts on $A_S$. Relabel the erased-set projectors from Lemma~\ref{lem:lift} by $\mathsf P_S\coloneqq P_{S^c}$. For distinct survivor sets $S,T$, the common receiver input is $A_{S\cap T}$, so Lemma~\ref{lem:overlap} gives
\begin{equation}\label{eq:survivor-overlap}
\norm{\mathsf P_S\mathsf P_T}\le\min\{1,d^{|S\cap T|}/M\}.
\end{equation}
Define the conditional survivor averages by
\begin{align}
\overline F^{\mathrm{surv}}_{n,s}
&\coloneqq\frac1{\binom ns}\sum_{|S|=s}F_{S^c},\\
\overline f^{\mathrm{surv}}_{n,s}
&\coloneqq\frac1{\binom ns}\sum_{|S|=s}f_{S^c}.
\label{eq:survivor-average-definitions}
\end{align}
Here $F_{S^c}$ and $f_{S^c}$ are the conditional fidelities in \eqref{eq:conditional-quantum-definition} and \eqref{eq:conditional-private-definition}.

Apply Lemma~\ref{lem:filter} to the $s$-subsets when $t<s\le p_0n$. Its support obeys $|S\cap T|\le t$, so \eqref{eq:survivor-overlap} supplies the overlap parameter $d^t/M$. Lemmas~\ref{lem:lift} and~\ref{lem:average} then give
\begin{equation}\label{eq:survivor-average}
\overline F^{\mathrm{surv}}_{n,s},\ 
\overline f^{\mathrm{surv}}_{n,s}
\le2n^{-2}+\ell_n\frac{d^t}M.
\end{equation}
If $\log_dM\ge h_n$, substitution of \eqref{eq:hn} shows that
\begin{align}
\ell_n\frac{d^t}M
&\le\ell_nd^{t-h_n},\\
&=d^{\log_d\ell_n+t-(t+\log_d\ell_n+4\log_dn)}
=n^{-4}.
\label{eq:survivor-threshold-algebra}
\end{align}
Hence both averages in \eqref{eq:survivor-average} are at most $3n^{-2}$ on the permitted range.

The survivor count $S_n$ has distribution $\Bin(n,q)$. The excluded probability
\begin{equation}\label{eq:survivor-exceptional-probability}
\tau_n^{\mathrm{surv}}\coloneqq
\Prb\{S_n\le t\}+\Prb\{S_n>p_0n\}
\end{equation}
is exponentially small by \eqref{eq:binomial-exponential}, because $t=o(n)$ and $q<p_0$. Averaging \eqref{eq:survivor-average} over $S_n$, and bounding every excluded conditional fidelity by one, gives
\begin{equation}\label{eq:survivor-total-converse}
F_{\et},\ F_{\eg},\ f_{\opt}
\le3n^{-2}+\tau_n^{\mathrm{surv}}
\quad\text{if }\log_dM\ge h_n.
\end{equation}
The private bound again uses Lemma~\ref{lem:flags} before taking the exact mixture. For fixed $\eps<1$, choose $n$ large enough that the right side of \eqref{eq:survivor-total-converse} is smaller than $1-\eps$. Every admissible code with the required fidelity must then satisfy
\begin{equation}\label{eq:survivor-size-bound}
\log_2M<h_n\log_2d=O((\ln n)^2).
\end{equation}
This proves the upper bounds for both quantum tasks and optimized private communication. The actual-marginal and maximal-message upper bounds follow from \eqref{eq:private-ordering}. The size-$1$ code is feasible with zero error for every task, so all optimal logarithmic sizes are nonnegative. Finally, $h_n/\sqrt n\to0$ gives the stated zero second-order coefficient.
\end{proof}

This proposition does not show that the optimal dimension remains bounded as $n$ grows. It shows that any growth is negligible on the $\sqrt n$ scale. Its proof uses the same subset estimate without a new assumption on the encoder. For the two quantum tasks and $0\le\eps<1/2$, the antidegradability of $\cE_{p,d}$ gives the stronger known bound $M_\star^*(n,\eps;p,d)\le(1-2\eps)^{-1}$ for every $p\ge1/2$~\cite[Corollary~2]{KDWW2021}. Proposition~\ref{prop:above-half} treats all fixed $0<\eps<1$ and also the private criteria; the cited quantum bound does not by itself provide those private conclusions.

\subsection{The balanced channel and the endpoints}
The condition $p<1/2$ enters the main converse through the auxiliary ratio $r=\alpha/(1-\alpha)$ with $\alpha=k/n$, as explained after Lemma~\ref{lem:filter}. A tensor component involving $i$ positions has eigenvalue $(-r)^i$. At $p=1/2$, the typical subset sizes approach $n/2$, so $r$ approaches one and no fixed $r_0<1$ bounds it. The present filter therefore does not prove a matching second-order formula at the balanced channel. There is also a real change in the available error regimes. The entanglement transmission bounds distinguishing errors below and above one half were already identified by Beigi, Datta, and Leditzky~\cite[Section~4]{BDL2016}. The dimension bound \eqref{eq:balanced-dimension} below is the specialization of Ref.~\cite[Corollary~2, Eq.~(198) in the arXiv version]{KDWW2021} to the balanced erasure channel. That result applies to arbitrary antidegradable channels and already includes both quantum tasks considered here. The protocol in Ref.~\cite[Section~IV.C]{KDWW2021} permits a state prepared locally by the sender and local encoding and decoding, so it includes arbitrary mixed reference--input states for entanglement generation as well as entanglement transmission. We retain a direct proof to make the boundary discussion self-contained and to connect it with the overlap lemma.

\begin{proposition}[A bound at the balanced channel]\label{prop:balanced}
Fix integers $n\ge1$, $d\ge2$, and $M\ge1$, and set $p\coloneqq1/2$. For an unassisted entanglement transmission or entanglement generation code of dimension $M$ over $\cE_{1/2,d}^{\otimes n}$, let $F_\star$ denote its Bell-target fidelity from \eqref{eq:transmission-fidelity} or \eqref{eq:generation-fidelity}, with $\star\in\{\et,\eg\}$. Then
\begin{equation}\label{eq:balanced-fidelity}
F_\star\le\frac12\left(1+\frac1M\right).
\end{equation}
Consequently, for $0\le\eps<1/2$,
\begin{equation}\label{eq:balanced-dimension}
M_\star^*(n,\eps;1/2,d)\le\frac1{1-2\eps}.
\end{equation}
Here $M_\star^*$ is defined in \eqref{eq:operational-M}. For fixed $1/2<\eps<1$, the optimal logarithmic dimension $Q_{n,\et}^{\eps}$ from \eqref{eq:operational-Q} instead satisfies
\begin{equation}\label{eq:balanced-lower}
Q_{n,\et}^{\eps}(1/2,d)
\ge\sqrt n\log_2d\,\Phi^{-1}(\eps)-O(1).
\end{equation}
\end{proposition}
Here $\Phi^{-1}$ is the inverse of the standard normal cumulative distribution function defined in \eqref{eq:normal-cdf}.
\begin{proof}
At $p=1/2$, the channel and its complement in \eqref{eq:channel-isometry} are identical after a relabeling. Apply the code's decoder separately to the receiver output and to the environment output in this isometric representation. These are two mathematical recovery tests, not simultaneous operations in the protocol. Their fidelities are equal for the original encoded state because the two marginal channels coincide.

The two tests have disjoint physical inputs and separate auxiliary systems. The common-input space in the proof of Lemma~\ref{lem:overlap} is therefore one dimensional. Their projectors $P_0,P_1$ obey $\norm{P_0P_1}\le1/M$. For two projectors,
\begin{equation}\label{eq:two-projectors}
\norm{P_0+P_1}\le1+\norm{P_0P_1}.
\end{equation}
For an explicit proof, define the linear operator
\begin{equation}\label{eq:two-projector-analysis-operator}
A\ket\psi\coloneqq(P_0\ket\psi,P_1\ket\psi).
\end{equation}
Then
\begin{align}
A^\dagger A&=P_0+P_1,\\
AA^\dagger&=
\begin{pmatrix}P_0&P_0P_1\\P_1P_0&P_1\end{pmatrix}.
\label{eq:two-projector-block-operators}
\end{align}
Set $c\coloneqq\norm{P_0P_1}$. For arbitrary vectors $\ket x,\ket y$ in the original Hilbert space, Cauchy--Schwarz and $2\norm{\ket x}\norm{\ket y}\le\norm{\ket x}^2+\norm{\ket y}^2$ imply
\begin{align}
\left\langle(\ket x,\ket y),AA^\dagger(\ket x,\ket y)\right\rangle
&=\bra xP_0\ket x+\bra yP_1\ket y
+2\operatorname{Re}\bra xP_0P_1\ket y,\\
&\le\norm{\ket x}^2+\norm{\ket y}^2+2c\norm{\ket x}\norm{\ket y},\\
&\le(1+c)(\norm{\ket x}^2+\norm{\ket y}^2).
\label{eq:two-projector-quadratic-estimate}
\end{align}
Thus $\norm{AA^\dagger}\le1+c$. Since $\norm{A^\dagger A}=\norm A^2=\norm{AA^\dagger}$, this proves \eqref{eq:two-projectors}.

Let $\widetilde\rho$ be the normalized encoded state after applying the channel isometry \eqref{eq:channel-isometry} to all $n$ inputs, enlarged by the separate decoder auxiliary systems. Each test has expectation $F_\star$ in this same state. Therefore
\begin{align}
2F_\star
&=\Tr[(P_0+P_1)\widetilde\rho],\\
&\le\norm{P_0+P_1}
\le1+\frac1M.
\label{eq:balanced-common-state-bound}
\end{align}
This establishes \eqref{eq:balanced-fidelity} for an arbitrary, possibly mixed reference--input state. If the infidelity is at most $\eps<1/2$, it follows that
\begin{align}
2(1-\eps)&\le1+\frac1M,\\
M&\le\frac1{1-2\eps}.
\label{eq:balanced-error-rearrangement}
\end{align}
Maximizing over admissible codes gives \eqref{eq:balanced-dimension}.

For $\eps>1/2$, the achievable calculation in Corollary~\ref{cor:normal-lower} remains valid at $p=1/2$, provided its chosen integer threshold is below $n/2$. Now $z_\eps\coloneqq\Phi^{-1}(1-\eps)<0$. Choose the positive constant $c_{\rm ach}$ by \eqref{eq:ach-C} with $p=1/2$, and define
\begin{equation}\label{eq:balanced-achievable-threshold}
u\coloneqq\left\lceil\frac n2+\frac{\sqrt n}{2}z_\eps+c_{\rm ach}\right\rceil.
\end{equation}
Since $z_\eps<0$, the negative term of order $\sqrt n$ eventually exceeds $c_{\rm ach}+1$ in absolute value. Therefore $0\le u<n/2$ for all sufficiently large $n$. Equations~\eqref{eq:ach-tail}--\eqref{eq:ach-error-bound} give infidelity at most $\eps$ for $M\coloneqq d^{n-2u}$. The ceiling in \eqref{eq:balanced-achievable-threshold} yields
\begin{align}
\log_2M
&=(n-2u)\log_2d,\\
&\ge-\sqrt n\,z_\eps\log_2d
-2(c_{\rm ach}+1)\log_2d,\\
&=\sqrt n\log_2d\,\Phi^{-1}(\eps)-O(1).
\label{eq:balanced-achievable-size}
\end{align}
This proves \eqref{eq:balanced-lower}.
\end{proof}

Thus the dimension is bounded at the balanced channel for every fixed infidelity below one half, while an order-$\sqrt n$ logarithmic dimension is achievable above one half. A matching upper bound in the latter range and the case $\eps=1/2$ remain open within the present analysis.

The remaining endpoints have elementary behavior. At $p=0$, a noiseless channel transmits dimension $d^n$ exactly. Equation~\eqref{eq:dimension-bound} gives the upper bound $M\le d^n/(1-\eps)$ even for entanglement generation. Therefore both tasks have
\begin{equation}\label{eq:noiseless}
Q_{n,\star}^{\eps}(0,d)=n\log_2d+O(1).
\end{equation}
At $p=1$, the receiver output is independent of the sender. If $\sigma_{\widehat S}$ is the decoder's output state, then the joint output is $\rho_R\otimes\sigma_{\widehat S}$. Expanding the maximally entangled vector gives
\begin{equation}\label{eq:product-bell-overlap}
\Tr[\Phi^M_{R\widehat S}(\rho_R\otimes\sigma_{\widehat S})]
=\frac1M\Tr(\rho_R^{\mathsf T}\sigma_{\widehat S}),
\end{equation}
where the transpose and the identification of $R$ with $\widehat S$ use the fixed Bell basis. In entanglement transmission, $\rho_R=I_R/M$, so \eqref{eq:product-bell-overlap} equals $1/M^2$ for every decoder output. In entanglement generation, $\rho_R$ is arbitrary. Since $0\le\sigma_{\widehat S}\le I_{\widehat S}$,
\begin{equation}\label{eq:product-bell-maximum}
\frac1M\Tr(\rho_R^{\mathsf T}\sigma_{\widehat S})\le\frac1M.
\end{equation}
Equality is attained when both states are the same real basis state. Requiring the appropriate maximal fidelity to be at least $1-\eps$ therefore gives
\begin{equation}\label{eq:replacer}
M_\et^*(n,\eps;1,d)=\left\lfloor(1-\eps)^{-1/2}\right\rfloor,
\qquad
M_\eg^*(n,\eps;1,d)=\left\lfloor(1-\eps)^{-1}\right\rfloor.
\end{equation}
Table~\ref{tab:scope} collects the parameter ranges without conflating these boundary statements with Theorem~\ref{thm:main}.

\begin{table}[!t]
\centering
\small
\renewcommand{\arraystretch}{1.2}
\begin{tabular}{@{}p{.17\textwidth}p{.45\textwidth}p{.30\textwidth}@{}}
\toprule
Erasure\newline probability & Statement at fixed $0<\eps<1$ & Result and qualification\\
\midrule
$p=0$ & $Q_{n,\star}^{\eps}=n\log_2d+O(1)$ & Equation~\eqref{eq:noiseless}\\
$0<p<1/2$ & $Q_{n,\star}^{\eps}=nC+\sqrt{nV}\,\Phi^{-1}(\eps)+O((\ln n)^2)$ & Theorem~\ref{thm:main}; lower-bound loss $O(1)$\\
$p=1/2$ & Bounded dimension for $\eps<1/2$; positive $\sqrt n$ achievable term for $\eps>1/2$ & Proposition~\ref{prop:balanced}; no matching formula for \mbox{$\eps\ge1/2$}\\
$1/2<p<1$ & $Q_{n,\star}^{\eps}=O((\ln n)^2)$ & Proposition~\ref{prop:above-half}; \mbox{$Q_{n,\star}^{\eps}/\sqrt n\to0$}\\
$p=1$ & Bounded dimensions, different for entanglement transmission and entanglement generation & Equation~\eqref{eq:replacer}\\
\bottomrule
\end{tabular}
\caption{Scope of the results for unassisted entanglement transmission and entanglement generation, with $\star\in\{\et,\eg\}$. All asymptotic statements hold for fixed parameters. In the positive-capacity range, $C$ and $V$ are given by \eqref{eq:qv}. The table does not assert uniformity near $p=1/2$ or near the error endpoints.}
\label{tab:scope}
\end{table}

\FloatBarrier

\end{document}